\documentclass{article}

\usepackage{stylesheet}

\usepackage[utf8]{inputenc} % allow utf-8 input
\usepackage[T1]{fontenc}    % use 8-bit T1 fonts
\usepackage{hyperref}       % hyperlinks
\usepackage{url}            % simple URL typesetting
\usepackage{booktabs}       % professional-quality tables
\usepackage{amsfonts}       % blackboard math symbols
\usepackage{nicefrac}       % compact symbols for 1/2, etc.
\usepackage{microtype}      % microtypography
\usepackage{lipsum}
\usepackage{fancyhdr}       % header
\usepackage{graphicx}       % graphics
\usepackage{caption}
\usepackage{subcaption}
\usepackage{listings}
\usepackage{float}
\usepackage{pdflscape}
\graphicspath{{media/}}     % organize your images and other figures under media/ folder
\usepackage{amssymb}% http://ctan.org/pkg/amssymb
\usepackage{pifont}% http://ctan.org/pkg/pifont
\newcommand{\cmark}{\ding{51}}%
\newcommand{\xmark}{\ding{55}}%
\newcommand{\NA}{---}%
\usepackage[usenames,dvipsnames]{xcolor}

\title{Taking Advice from ChatGPT
\thanks{\textit{\underline{Citation}}: 
\textbf{Zhang, P. (2023). Taking Advice from ChatGPT.}} 
}

\author{
  Peter Zhang \thanks{\textit{\underline{Advised by}}: 
Professor Don Moore, Haas School of Business
}\\
  UC Berkeley \\
  Berkeley, CA\\
  \texttt{petez@berkeley.edu} \\
}

\newcommand{\WoA}{\mathrm{WoA}}
\newcommand{\BS}{\mathrm{BS}}
\newcommand{\ECE}{\mathrm{ECE}}
\newcommand{\AC}{\mathrm{AC}}

\begin{document}
\maketitle

%%%%%%%%%%%%%%%%%%%%%%%
% Abstract
%%%%%%%%%%%%%%%%%%%%%%%

\begin{abstract}
A growing literature studies how humans incorporate advice from algorithms. This study examines an algorithm with millions of daily users: ChatGPT. In a preregistered study, 118 student participants answer 2,828 multiple-choice questions across 25 academic subjects. Participants receive advice from a GPT model and can update their initial responses. The advisor’s identity (``AI chatbot'' versus a human ``expert''), presence of a written justification, and advice correctness do not significantly affect weight on advice. Instead, participants weigh advice more heavily if they (1) are unfamiliar with the topic, (2) used ChatGPT in the past, or (3) received more accurate advice previously. The last two effects---algorithm familiarity and experience---are stronger with an AI chatbot as the advisor. Participants that receive written justifications are able to discern correct advice and update accordingly. Student participants are miscalibrated in their judgements of ChatGPT advice accuracy; one reason is that they significantly misjudge the accuracy of ChatGPT on 11/25 topics. Participants \textit{under-weigh} advice by over 50\% and can score better by trusting ChatGPT more.
\end{abstract}

% keywords can be removed

%%%%%%%%%%%%%%%%%%%%%%%
% Introduction
%%%%%%%%%%%%%%%%%%%%%%%

\section{Introduction}
\label{sec:intro}

In late 2022, ChatGPT showed the world the power of large language models (LLMs) \cite{zhang2023one}. ChatGPT is a generative pretrained language model developed by OpenAI, an AI research lab. AI chatbots like ChatGPT and its cousins (BingChat, Bard, Jasper) achieve ``surprisingly superior performance'' \cite{zhou2023comprehensive} due to an instruction-tuning process that teaches them to do what humans want \cite{stiennon2020learning, christiano2017deep}. Combined with pre-training at scale, LLMs are powerful interfaces for accessing knowledge \cite{kasneci2023chatgpt, razniewski2021language}.

The most recent model GPT-4, which now underlies ChatGPT Plus, is much more powerful \cite{bubeck2023sparks} and has been rigorously benchmarked on a variety of academic tests. According to OpenAI’s internal testing, GPT-4 outperforms the median human test-taker on SATs, LSATs, GREs, and several AP exams \cite{openai2023gpt4}. Other researchers have found that ChatGPT can pass the bar \cite{choi2023chatgpt}, achieve medical certifications \cite{gilson2023does, mbakwe2023chatgpt, fijavcko2023can, kemp2023chatgpt}, and even complete a college physics class \cite{kortemeyer2023could}. 

The novel accessibility and broad capabilities of AI chatbots are likely to reshape education \cite{kasneci2023chatgpt}. Many educators are scrambling to reconcile with ChatGPT with responses ranging from outright bans in school to welcome integration into curricula \cite{tlili2023if}. Some point towards risks to testing integrity \cite{susnjak2022chatgpt} and plagiarism \cite{cotton2023chatting}, while others argue that it provides personalized and immediate information \cite{sallam2023chatgpt, firat2023chat}. A recent meta-analysis finds that ``the number of papers that see ChatGPT as a threat is almost equal to the number of those that view it as opportunity'' \cite{leiter2023chatgpt}. Others still are rethinking traditional views of academic integrity and encouraging uses such as co-authorship \cite{dwivedi2023so, polonsky2023should, anders2023using}.

The multiple choice (MC) exam is particularly vulnerable to a rethinking. MC questions continue to be a predominant format for assessing understanding, analysis and recall \cite{kumar2023novel}. The strength of AI chatbots on multiple choice exams is worrying \cite{newton2023chatgpt} because students most commonly cheat by consulting online sources \cite{burgason2019cheating}. While some have suggested workarounds \cite{cotton2023chatting}, the fast-paced evolution of the underlying LLMs means that it ``may not be long before [these] models become so intelligent that we can no longer exploit their weaknesses'' \cite{gonsalves2023chatgpt}.

This study seeks to document how students use information from ChatGPT on MC tests, contributing to a largely qualitative literature on how students empirically interact with AI chatbots \cite{haensch2023seeing}. While it takes MC tests as a starting point, the work has implications for broader research on algorithm aversion and appreciation, as well as on human-AI collaboration. The study is guided by two questions: First, what influences the weight humans place on chatbot advice? Second, are humans good at judging when AI chatbot advice is correct?

%%%%%%%%%%%%%%%%%%%%%%%
% Literature review
%%%%%%%%%%%%%%%%%%%%%%%

\section{Literature Review}
\label{sec:litreview}

A rich literature examines how people take advice from algorithms. Two core competing findings are algorithm aversion \cite{dietvorst2015algorithm} (a tendency to disproportionately punish algorithms when they err) and algorithm appreciation \cite{logg2019algorithm} (a tendency to prefer algorithm advice prima facie). Numerous studies have explored mediating mechanisms, including task objectivity \cite{castelo2019task}, perceived competence \cite{hou2021expert}, human input \cite{dietvorst2018overcoming}, learning \cite{berger2021watch, reich2022overcome}, and time pressure \cite{jung2021towards}, among others \cite{castelo2019task}. One literature review categorizes these effects into algorithm characteristics (agency, performance, capabilities, and human involvement) and human characteristics (expertise and social distance) \cite{jussupow2020we}. Another analyzes broad themes of expectations and expertise, decision autonomy, incentivization, cognitive compatibility, and divergent rationalities \cite{burton2020systematic}. Five types of explanations are relevant to this study.

This study is a direct test of explanations about \textit{social distance}. If algorithm aversion is truly a preference for humans, a natural remedy is to make algorithms more human-like \cite{morewedge2022preference}. Both adjacent literature \cite{hessler2022self} and experimental evidence suggests that people are more likely to accept advice from an anthropomorphized algorithm \href{ochmann2020influence}. In the business world, AI chatbots are now successful consumer-facing assistants \cite{luo2019frontiers}, and the their perceived human-likeness is important to their success \cite{schanke2021estimating}. At the same time, other studies suggest that appearing too human can induce aversion if algorithms traverse into an uncanny valley \cite{strait2015too}. ChatGPT’s ability to provide natural language explanations comparable to humans \cite{saha2022hard} may cause humans to treat ChatGPT similarly to a human advisor and distinctly from other algorithms.

Three other explanations are commonly cited. The first, \textit{task difficulty}, suggests that increasing task difficulty causes people to rely more heavily on (algorithmic) advice \cite{gino2007effects, bogert2021humans} and is supported by real-world evidence on teachers \cite{kaufmann2021algorithm}. In this study, familiarity in the question topic an approximate measure of task difficulty. A second explanation, \textit{algorithm familiarity}, reasons that people who are more familiar with using algorithms for some task will be less averse to the advice \cite{castelo2019task, liu2003algorithm}, an effect that was confirmed in a real-world medical context. This study measures algorithm familiarity by asking questions about past usage. The third explanation, \textit{experience}, argues that participants are rationally updating their beliefs about algorithm competence and that presenting their performance can reduce aversion and develop trust over time \cite{you2022algorithmic, filiz2021reducing, cabiddu2022users}, although some studies find that accuracy matters less than expected \cite{alexander2018trust}. This study uses a simple model of participant beliefs about advice accuracy as a measure of experience.

Finally, a developing literature studies the effect of algorithm \textit{interpretability} on aversion. Interpretability is theorized as allowing ``the user to rapidly calibrate their trust in the system's outputs, spotting flaws in its reasoning or seeing when it is unsure'' \cite{tomsett2020rapid}. Studies have found mixed effects of output interpretability \cite{destefano2022providing, altintas2023effect, ahn2021will, ben2021explainable} and model transparency \cite{schmidt2020calibrating, lehmann2022risk}, although field experiments on physicians find that they benefit from explainable AI advice \cite{gaube2023non, panigutti2022understanding}. In this study, providing GPT model's text reasoning enables a test of whether interpretability makes a difference.

Surprisingly few studies have examined the role of ChatGPT as an adviser. Some studies have explored potential problems with using ChatGPT for advice on health \cite{oviedo2023risks, howard2023chatgpt, xie2023aesthetic, nastasi2023does}, investing \cite{george2023review}, and education \cite{firat2023chat}. Empirical studies have documented a corrupting effect of moral advice generated by GPT-2 \cite{leib2023corrupted}, GPT-3\cite{momentrusting}, and ChatGPT \cite{krugel2023chatgpt}. On Twitter, GPT-3 generated texts appear to be more effective at convincing humans to believe (accurate and inaccurate) information \cite{spitale2023ai}. Finally, humans appear to trust robots more with ChatGPT as an interface \cite{ye2023improved}. One recent study studies ChatGPT in the algorithm aversion context on a essay-writing task \cite{bohm2023content}. The authors find that while people may devalue the outputs of ChatGPT relative to a human author, they judge the content equally and are not deterred from sharing.

Little is known about how humans judge the accuracy of ChatGPT. A plethora of studies have shown that humans tend to overestimate their own abilities \cite{moore2008trouble} and misjudge the abilities of others \cite{moore2007overconfidence}. Similarly, studies suggest that LLMs calibration is good after pre-training \cite{chen2022close, kadavath2022language, jiang2021can} but degrades after learning from human feedback \cite{bai2022training}. Yet studies have not evaluated whether humans accurately estimate the accuracy of LLM outputs. One study suggests that humans may become miscalibrated on AI feedback because of misallocation of blame \cite{chong2022human}, the this study and others fail to explicitly document the level and nature of (mis)calibration. This study seeks to fill that gap by evaluating human confidence in LLM outputs on a broad set of topic areas.

%%%%%%%%%%%%%%%%%%%%%%%
% Procedure
%%%%%%%%%%%%%%%%%%%%%%%

\section{Procedure}
\label{sec:procedure}

\begin{figure}
  \centering
  \includegraphics[width=.8\textwidth]{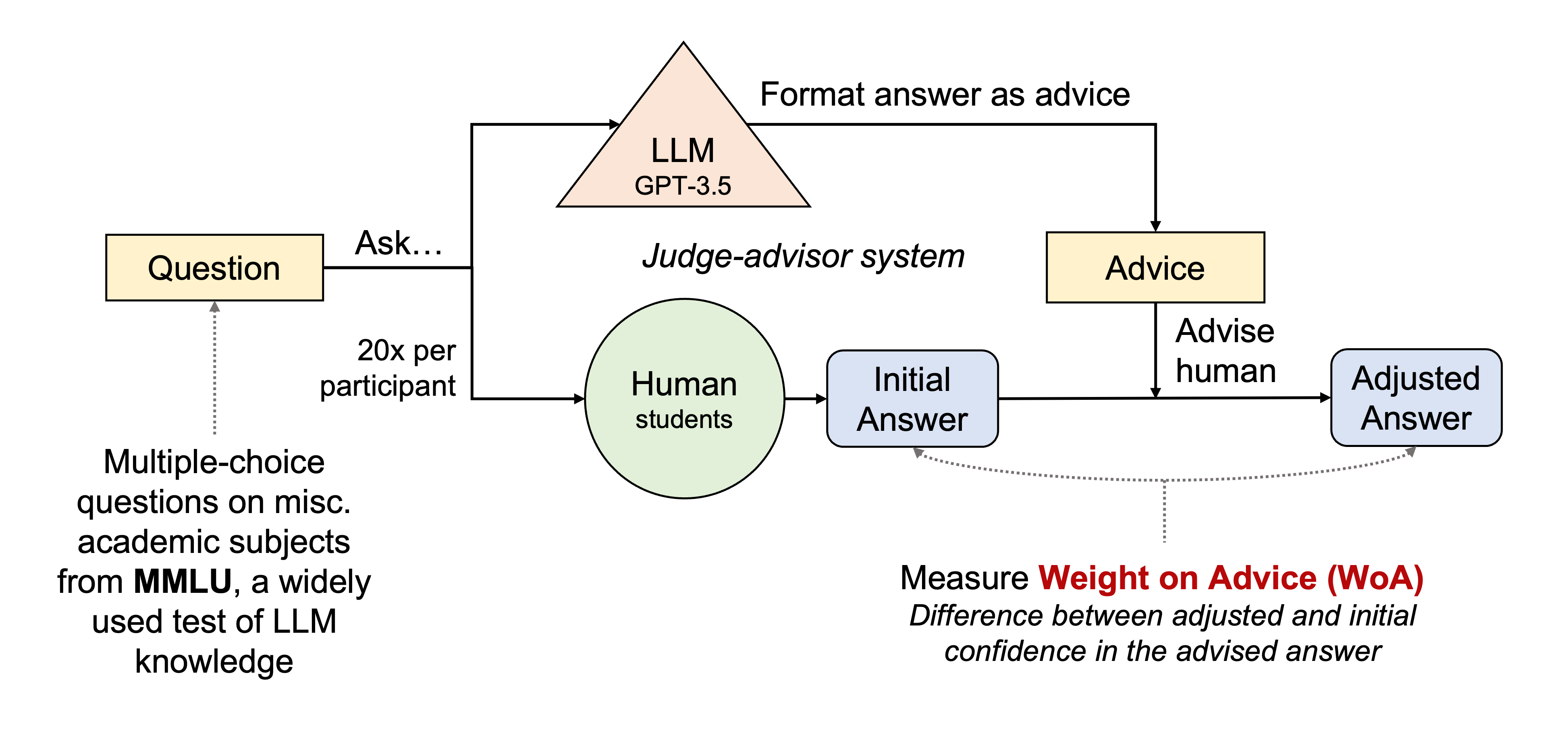}
  \caption{\textbf{Overview of study design.} Both LLMs and human participants answer questions. The study focuses on how humans take LLM advice. }
  \label{fig:study_design}
\end{figure}

\paragraph{Overview}
The study simulates an environment in which students receive aid from ChatGPT. MC questions are sourced from real academic tests and original outputs are obtain by quering GPT models. Participants attempt to make calibrated guesses before and after seeing the advice that is generated. An overview of the study design is displayed in Figure \ref{fig:study_design}. All code and data except for survey responses are documented in the accompanying \href{https://github.com/petezh/ChatGPT-Advice/}{GitHub} repository. The study methods are preregistered on AsPredicted under predictions \href{https://aspredicted.org/K7Y_VPW}{\#122800} and \href{https://aspredicted.org/KW3_8W2}{\#126040}.

\paragraph{Dataset}
Answers are drawn from the Massive Multitask Language Understanding (MMLU) dataset \cite{hendrycks2020measuring}, a widely used benchmark \cite{openai2023gpt4} of LLM knowledge understanding that broadly encompasses academics. The dataset consists entirely of MC questions and draws from real tests such as the Advanced Placement exams. Participants answer questions from only 25 of the original 57 topics, topics which college students are expected to have a reasonable chance to succeed. A total of 688 questions are sampled from the topics. See Appendix \ref{sec:appendix_dataset} for descriptions of the 25 topics and selection procedure.

\begin{table}
    \centering
    \caption{MMLU topics included in this study.}
    \begin{tabular}{lp{0.4\textwidth}p{0.3\textwidth}l}
    \toprule
    Supercategory   & Topics        & Example Question \\
    \midrule
    STEM            & Clinical Knowledge, Physics, Elementary Mathematics, Formal Logic, APs (Biology, Chemistry, Comp. Sci, Physics, Statistics), Human Aging & In which situation can the expression 64 + 8 be used? \\
    Social Science & APs (Human Geo, Government, Macro/Micro, Psych), Sociology, U.S. Foreign Policy, Global Facts & What does Berger (1963) describe as a metaphor for social reality? \\
    Humanities      & APs (US/World/European History), Philosophy, Misc. topics & Descartes argues against trusting the senses on the grounds that \_\_\_\_\_. \\
    \bottomrule
\end{tabular}
    \label{tab:mmlu_topics}
\end{table}

\paragraph{Model evaluation}
The advice is generated by GPT-3.5, a LLM by OpenAI fine-tuned to follow human instructions, on the constructed dataset \cite{ouyang2022training}. Specifically, calls are made to the Completions API with \verb|text-davinci-003| as the engine \footnote{See the \href{https://platform.openai.com/docs/api-reference/completions}{API documentation}.} Models are prompted with standard and chain-of-thought (CoT) prompts \cite{wei2022chain}.  CoT prompts yield the same accuracy but better explanations. The advice used in the survey is generated using a zero-shot CoT prompt. See Appendix \ref{sec:appendix_model_evaluation} for a comparison to standard prompting and illustration of the prompt text.

\begin{figure}
  \centering
  \includegraphics[width=.8\textwidth]{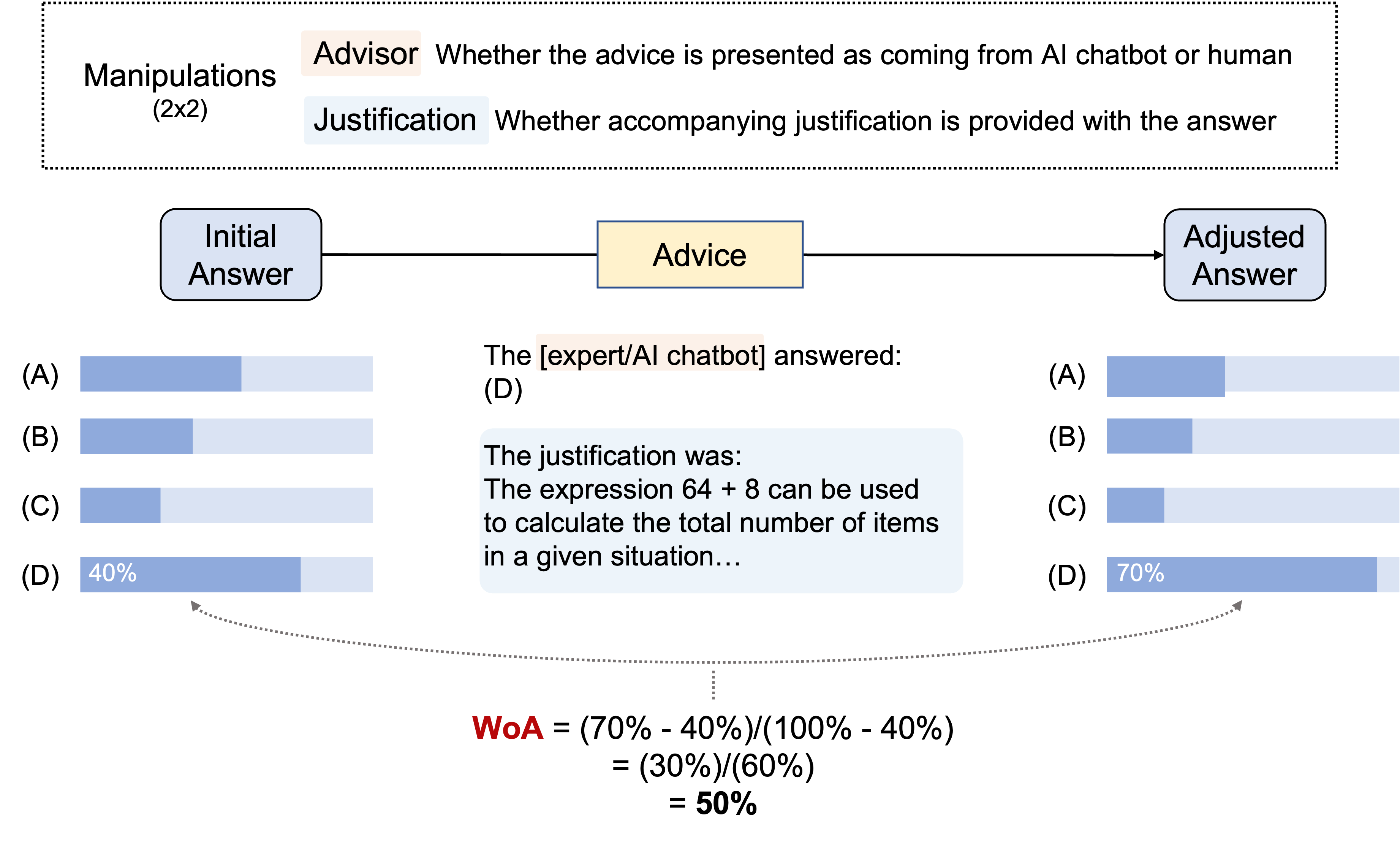}
  \caption{\textbf{Judge-advisor system.} Participants provide judgements about the probability that each answer is true. $WoA$ is a measure of how advice changes the probability allocated to the advised answer.}
  \label{fig:judge_advisor_system}
\end{figure}

\paragraph{Lab experiment}
Participant use this advice is a survey-based lab experiment.\footnote{The experiment is approved under UC Berkeley CPHS Protocol \#2023-03-16125.} The setup models the well-studied judge-advisor system \cite{sniezek2001trust}. Participants are shown randomly selected questions and report their confidence in each answer choice before and after receiving advice. The advice is manipulated by varying the advisor’s identity and selectively providing justifications. Participants receive advice from an advisor randomly identified as a generic ``expert'' or an ``AI chatbot''. They are also randomly assigned to receive a justification in addition to the answer. The manipulations are displayed in Figure \ref{fig:judge_advisor_system}.

\begin{figure}
  \centering
  \includegraphics[width=.8\textwidth]{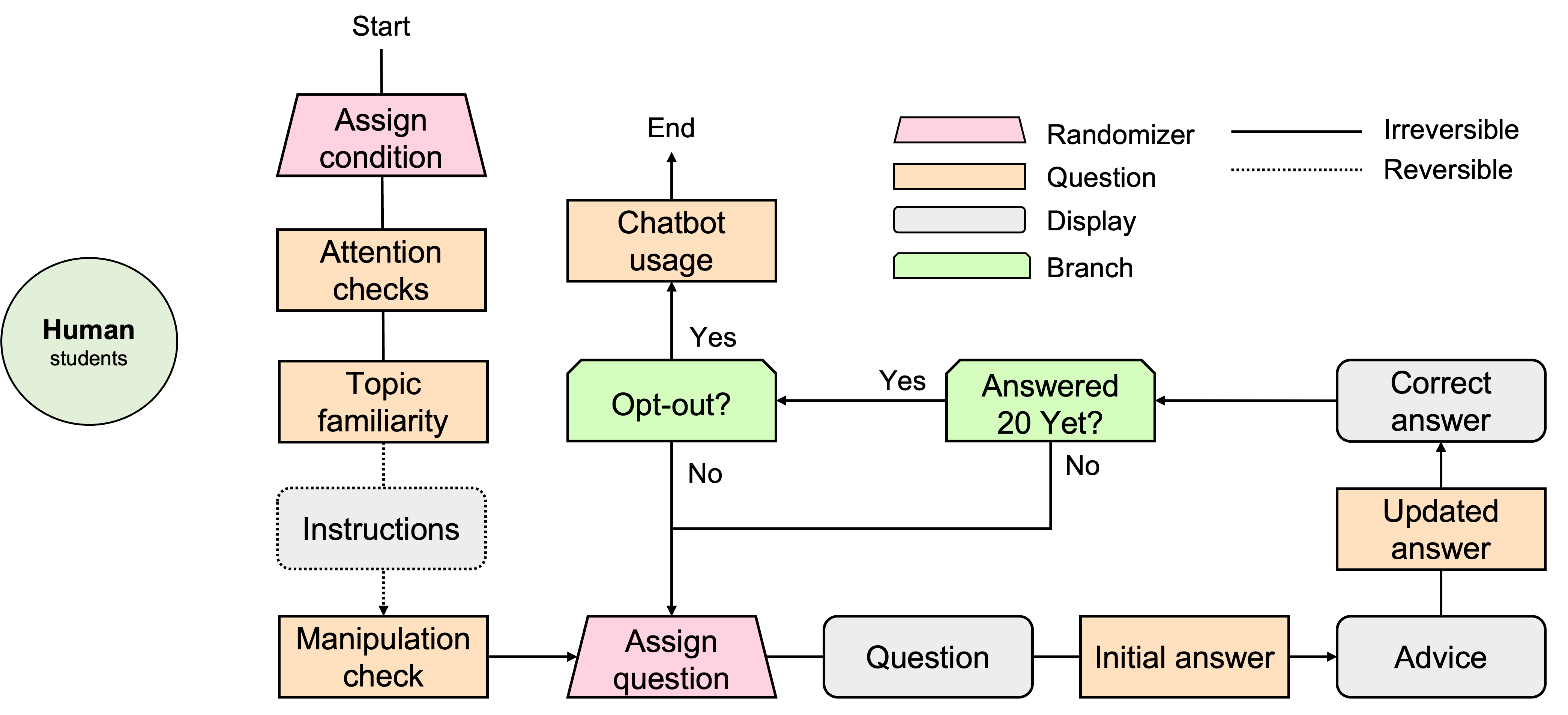}
  \caption{\textbf{Qualtrics survey flow.} Survey blocks are color coded by element type and line pattern corresponds to reversibility. The survey begins with a consent notice and ends with a debriefing. Note that participants are \textit{required} to pass the attention and manipulation checks. Participants may return to previous instruction pages to pass the manipulation check.}
  \label{fig:survey_flow}
\end{figure}

The experiment is administered via a Qualtrics survey. A \href{https://berkeley.qualtrics.com/jfe/form/SV_b3hEwR3snc6Veh8}{live link} and \href{https://drive.google.com/file/d/1_DTL1K_MHPIrn8Bp_7OZmNXczZBTohaV/view?usp=share_link}{full printout} of the survey are available for readers. Participants
\begin{itemize}
    \item are assigned to the conditions;
    \item must pass a simple attention check;
    \item provide their level of familiarity (“comfortable”, “neutral”, or “uncomfortable”) with 8 topic areas that are constructed by grouping topics, as well as their major(s);
    \item complete an example that explains the judge-advisor setup, the concept of confidence, and identifies the advice format (advisor identity and presence of justification);
    \item pass a manipulation check that reinforces the advisor identity;
    \item complete at least 20 questions in which they:
    \begin{itemize}
        \item are assigned a random question and provide an initial answer;
        \item receive advice and update their answer;
        \item discover the correct answer and the points they have earned; and
        \item have the opportunity to out-opt once they have completed 20;
    \end{itemize}
    \item fill out a questionnaire about their usage of ChatGPT; and finally
    \item exit the survey.
\end{itemize}
The survey flow is displayed in Figure \ref{fig:survey_flow}.

\paragraph{Scoring and compensation}
Participants are scored by a point system that rewards accurate and calibrated answers with cash prizes. The system is based on Brier scores, a widely used scoring rule for encouraging both accuracy and calibration \cite{rufibach2010use}. Let $f_{X, \mathrm{init}}$. denote the initial confidence and $f_{X, \mathrm{adj}}$ denote the adjusted confidence for each answer choice $X \in \{A,B,C,D\}$. Let $o_X$ be an indicator for whether $X$ is correct. Then, the score is:

$$\BS(f) = \sum_{X \in \{A,B,C,D\}} \left[(f_X - o_X)^2 \right]$$

The Brier score is scaled to give 0 points to a uniform (25\% across choices) distribution and \textcolor{OliveGreen}{750 points} for a full-confidence correct answer. The score is asymmetric insofar as it penalizes a full-confidence incorrect answer by \textcolor{Maroon}{-1250 points}. The score is centered at 0 so that participants are not able to earn points by merely completing more questions with uniform distributions. The re-scaled scoring rule evenly weights the initial and adjusted forecast:

$$\textrm{Score} = \sum_{f \in \{f_{X, \mathrm{init}}, f_{X, \mathrm{adj}}\}} 750 - 1000 \cdot \BS(f)$$

Participants can earn prizes by (1) placing among the top 5 scorers and earning 10 USD or (2) through a random drawing for 50 USD with tickets proportional to score. The former is designed to reward effort\footnote{Anecdotally, scoring dramatically improves participant engagement. Subjects reported feeling more interested and invested in the questions, particularly compared to a setup that does not reveal the correct answer. The effect appear to be even stronger when the reward is deterministic.} while the latter ensures that payout remains somewhat proportionate to score \cite{roulston2007performance}.

\paragraph{Participants}
A total of 142 undergraduate students at UC Berkeley are recruited through the Research Participant Pool (RPP) at the Haas School of Business. The participants are primarily business majors in their third and fourth years. Six small pilot sessions were conducted from 04/04/2023 to 04/11/2023 to debug the survey. The 12 sessions comprising the study dataset were administered from 04/13/2023 to 04/25/2023 and included 118 participants. All sessions were conducted at the Experimental Social Science Laboratory. Participants were compensated with course credit and performance-based monetary awards.

\paragraph{Data Processing}
Letting $\hat{X}$ denote choice of the advisor, weight on advice $\WoA$ is computed as
$$\WoA = \frac{f_{\hat{X}, \mathrm{adj}} - f_{\hat{X}, \mathrm{init}}}{1 - f_{\hat{X}, \mathrm{init}}}$$

The winsorization procedure replaces negative values with zeroes:

$$\WoA \leftarrow \max(0, \WoA)$$

The term ``advice confidence'' denotes $\AC=f_{\hat{X}, \mathrm{adj}}$, the adjusted confidence in the advisor’s answer.  Categorical variables (topic familiarity, chatbot usage) are converted to integer values using basic rules. A Beta-Bernoulli process is used to model beliefs in advice correctness \cite{gunes2021strategic}. See Appendix \ref{sec:appendix_processing} for some limitations of the approach and details of these decisions.

%%%%%%%%%%%%%%%%%%%%%%%
% Analysis
%%%%%%%%%%%%%%%%%%%%%%%

\newcommand{\A}{\textbf{\textcolor{Maroon}{A}}}
\newcommand{\B}{\textbf{\textcolor{Blue}{B}}}
\newcommand{\C}{\textbf{\textcolor{OliveGreen}{C}}}
\newcommand{\D}{\textbf{\textcolor{BurntOrange}{D}}}
\newcommand{\E}{\textbf{\textcolor{Plum}{E}}}

\section{Analysis}
\label{sec:analysis}

Participant answered 2,828 questions with an average of 23.97 questions per participant. After computing $\WoA$, 166 questions (5.87\%) with negative weight on advice are winsorized. Descriptive statistics are available in Appendix \ref{sec:appendix_descriptive_stats}. Qualitative findings about ChatGPT usage are presented in Appendix \ref{sec:appendix_qualitative}.

\subsection{Weight on Advice}

\paragraph{Hypothesis}
AsPredicted \href{https://aspredicted.org/KW3_8W2}{\#126040}) predicts that participants place greater weight on advice when the advisor is identified as an “AI chatbot.”

\begin{figure}
  \centering
  \includegraphics[width=.9\textwidth]{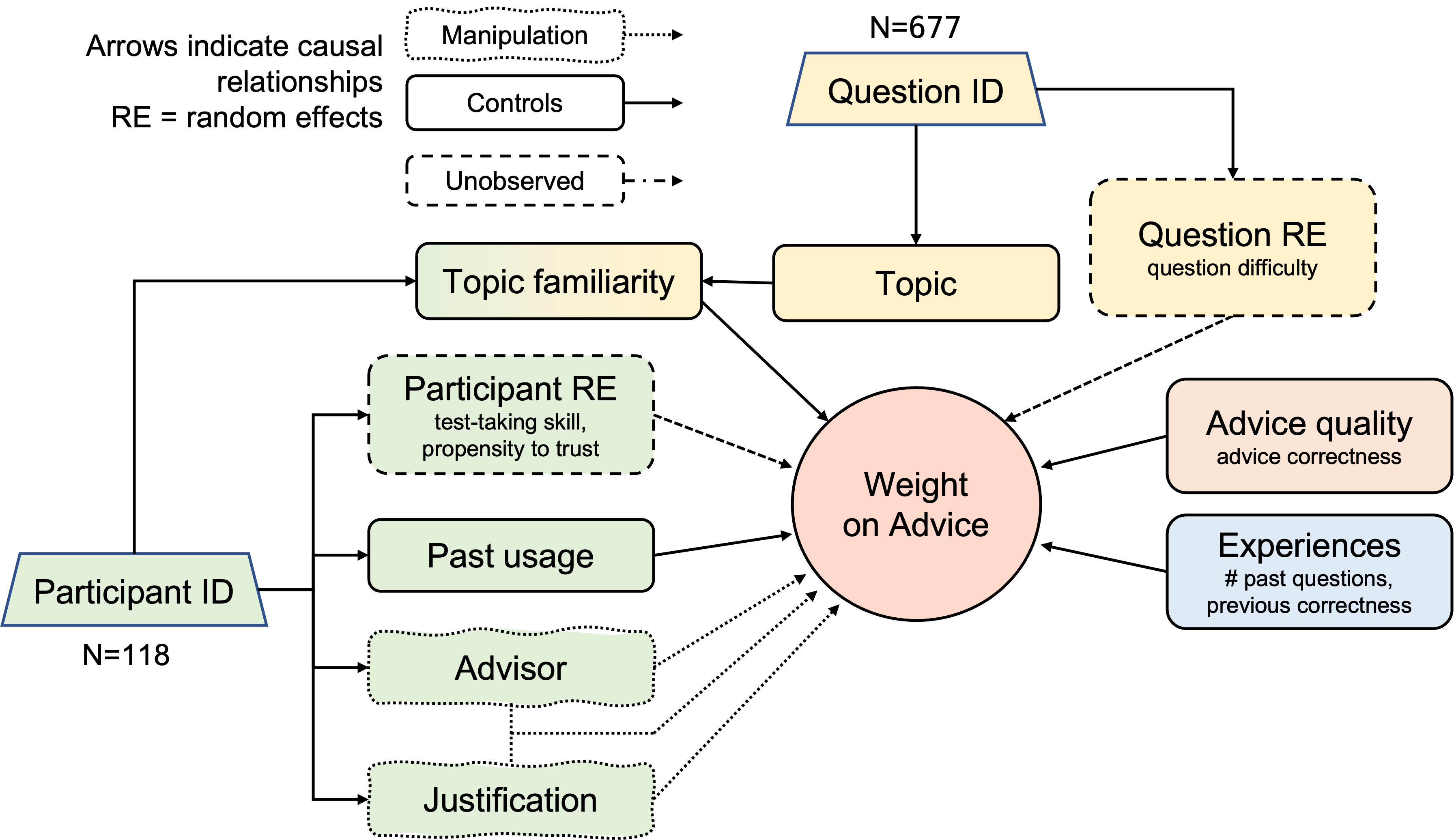}
  \caption{\textbf{Proposed causal mechanisms.} The study examines several predictors of weight on advice. The random assignment of \textcolor{LimeGreen}{\textbf{participant}} / \textcolor{Dandelion}{\textbf{question}}, \textcolor{CadetBlue}{\textbf{experience}} over several questions, and varying \textcolor{Melon}{\textbf{advice quality}} permits a decomposition of what predicts \textcolor{Salmon}{\textbf{weight on advice}}.}
  \label{fig:causal_diagram}
\end{figure}

\paragraph{Method}
Weight on advice is progressively regressed on a broader set of variables in each specification (see Figure  \ref{fig:causal_diagram}). Weight on advice is regressed on the advisor identity (\verb|advisor|), justification (\verb|give_justification|), and their interaction in \A. Controls, including topic familiarity (\verb|topic_familiarity|) are included in \B, past usage (\verb|usage_level|) in \C, advice quality (\verb|advice_is_correct|) in \D, and experiences (\verb|advice_accuracy_belief|, \verb|question_num|) in \E. (AsPredicted \href{https://aspredicted.org/KW3_8W2}{\#126040}) preregisters regressions \A-\B\ and conduct regressions \C-\E\ as a non-preregistered exploratory analysis. All regressions include random effects for participants (\verb|participant_id|) to account for unobservable differences in participants such as test-taking skill, propensity to trust, calibration skill, etc. Random effects are included for questions (\verb|question_id|) to account for within-topic differences in question difficulty. For concision, additional analyses are reserved for Appendix \ref{sec:appendix_analysis}.

\paragraph{Results}
The results of the regression are displayed in Table \ref{tab:main_regression_results}. From specification \A, there is no support for the initial hypothesis that $\WoA$ is greater if the advisor is identified as a chatbot. The coefficients are directionally correct but not statistically significant. In this and other specifications, the random effects for participants and questions are highly significant. Appendix \ref{sec:appendix_optionality} explores whether participant engagement might mediate the effect.

After including topic familiarity in specification \B, there is a highly significant increase in weight of advice when the participant is uncomfortable with the topic. Compared to a baseline of comfort in the topic, weight on advice is 6.1\%, 95\% CI [2.19-10.08\%] higher when a participant is uncomfortable in the topic. The effect is persistent and similar in size across specifications \C-\E. Robustness checks are conducted in Appendix \ref{sec:appendix_topic_familiarity}.

In specification \C, past usage of chatbots has a marginally significant effect for participants in the “AI chatbot” advisor condition. The interpretation is that for each step increase in usage level (e.g. having used AI chatbots instead of merely hearing about them), weight on advice increases by 5.0
\%, 95\% CI [0.0\%, 9.9\%]. For participants in the “expert” advisor condition, the effect is not significant, suggesting that the result is driven by participant understanding of the advisor’s capabilities. This effect is persistent and similar in size across specifications \D-\E. Appendix \ref{sec:appendix_past_usage} suggests that the result may mostly be driven by whether or not participants have used chatbots before.

On face, specification \D\ reveals a surprising absence of a direct effect from the quality of advice, measured as whether the advice is actually correct. The coefficient becomes significant in specification \E, suggesting that participants place 2.9\% more weight on advice if it is true. Additional exploratory analyses are performed in Appendix \ref{sec:appendix_advice_quality} controlling for initial confidence; the results reveal a strong effect that is mediated by giving justifications.

Finally, specification \E\ identifies significant effect of experience that may be mediated by the advisor’s identity. For every 10\% increase in believed advice accuracy, participants with an AI chatbot advisor place 6.02\% 95\% CI [4.29\%, 7.75\%] greater weight on advice. If the advisor is a generic expert, the coefficient is 5.05\% 95\% CI [3.43\%, 6.66\%] per 10\% increase in belief. Participants appear to place less weight on advice over time. For each additional question completed, participants place 0.4\% 95\% CI[0.362\%, 0.6\%] less weight on advice. Appendix \ref{sec:appendix_experience} shows that the deflated coefficient in the human expert condition is robust to beliefs. Moreover, there is a significant effect of the last advice's correctness that is mediated by advisor identity.

\begin{table}
\caption{Results of regression analyses examining weight on advice.}
\label{tab:main_regression_results}
\begin{center}
\begin{tabular}{llllll}
\toprule
                                             & \A        & \B        & \C        & \D        & \E         \\
\midrule

Intercept                                    & 0.329*** & 0.302*** & 0.169**  & 0.161**  & -0.159*    \\
                                             & (0.044)  & (0.045)  & (0.081)  & (0.082)  & (0.096)    \\
advice\_accuracy\_belief                     &          &          &          &          & 0.602***   \\
                                             &          &          &          &          & (0.088)    \\
advice\_accuracy\_belief:advisor[T.expert]   &          &          &          &          & -0.097     \\
                                             &          &          &          &          & (0.120)    \\
advice\_is\_correct[T.True]                  &          &          &          & 0.013    & 0.029**    \\
                                             &          &          &          & (0.015)  & (0.014)    \\
advisor[T.expert]                            & -0.027   & -0.028   & 0.056    & 0.054    & 0.067      \\
                                             & (0.059)  & (0.059)  & (0.118)  & (0.118)  & (0.137)    \\
advisor[T.expert]:give\_justification[T.yes] & -0.027   & -0.021   & -0.025   & -0.025   & -0.023     \\
                                             & (0.079)  & (0.079)  & (0.079)  & (0.079)  & (0.076)    \\
give\_justification[T.yes]                   & 0.057    & 0.056    & 0.064    & 0.064    & 0.074      \\
                                             & (0.058)  & (0.058)  & (0.057)  & (0.057)  & (0.055)    \\
participant\_id Var                          & 0.364*** & 0.362*** & 0.354*** & 0.353*** & 0.337***   \\
                                             & (0.056)  & (0.055)  & (0.055)  & (0.055)  & (0.052)    \\
question\_id Var                             & 0.044**  & 0.040**  & 0.040**  & 0.040**  & 0.045**    \\
                                             & (0.018)  & (0.018)  & (0.018)  & (0.018)  & (0.018)    \\
question\_num                                &          &          &          &          & -0.004***  \\
                                             &          &          &          &          & (0.001)    \\
topic\_familiarity[T.Neutral]                &          & 0.026    & 0.027    & 0.027    & 0.026      \\
                                             &          & (0.018)  & (0.018)  & (0.018)  & (0.017)    \\
topic\_familiarity[T.Uncomfortable]          &          & 0.061*** & 0.062*** & 0.062*** & 0.063***   \\
                                             &          & (0.020)  & (0.020)  & (0.020)  & (0.020)    \\
usage\_level                                 &          &          & 0.050*   & 0.049*   & 0.045*     \\
                                             &          &          & (0.025)  & (0.025)  & (0.024)    \\
usage\_level:advisor[T.expert]               &          &          & -0.033   & -0.033   & -0.019     \\
                                             &          &          & (0.036)  & (0.036)  & (0.035)    \\
\bottomrule
\end{tabular}
\end{center}
\end{table}

\subsection{Advice Confidence}

\paragraph{Hypothesis}
(AsPredicted \href{https://aspredicted.org/KW3_8W2}{\#126040}) predicts that (1) student's advice confidence will display overconfidence in language model accuracy and that (2) the overconfidence is mitigated by feedback.

\paragraph{Method}
For choice $X$ and question $j$, let $t_{j,X}$ denote whether $X$ is correct and $f_{j,X}$ denote the participant’s confidence in the choice. Calibration curves are constructed by partitioning advice confidences over $[0, 1]$ into 10 equal-width bins and plotting the average advice confidence $e_i  \equiv \mathbb{E}_{j \in I}[f_{j,X}]$ and accuracy $o_i \equiv \mathbb{E}_{j \in I}[t_{j,X}]$ for each bin $i$.

This section focuses on the advised choice $\hat{X}$ with the goal of evaluating how well the participants evaluate advice accuracy. Setting $e_i  \equiv \mathbb{E}_{j \in I}[f_{j,\hat{X}, \mathrm{adj}}]$ and accuracy $o_i  \equiv \mathbb{E}_{j \in I}[t_{j,\hat{X}}]$, a calibration curve is constructed for participant’s confidence in the advisor’s answers. Miscalibration is measured by expected calibration error ($\ECE$), the average deviation from ideal calibration weighted by sample size. Letting $P(i)$ denote the proportion of samples in bin $i$, $ECE$ is defined as:
$$\ECE = \sum_{i=1}^{10} P(i) \cdot |o_i - e_i|$$

To evaluate whether participants become more calibrated over time, $ECE$ is computed over groups of 5 question numbers. To calculate standard errors, $ECE$ is bootstrapped on 1000 samples within each question group.

As preregistered, AC is measured for each topic and compared to the actual accuracy of the model on a larger evaluation set on the topic (discussed in Appendix \ref{sec:appendix_model_evaluation}). Mistaken beliefs about per-topic confidence are identified by a Brunner-Munzel \cite{brunner2000nonparametric} comparisons between the mean accuracy $T=\{t_j\}$ and confidence $F=\{f_{j,X,\mathrm{adj}}\}$. The false discovery rate is controlled to 0.05 by using the Benjamin-Hochberg procedure \cite{benjamini1995controlling}.

Finally, participant performance is compared to a simple, proportionate update under different values of $\WoA$. The optimal weight on advice $\WoA^* \in [0,1]$ is the value that minimizes Brier Score.

\begin{figure}
  \centering
\begin{subfigure}{.5\textwidth}
  \includegraphics[width=\textwidth]{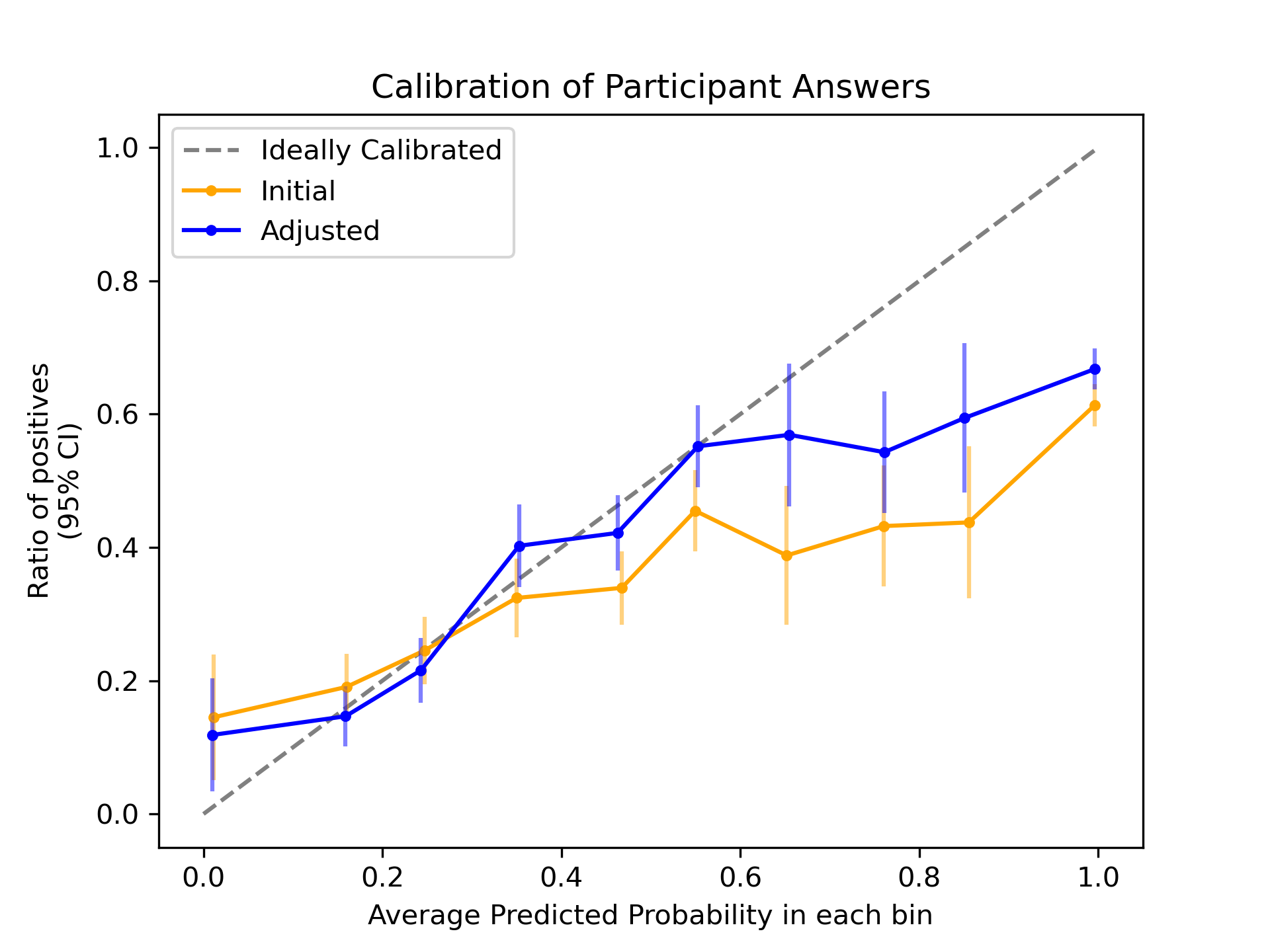}
  \caption{All participant answers.}
  \label{fig:calibration_curve_participant}
\end{subfigure}%
\begin{subfigure}{.5\textwidth}
  \includegraphics[width=\textwidth]{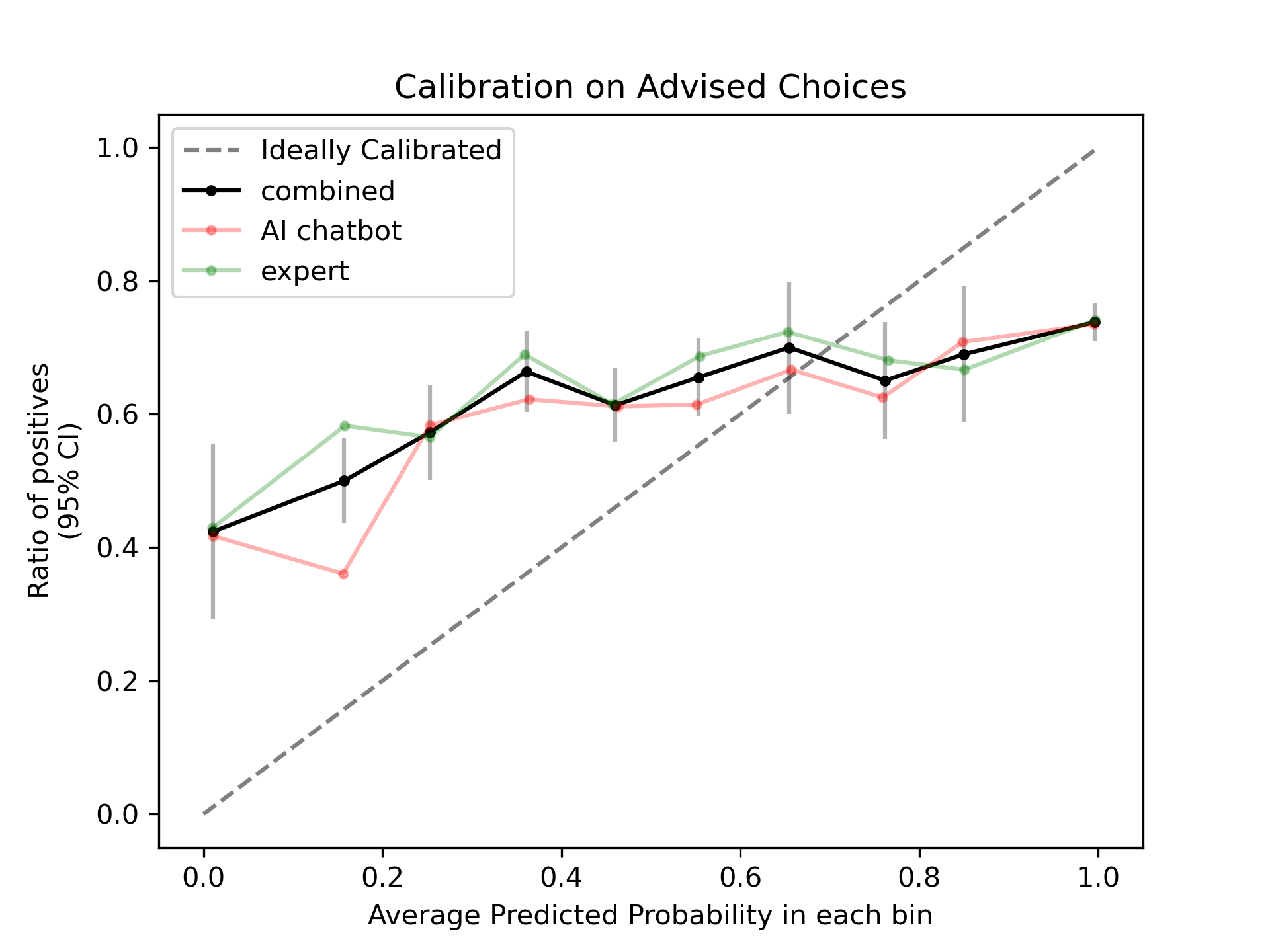}
  \caption{Advisor answers only.}
  \label{fig:calibration_curve_advice}
\end{subfigure}
  \caption{\textbf{Calibration curves.} Error bars are 95\% confidence intervals. Dotted line corresponds to a theoretically ideal calibration curve in which average predicted probability exactly equals ratio of positives.}
\end{figure}

\paragraph{Results}
Consistent with an extensive prior literature \cite{moore2008trouble}, participants are overconfident in their own answers. Figure \ref{fig:calibration_curve_participant} reveals (1) significant overconfidence in initial answers ($\ECE_{\mathrm{init}} = 0.183$) for confidence levels greater than 0.5 and (2) attenuated but persistent overconfidence in adjusted answers ($\ECE_{\mathrm{adj}} = 0.137$). Receiving advice appears to “lift” the actual accuracy of high-confidence answers and improve calibration.

Considering only adjusted confidence in advised choices, participants are significantly more miscalibrated ($\ECE_{\mathrm{advised}}=0.201$), displaying both overconfidence and underconfidence at different confidence levels. Participants dramatically underestimate the accuracy of advised choices at confidence levels below $0.5$. For example, when participants place 0 to 10\% confidence in the advisor’s answer, the answer is actually correct 42.4\%, 95\% CI [29.2\%, 55.5\%] of the time. Moreover, the effect applies across both advisor conditions (Figure \ref{fig:calibration_curve_advice}) and is persistent (see Appendix \ref{sec:appendix_miscalibration}).

\begin{figure}
  \centering
  \includegraphics[width=\textwidth]{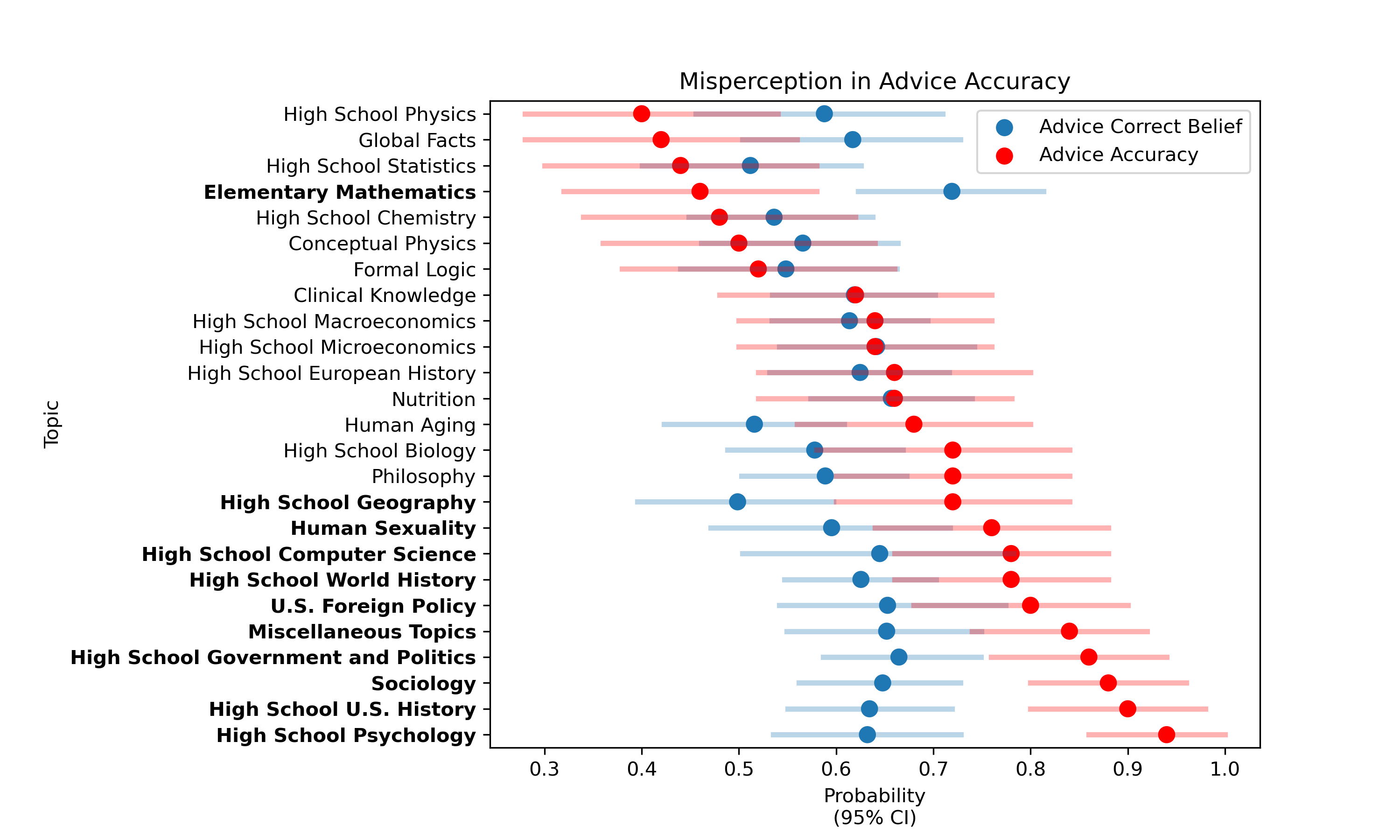}
  \caption{\textbf{Believed and actual advice accuracy.} Error bars are 95\% confidence intervals. Blue corresponds average participant confidence in the correctness advised answer. Red corresponds to the results from an evaluation on 50 questions in each subject. \textbf{Bolded} topics have significantly different advice correct beliefs (blue) and advice accuracy (red), suggesting a discrepancy between participant beliefs and reality.}
  \label{fig:misperception_in_advice_accuracy}
\end{figure}

\begin{figure}
  \centering
\begin{subfigure}{.5\textwidth}
  \includegraphics[width=\textwidth]{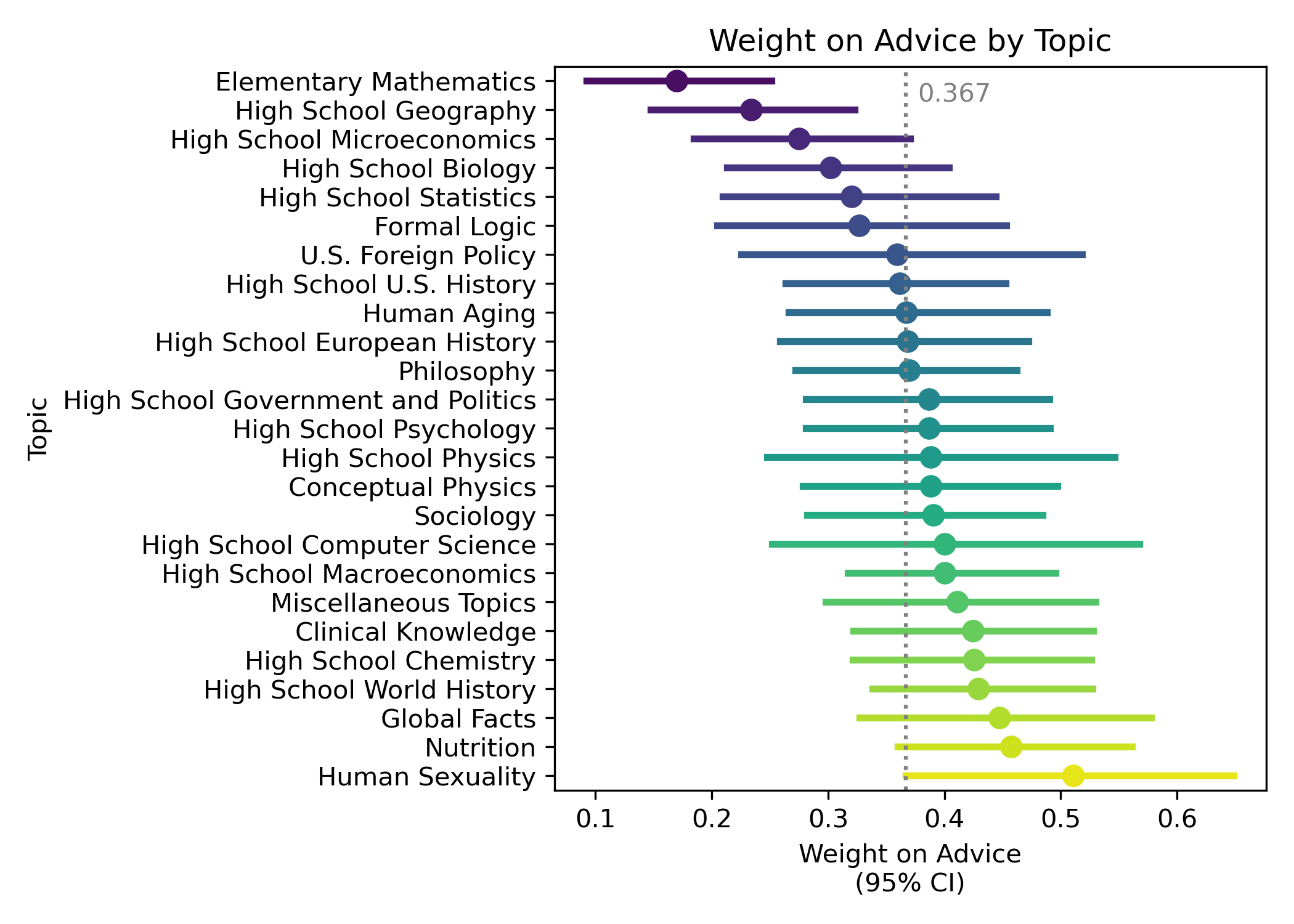}
  \caption{\textbf{Weight on advice by topic}. Error bars are 95\% confidence intervals. The average weight on advice is plotted as a dotted line.}
  \label{fig:woa_by_topic}
\end{subfigure}\hfill%
\begin{subfigure}{.45\textwidth}
  \includegraphics[width=\textwidth]{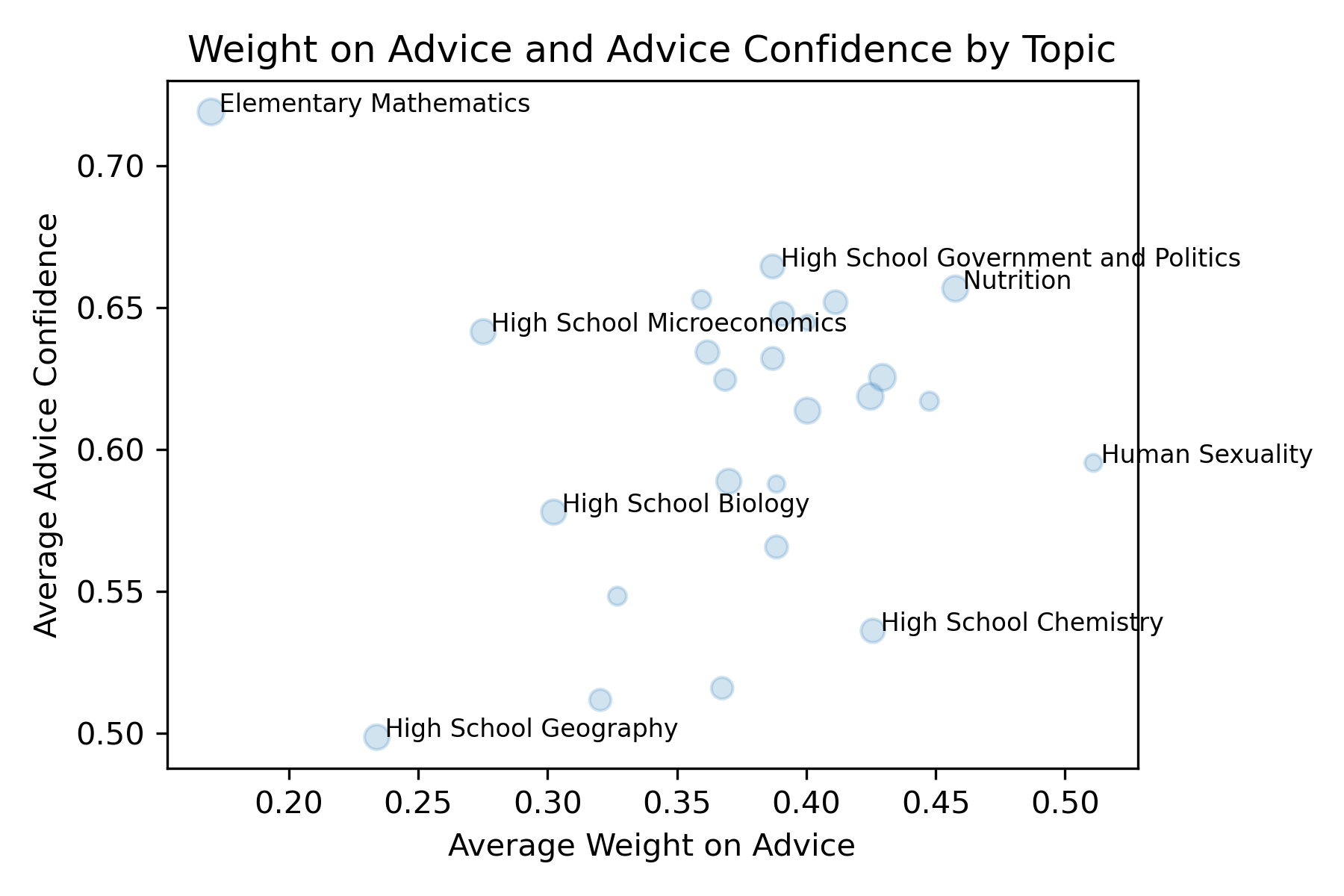}
  \caption{\textbf{Weight on advice and correctness}. Point size corresponds to the number of samples in the topic.}
  \label{fig:woa_and_correctness}
\end{subfigure}
  \caption{Figure \ref{fig:woa_by_topic} examines weight on advice for each topic and Figure \ref{fig:woa_and_correctness} displays its relationship with advice confidence.}
\end{figure}

The remainder of this section considers only participants in the AI chatbot condition. These participants are limited in their ability to predict differences in ChatGPT's advice quality across topics. Participants overestimate advice accuracy on Elementary Mathematics questions and underestimate accuracy for 10 topics displayed in Figure \ref{fig:misperception_in_advice_accuracy}. Average beliefs in advice accuracy are only moderately correlated (Pearson’s $r$=0.339, $p$=0.097) with actual advice accuracy when grouped by topic (Figure \ref{fig:woa_and_correctness}).

These errors are significant for knowing when to place weight on advice. $WoA$ and $AC$ are highly correlated at the question level (Pearson’s $r$=0.639, $p$=1.69e-150) and participants place significantly more (Human Sexuality) or less weight (Elementary Mathematics) on some topics compared to others (Figure \ref{fig:woa_by_topic}).

\begin{figure}
  \centering
  \includegraphics[width=\textwidth]{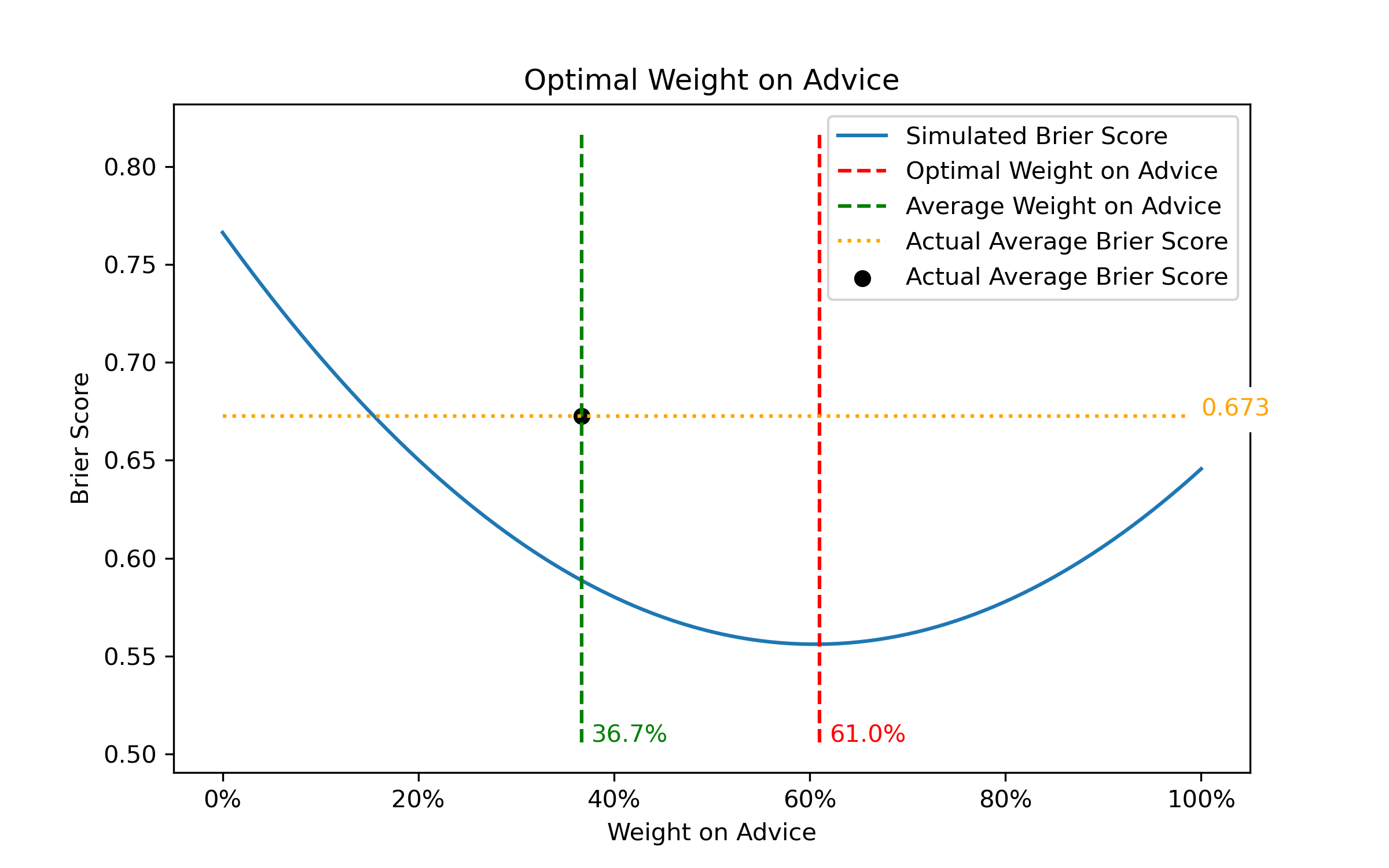}
  \caption{\textbf{Brier score-optimal weight on advice.} The dot marks actual participant weight on advice and Brier score. Participants significantly underweigh advice (optimal $61.0\%$ versus actual $36.7\%$) and under-perform the baseline even at the sub-optimal weight on advice.}
  \label{fig:woa_brier_score}
\end{figure}

Overall, participants do not place enough weight on advice (Figure \ref{fig:woa_brier_score}). The Brier score is minimized at $\WoA^*=0.61$, achieving an average score of $\BS=0.556$. The optimal weight on advice is over 50\% higher than the average participant weight on advice, $\overline{\WoA}=0.367$.

Notably, participants also score ($\BS=0.673$) significantly worse in practice if they uniformly applied the same weight on advice with a proportionate update ($\BS=0.588$). The poor performance compared to a uniform baseline is due to (1) misallocation, the suboptimal proportioning of confidence to other answer choices and (2) extremism, participants' tendency to place no weight on advice or too much weight on advice, leading to excessive overconfidence. 
Appendix \ref{sec:appendix_inefficiency} analyzes both effects and concludes that extremism is a larger factor.

%%%%%%%%%%%%%%%%%%%%%%%
% Discussion
%%%%%%%%%%%%%%%%%%%%%%%

\section{Discussion}
\label{sec:discussion}

This study contributes to literature on algorithm aversion and human computer interaction by performing a study of how students incorporate advice from ChatGPT on MC tests.

\paragraph{Weight on Advice}
Results are summarized in Table \ref{tab:result_summary}. Advisor identity, presence of justification, and their interaction do not significantly affect weight on advice. The effect continues to be insignificant at a 95\% level after including various controls. This result joins several other studies in finding a no effect of the advisor’s identity on weight on advice \cite{jussupow2020we} and agrees with another study on ChatGPT \cite{bohm2023content}. At the participant level, the sample is powered to detect a median effect size comparable found previously \cite{logg2019algorithm}. At the measured effect size, identifying an effect for advisor identity or presence of justification requires a sample size about four times larger (Appendix \ref{sec:appendix_power}). This might suggest that algorithm aversion or appreciation is less significant for ChatGPT than for other algorithms. Alternatively, the  null result may be an artifact of the description of the non-algorithm advisor, which is vaguely introduced as an ``expert'' \cite{hou2021expert}. Moreover, in Appendix \ref{sec:appendix_optionality}, the effects are recovered by including an interaction term for optional questions, suggesting that participant engagement might be adding noise to the results. Limitations to generalizing the result are discussed the limitations section below.

The analysis of controls confirms several prior results. Task difficulty in the form of topic familiarity mediates advice usage \cite{gino2007effects}. Harder tasks increase weight on advice more so for AI than human advice, but not significantly so (Appendix \ref{sec:appendix_topic_familiarity}).

In exploratory analyses, prior usage of chatbots predicts greater weight on advice in the chatbot condition. The result agrees with prior studies showing that familiarity with the algorithm predicts greater usage of algorithms \cite{castelo2019task}. Given the relatively recent release of ChatGPT, students and educators still have different levels of familiarity with the technology \cite{tlili2023if}. These results could explain for why prior usage and ChatGPT are highly correlated \cite{choudhury2023investigating}: new users underestimate the performance of the tool and gradually learn to trust and use it more. This reinforcing dynamic may predict how different people will use ChatGPT in the classroom and beyond.

In a further exploratory analysis, performance on previous questions and corresponding beliefs about advice accuracy are predictive of WoA, agreeing with prior work \cite{you2022algorithmic, filiz2021reducing, cabiddu2022users}. The result is stronger for AI chatbots under different models of beliefs, particularly when examining the effect of the last piece of advice (see Appendix \ref{sec:appendix_experience}). If the result holds, it suggests that people may be more perceptive or critical of AI performance.

Finally, there is initially a small effect of advice correctness on weight on advice.  Appendix \ref{sec:appendix_advice_quality} finds that the effect is much larger after correcting for the participant's initial answer. Moreover, there is a satisfying explanation for how participants identify correct advice: the effect is only significant for participants in the condition that receives justifications. These results the theory that interpretability improves human adoption of algorithms and suggests that natural language reasoning is an effect medium for interpretability \cite{tomsett2020rapid}.

\begin{table}[H]
    \centering
    \caption{Summary of weight on advice results.}
    \begin{tabular}{llccccc}
\toprule
\multicolumn{2}{c}{\textit{Explanation}} & \multicolumn{3}{c}{\textit{Analysis}} & \multicolumn{2}{c}{\textit{Result}} \\
\cmidrule(r){1-2} \cmidrule(r){3-5} \cmidrule(r){6-7}
Explanation type & Study metric & Prereg? & Spec. & Appendix & WOA Effect & Interaction?\\
\midrule
\textit{Social distance} & Advisor identity & \cmark & \A & \NA & \NA & \NA \\
\textit{Task difficulty} & Topic familiarity & \cmark & \B & \ref{sec:appendix_topic_familiarity} & + & \NA \\
\textit{Algorithm familiarity} & ChatGPT usage & \xmark & \C & \ref{sec:appendix_past_usage} & + & + Advisor identity \\
\textit{Experience} & Past accuracy & \xmark & \E & \ref{sec:appendix_experience} & + & + Advisor identity \\
\textit{Interpretability} & Justification & \cmark & \A & \ref{sec:appendix_advice_quality} & \NA & + Advice quality \\
\bottomrule
\end{tabular}
    \label{tab:result_summary}
\end{table}

\paragraph{Advice Confidence}
While participants are indeed overconfident in their own answers, they err even worse in judging the correctness of AI advice. The calibration error is significant and persistent across 40 rounds of feedback. 

Exploring one potential explanation, participants misjudge advisor accuracy across topics and underestimate accuracy on 10 out of 25 topics. These misjudgements contribute to calibration error and affect weight on advice. Participants broadly misunderstand which tasks the AI advisor is good at. Participants generally overestimate accuracy on procedural topics such as mathematics and physics, while underestimating accuracy on social science topics such history, government, geography, and so on.

Furthermore, by increasing the average weight of advice by over 50\% and uniformly adjusting their answers, participants could have improved their score significantly. Participants don’t use advice enough and are inefficient when they do, providing an important caveat to the results of \cite{bowman2022measuring}: in order study, participants were not able to perform better than ChatGPT alone when they were unable to interact with the model (see Appendix \ref{sec:appendix_topic_familiarity}). Overall, students do not place enough trust in algorithms like ChatGPT.

\paragraph{Limitations}
There are several limitations to this study. First, the study setup may be inefficient for the attempted regressions. Sampling from hundreds of questions introduces substantial variance to the data collection process that limits the efficiency of the estimators. Reducing randomization, for example by fixing a set of equally-difficult questions, may lead to better estimates. Otherwise, a study with a similar setup may require a larger sample size to identify the same effect.

Simultaneously, the lengthy survey required by the design might limit participant engagement. Appendix \ref{sec:appendix_optionality} suggests the possibility that the effects are real, but only for participants willing to do optional questions.

The advice could have been improved. The generated advice may not be representative of “ChatGPT’s output,” as suggested to participants. Model output is quite sensitive to prompting format, particularly across topics (see Appendix \ref{sec:appendix_model_evaluation}). The study design was constrained to using a similarly powerful InstructGPT model instead of OpenAI’s ChatGPT API, which was released after collecting generations.

Moreover, the advisor identity could benefit from clarification. Previous work shows that algorithm aversion and appreciation effects are sensitive to the description of the advisor \cite{hou2021expert}. It may have been worthwhile to more explicitly clarify the identity of the human expert, although Appendix \ref{sec:appendix_manipulation} addresses this criticism.

The incentives could also be improved. By providing an additional payout to the top performers, the rewards could encourage excessive confidence for the sake of scoring higher on the leaderboard, a phenomenon documented in forecasting tournaments \cite{sempere2021alignment}.

Finally, the study population limits the external validity of the results. These findings apply to a set of business students at UC Berkeley who might be much more informed on tests and familiar with ChatGPT compared to even the average student. 

\paragraph{Future Work}
Extensions of this study could begin by addressing its limitations. First, introducing more advisor identity conditions could help contrast how people judge ChatGPT's advice compared to other algorithms or other human advisors. For example, identifying an advisor as ``a previous participant'' or a generic ``statistical model''  \cite{logg2019algorithm} might cause participants to weight the advice less than that of an expert or ChatGPT.

The advice can be improved with better prompting techniques such as iterative bootstrapping \cite{sun2023enhancing} or self-consistency \cite{wang2022self} for chain-of-thought. To further study mediators of weight on advice, the advice might include model output probabilities over answer choices \cite{kadavath2022language} (with varying levels of calibration \cite{vodrahalli2022uncalibrated, tomsett2020rapid}), enable live user interaction \cite{bowman2022measuring}, or support modification of prompts \cite{dietvorst2018overcoming}. In addition to using the ChatGPT API, a comparison of ChatGPT output and ChatGPT Plus outputs could illuminate the relative impact of different AI feedback.

Researchers could study a much wider variety of tasks. Aside from the other 30+ topics in the MMLU benchmark, participants could provide quantitative answers, answer free response problems, or complete other natural language tasks (e.g. from BIG-Bench \cite{srivastava2022beyond}). Less structured responses would require new and untested metrics of distance between answers such BLEU scores \cite{papineni2002bleu}. Moreover, multi-modal models are able to perform natural language reasoning over images \cite{ghosh2019generating, salewski2022clevr} and videos \cite{chadha2021ireason, yu2020long}, creating new opportunities for study.

Further research might document how different populations take advice from LLM-based tools. For example, students from various grade levels or courses of study might take advice in different ways. Beyond education, many occupations will likely involve increasing collaboration with AI tools \cite{zarifhonarvar2023economics, eloundou2023gpts}. Previous studies have documented differences in algorithm aversion across populations \cite{liu2003algorithm}. As LLMs influence a greater number of human decisions, a nuanced understanding of how different people take their advice will be increasingly important.

\section*{Conflicts of Interest}
The authors declare that they have no conflict of interest.

\section*{Acknowledgments}
The technical execution of the project was completed in close collaboration with Isabella Borkovic. The project was advised by Professor Don Moore, who provided helpful direction throughout; I placed great weight on his advice. Thanks to Sandy Campbell, Karin Garrett, and Sophia Li for feedback. Karin Garrett and Isabella Borkovic each led laboratory sessions. Members of the Moore Accuracy Lab tested and provided feedback on the survey. Critical support for running experiments was provided by the Research Participant Program and Experimental Social Science Laboratory. Funding for prizes was provided by the Michael and Chris Boskin Scholarship.

%%%%%%%%%%%%%%%%%%%%%%%
% Bibliography
%%%%%%%%%%%%%%%%%%%%%%%
\bibliographystyle{unsrt}  
\bibliography{references}

\begin{thebibliography}{100}

\bibitem{zhang2023one}
Chaoning Zhang, Chenshuang Zhang, Chenghao Li, Yu~Qiao, Sheng Zheng,
  Sumit~Kumar Dam, Mengchun Zhang, Jung~Uk Kim, Seong~Tae Kim, Jinwoo Choi,
  et~al.
\newblock One small step for generative ai, one giant leap for agi: A complete
  survey on chatgpt in aigc era.
\newblock {\em arXiv preprint arXiv:2304.06488}, 2023.

\bibitem{zhou2023comprehensive}
Ce~Zhou, Qian Li, Chen Li, Jun Yu, Yixin Liu, Guangjing Wang, Kai Zhang, Cheng
  Ji, Qiben Yan, Lifang He, et~al.
\newblock A comprehensive survey on pretrained foundation models: A history
  from bert to chatgpt.
\newblock {\em arXiv preprint arXiv:2302.09419}, 2023.

\bibitem{stiennon2020learning}
Nisan Stiennon, Long Ouyang, Jeffrey Wu, Daniel Ziegler, Ryan Lowe, Chelsea
  Voss, Alec Radford, Dario Amodei, and Paul~F Christiano.
\newblock Learning to summarize with human feedback.
\newblock {\em Advances in Neural Information Processing Systems},
  33:3008--3021, 2020.

\bibitem{christiano2017deep}
Paul~F Christiano, Jan Leike, Tom Brown, Miljan Martic, Shane Legg, and Dario
  Amodei.
\newblock Deep reinforcement learning from human preferences.
\newblock {\em Advances in neural information processing systems}, 30, 2017.

\bibitem{kasneci2023chatgpt}
Enkelejda Kasneci, Kathrin Se{\ss}ler, Stefan K{\"u}chemann, Maria Bannert,
  Daryna Dementieva, Frank Fischer, Urs Gasser, Georg Groh, Stephan
  G{\"u}nnemann, Eyke H{\"u}llermeier, et~al.
\newblock Chatgpt for good? on opportunities and challenges of large language
  models for education.
\newblock {\em Learning and Individual Differences}, 103:102274, 2023.

\bibitem{razniewski2021language}
Simon Razniewski, Andrew Yates, Nora Kassner, and Gerhard Weikum.
\newblock Language models as or for knowledge bases.
\newblock {\em arXiv preprint arXiv:2110.04888}, 2021.

\bibitem{bubeck2023sparks}
S{\'e}bastien Bubeck, Varun Chandrasekaran, Ronen Eldan, Johannes Gehrke, Eric
  Horvitz, Ece Kamar, Peter Lee, Yin~Tat Lee, Yuanzhi Li, Scott Lundberg,
  et~al.
\newblock Sparks of artificial general intelligence: Early experiments with
  gpt-4.
\newblock {\em arXiv preprint arXiv:2303.12712}, 2023.

\bibitem{openai2023gpt4}
OpenAI.
\newblock Gpt-4 technical report.
\newblock {\em arXiv:2303.08774}, 2023.

\bibitem{choi2023chatgpt}
Jonathan~H Choi, Kristin~E Hickman, Amy Monahan, and Daniel Schwarcz.
\newblock Chatgpt goes to law school.
\newblock {\em Available at SSRN}, 2023.

\bibitem{gilson2023does}
Aidan Gilson, Conrad~W Safranek, Thomas Huang, Vimig Socrates, Ling Chi,
  Richard~Andrew Taylor, David Chartash, et~al.
\newblock How does chatgpt perform on the united states medical licensing
  examination? the implications of large language models for medical education
  and knowledge assessment.
\newblock {\em JMIR Medical Education}, 9(1):e45312, 2023.

\bibitem{mbakwe2023chatgpt}
Amarachi~B Mbakwe, Ismini Lourentzou, Leo~Anthony Celi, Oren~J Mechanic, and
  Alon Dagan.
\newblock Chatgpt passing usmle shines a spotlight on the flaws of medical
  education, 2023.

\bibitem{fijavcko2023can}
Nino Fija{\v{c}}ko, Lucija Gosak, Gregor {\v{S}}tiglic, Christopher~T Picard,
  and Matthew~John Douma.
\newblock Can chatgpt pass the life support exams without entering the american
  heart association course?
\newblock {\em Resuscitation}, 185, 2023.

\bibitem{kemp2023chatgpt}
Matthew~W Kemp, Susan~JS Logan, Pooja~Sharma Dimri, Navkaran Singh, Citra~NZ
  Mattar, Pradip Dashraath, Harshaana Ramlal, Aniza~P Mahyuddin, Suren Kanayan,
  Sean~WD Carter, et~al.
\newblock Chatgpt outscored human candidates in a virtual objective structured
  clinical examination (osce) in obstetrics and gynecology.
\newblock {\em American Journal of Obstetrics and Gynecology}, 2023.

\bibitem{kortemeyer2023could}
Gerd Kortemeyer.
\newblock Could an artificial-intelligence agent pass an introductory physics
  course?
\newblock {\em arXiv preprint arXiv:2301.12127}, 2023.

\bibitem{tlili2023if}
Ahmed Tlili, Boulus Shehata, Michael~Agyemang Adarkwah, Aras Bozkurt, Daniel~T
  Hickey, Ronghuai Huang, and Brighter Agyemang.
\newblock What if the devil is my guardian angel: Chatgpt as a case study of
  using chatbots in education.
\newblock {\em Smart Learning Environments}, 10(1):15, 2023.

\bibitem{susnjak2022chatgpt}
Teo Susnjak.
\newblock Chatgpt: The end of online exam integrity?
\newblock {\em arXiv preprint arXiv:2212.09292}, 2022.

\bibitem{cotton2023chatting}
Debby~RE Cotton, Peter~A Cotton, and J~Reuben Shipway.
\newblock Chatting and cheating: Ensuring academic integrity in the era of
  chatgpt.
\newblock {\em Innovations in Education and Teaching International}, pages
  1--12, 2023.

\bibitem{sallam2023chatgpt}
Malik Sallam.
\newblock Chatgpt utility in healthcare education, research, and practice:
  Systematic review on the promising perspectives and valid concerns.
\newblock In {\em Healthcare}, volume~11, page 887. MDPI, 2023.

\bibitem{firat2023chat}
Mehmet Firat.
\newblock How chat gpt can transform autodidactic experiences and open
  education.
\newblock {\em Department of Distance Education, Open Education Faculty,
  Anadolu Unive}, 2023.

\bibitem{leiter2023chatgpt}
Christoph Leiter, Ran Zhang, Yanran Chen, Jonas Belouadi, Daniil Larionov,
  Vivian Fresen, and Steffen Eger.
\newblock Chatgpt: A meta-analysis after 2.5 months.
\newblock {\em arXiv preprint arXiv:2302.13795}, 2023.

\bibitem{dwivedi2023so}
Yogesh~K Dwivedi, Nir Kshetri, Laurie Hughes, Emma~Louise Slade, Anand Jeyaraj,
  Arpan~Kumar Kar, Abdullah~M Baabdullah, Alex Koohang, Vishnupriya Raghavan,
  Manju Ahuja, et~al.
\newblock “so what if chatgpt wrote it?” multidisciplinary perspectives on
  opportunities, challenges and implications of generative conversational ai
  for research, practice and policy.
\newblock {\em International Journal of Information Management}, 71:102642,
  2023.

\bibitem{polonsky2023should}
Michael~Jay Polonsky and Jeffrey~D Rotman.
\newblock Should artificial intelligent agents be your co-author? arguments in
  favour, informed by chatgpt, 2023.

\bibitem{anders2023using}
Brent~A Anders.
\newblock Is using chatgpt cheating, plagiarism, both, neither, or forward
  thinking?
\newblock {\em Patterns}, 4(3), 2023.

\bibitem{kumar2023novel}
Archana~Praveen Kumar, Ashalatha Nayak, Manjula Shenoy, Shashank Goyal, et~al.
\newblock A novel approach to generate distractors for multiple choice
  questions.
\newblock {\em Expert Systems with Applications}, page 120022, 2023.

\bibitem{newton2023chatgpt}
Philip~Mark Newton.
\newblock Chatgpt performance on mcq-based exams.
\newblock 2023.

\bibitem{burgason2019cheating}
Kyle~A Burgason, Ophir Sefiha, and Lisa Briggs.
\newblock Cheating is in the eye of the beholder: An evolving understanding of
  academic misconduct.
\newblock {\em Innovative Higher Education}, 44:203--218, 2019.

\bibitem{gonsalves2023chatgpt}
Chahna Gonsalves.
\newblock On chatgpt: what promise remains for multiple choice assessment?
\newblock {\em Journal of Learning Development in Higher Education}, (27),
  2023.

\bibitem{haensch2023seeing}
Anna-Carolina Haensch, Sarah Ball, Markus Herklotz, and Frauke Kreuter.
\newblock Seeing chatgpt through students' eyes: An analysis of tiktok data.
\newblock {\em arXiv preprint arXiv:2303.05349}, 2023.

\bibitem{dietvorst2015algorithm}
Berkeley~J Dietvorst, Joseph~P Simmons, and Cade Massey.
\newblock Algorithm aversion: people erroneously avoid algorithms after seeing
  them err.
\newblock {\em Journal of Experimental Psychology: General}, 144(1):114, 2015.

\bibitem{logg2019algorithm}
Jennifer~M Logg, Julia~A Minson, and Don~A Moore.
\newblock Algorithm appreciation: People prefer algorithmic to human judgment.
\newblock {\em Organizational Behavior and Human Decision Processes},
  151:90--103, 2019.

\bibitem{castelo2019task}
Noah Castelo, Maarten~W Bos, and Donald~R Lehmann.
\newblock Task-dependent algorithm aversion.
\newblock {\em Journal of Marketing Research}, 56(5):809--825, 2019.

\bibitem{hou2021expert}
Yoyo Tsung-Yu Hou and Malte~F Jung.
\newblock Who is the expert? reconciling algorithm aversion and algorithm
  appreciation in ai-supported decision making.
\newblock {\em Proceedings of the ACM on Human-Computer Interaction},
  5(CSCW2):1--25, 2021.

\bibitem{dietvorst2018overcoming}
Berkeley~J Dietvorst, Joseph~P Simmons, and Cade Massey.
\newblock Overcoming algorithm aversion: People will use imperfect algorithms
  if they can (even slightly) modify them.
\newblock {\em Management science}, 64(3):1155--1170, 2018.

\bibitem{berger2021watch}
Benedikt Berger, Martin Adam, Alexander R{\"u}hr, and Alexander Benlian.
\newblock Watch me improve—algorithm aversion and demonstrating the ability
  to learn.
\newblock {\em Business \& Information Systems Engineering}, 63(1):55--68,
  2021.

\bibitem{reich2022overcome}
Taly Reich, Alex Kaju, and Sam~J Maglio.
\newblock How to overcome algorithm aversion: Learning from mistakes.
\newblock {\em Journal of Consumer Psychology}, 2022.

\bibitem{jung2021towards}
Markus Jung and Mischa Seiter.
\newblock Towards a better understanding on mitigating algorithm aversion in
  forecasting: An experimental study.
\newblock {\em Journal of Management Control}, 32(4):495--516, 2021.

\bibitem{jussupow2020we}
Ekaterina Jussupow, Izak Benbasat, and Armin Heinzl.
\newblock Why are we averse towards algorithms? a comprehensive literature
  review on algorithm aversion.
\newblock 2020.

\bibitem{burton2020systematic}
Jason~W Burton, Mari-Klara Stein, and Tina~Blegind Jensen.
\newblock A systematic review of algorithm aversion in augmented decision
  making.
\newblock {\em Journal of Behavioral Decision Making}, 33(2):220--239, 2020.

\bibitem{morewedge2022preference}
Carey~K Morewedge.
\newblock Preference for human, not algorithm aversion.
\newblock {\em Trends in Cognitive Sciences}, 2022.

\bibitem{hessler2022self}
Pascal~Oliver He{\ss}ler, Jella Pfeiffer, and Sebastian Hafenbr{\"a}dl.
\newblock When self-humanization leads to algorithm aversion: what users want
  from decision support systems on prosocial microlending platforms.
\newblock {\em Business \& Information Systems Engineering}, 64(3):275--292,
  2022.

\bibitem{luo2019frontiers}
Xueming Luo, Siliang Tong, Zheng Fang, and Zhe Qu.
\newblock Frontiers: Machines vs. humans: The impact of artificial intelligence
  chatbot disclosure on customer purchases.
\newblock {\em Marketing Science}, 38(6):937--947, 2019.

\bibitem{schanke2021estimating}
Scott Schanke, Gordon Burtch, and Gautam Ray.
\newblock Estimating the impact of “humanizing” customer service chatbots.
\newblock {\em Information Systems Research}, 32(3):736--751, 2021.

\bibitem{strait2015too}
Megan Strait, Lara Vujovic, Victoria Floerke, Matthias Scheutz, and Heather
  Urry.
\newblock Too much humanness for human-robot interaction: exposure to highly
  humanlike robots elicits aversive responding in observers.
\newblock In {\em Proceedings of the 33rd annual ACM conference on human
  factors in computing systems}, pages 3593--3602, 2015.

\bibitem{saha2022hard}
Swarnadeep Saha, Peter Hase, Nazneen Rajani, and Mohit Bansal.
\newblock Are hard examples also harder to explain? a study with human and
  model-generated explanations.
\newblock {\em arXiv preprint arXiv:2211.07517}, 2022.

\bibitem{gino2007effects}
Francesca Gino and Don~A Moore.
\newblock Effects of task difficulty on use of advice.
\newblock {\em Journal of Behavioral Decision Making}, 20(1):21--35, 2007.

\bibitem{bogert2021humans}
Eric Bogert, Aaron Schecter, and Richard~T Watson.
\newblock Humans rely more on algorithms than social influence as a task
  becomes more difficult.
\newblock {\em Scientific reports}, 11(1):1--9, 2021.

\bibitem{kaufmann2021algorithm}
Esther Kaufmann.
\newblock Algorithm appreciation or aversion? comparing in-service and
  pre-service teachers’ acceptance of computerized expert models.
\newblock {\em Computers and Education: Artificial Intelligence}, 2:100028,
  2021.

\bibitem{liu2003algorithm}
Nicole Tsz~Yeung Liu, Samuel~N. Kirshner, and Eric~T.K. Lim.
\newblock Is algorithm aversion weird? a cross-country comparison of
  individual-differences and algorithm aversion.
\newblock {\em Journal of Retailing and Consumer Services}, 72:103259, 2023.

\bibitem{you2022algorithmic}
Sangseok You, Cathy~Liu Yang, and Xitong Li.
\newblock Algorithmic versus human advice: Does presenting prediction
  performance matter for algorithm appreciation?
\newblock {\em Journal of Management Information Systems}, 39(2):336--365,
  2022.

\bibitem{filiz2021reducing}
Ibrahim Filiz, Jan~Ren{\'e} Judek, Marco Lorenz, and Markus Spiwoks.
\newblock Reducing algorithm aversion through experience.
\newblock {\em Journal of Behavioral and Experimental Finance}, 31:100524,
  2021.

\bibitem{cabiddu2022users}
Francesca Cabiddu, Ludovica Moi, Gerardo Patriotta, and David~G Allen.
\newblock Why do users trust algorithms? a review and conceptualization of
  initial trust and trust over time.
\newblock {\em European Management Journal}, 2022.

\bibitem{alexander2018trust}
Veronika Alexander, Collin Blinder, and Paul~J Zak.
\newblock Why trust an algorithm? performance, cognition, and neurophysiology.
\newblock {\em Computers in Human Behavior}, 89:279--288, 2018.

\bibitem{tomsett2020rapid}
Richard Tomsett, Alun Preece, Dave Braines, Federico Cerutti, Supriyo
  Chakraborty, Mani Srivastava, Gavin Pearson, and Lance Kaplan.
\newblock Rapid trust calibration through interpretable and uncertainty-aware
  ai.
\newblock {\em Patterns}, 1(4):100049, 2020.

\bibitem{destefano2022providing}
Timothy DeStefano, Katherine Kellogg, Michael Menietti, and Luca Vendraminelli.
\newblock Why providing humans with interpretable algorithms may,
  counterintuitively, lead to lower decision-making performance.
\newblock 2022.

\bibitem{altintas2023effect}
Onur Altintas, Abraham Seidmann, and Bin Gu.
\newblock The effect of interpretable artificial intelligence on repeated
  managerial decision-making under uncertainty.
\newblock {\em Available at SSRN 4331145}, 2023.

\bibitem{ahn2021will}
Daehwan Ahn, Abdullah Almaatouq, Monisha Gulabani, and Kartik Hosanagar.
\newblock Will we trust what we don't understand? impact of model
  interpretability and outcome feedback on trust in ai.
\newblock {\em arXiv preprint arXiv:2111.08222}, 2021.

\bibitem{ben2021explainable}
Daniel Ben~David, Yehezkel~S Resheff, and Talia Tron.
\newblock Explainable ai and adoption of financial algorithmic advisors: an
  experimental study.
\newblock In {\em Proceedings of the 2021 AAAI/ACM Conference on AI, Ethics,
  and Society}, pages 390--400, 2021.

\bibitem{schmidt2020calibrating}
Philipp Schmidt and Felix Biessmann.
\newblock Calibrating human-ai collaboration: Impact of risk, ambiguity and
  transparency on algorithmic bias.
\newblock In {\em Machine Learning and Knowledge Extraction: 4th IFIP TC 5, TC
  12, WG 8.4, WG 8.9, WG 12.9 International Cross-Domain Conference, CD-MAKE
  2020, Dublin, Ireland, August 25--28, 2020, Proceedings 4}, pages 431--449.
  Springer, 2020.

\bibitem{lehmann2022risk}
Cedric~A Lehmann, Christiane~B Haubitz, Andreas F{\"u}gener, and Ulrich~W
  Thonemann.
\newblock The risk of algorithm transparency: How algorithm complexity drives
  the effects on the use of advice.
\newblock {\em Production and Operations Management}, 31(9):3419--3434, 2022.

\bibitem{gaube2023non}
Susanne Gaube, Harini Suresh, Martina Raue, Eva Lermer, Timo~K Koch,
  Matthias~FC Hudecek, Alun~D Ackery, Samir~C Grover, Joseph~F Coughlin, Dieter
  Frey, et~al.
\newblock Non-task expert physicians benefit from correct explainable ai advice
  when reviewing x-rays.
\newblock {\em Scientific reports}, 13(1):1383, 2023.

\bibitem{panigutti2022understanding}
Cecilia Panigutti, Andrea Beretta, Fosca Giannotti, and Dino Pedreschi.
\newblock Understanding the impact of explanations on advice-taking: a user
  study for ai-based clinical decision support systems.
\newblock In {\em Proceedings of the 2022 CHI Conference on Human Factors in
  Computing Systems}, pages 1--9, 2022.

\bibitem{oviedo2023risks}
Oscar Oviedo-Trespalacios, Amy~E Peden, Thomas Cole-Hunter, Arianna Costantini,
  Milad Haghani, Sage Kelly, Helma Torkamaan, Amina Tariq, James David~Albert
  Newton, Timothy Gallagher, et~al.
\newblock The risks of using chatgpt to obtain common safety-related
  information and advice.
\newblock {\em Available at SSRN 4346827}, 2023.

\bibitem{howard2023chatgpt}
Alex Howard, William Hope, and Alessandro Gerada.
\newblock Chatgpt and antimicrobial advice: the end of the consulting infection
  doctor?
\newblock {\em The Lancet Infectious Diseases}, 23(4):405--406, 2023.

\bibitem{xie2023aesthetic}
Yi~Xie, Ishith Seth, David~J Hunter-Smith, Warren~M Rozen, Richard Ross, and
  Matthew Lee.
\newblock Aesthetic surgery advice and counseling from artificial intelligence:
  A rhinoplasty consultation with chatgpt.
\newblock {\em Aesthetic Plastic Surgery}, pages 1--9, 2023.

\bibitem{nastasi2023does}
Anthony~J Nastasi, Katherine~R Courtright, Scott~D Halpern, and Gary~E
  Weissman.
\newblock Does chatgpt provide appropriate and equitable medical advice?: A
  vignette-based, clinical evaluation across care contexts.
\newblock {\em medRxiv}, pages 2023--02, 2023.

\bibitem{george2023review}
A~Shaji George and AS~Hovan George.
\newblock A review of chatgpt ai's impact on several business sectors.
\newblock {\em Partners Universal International Innovation Journal},
  1(1):9--23, 2023.

\bibitem{leib2023corrupted}
Margarita Leib, Nils K{\"o}bis, Rainer~Michael Rilke, Marloes Hagens, and Bernd
  Irlenbusch.
\newblock Corrupted by algorithms? how ai-generated and human-written advice
  shape (dis) honesty.
\newblock {\em arXiv preprint arXiv:2301.01954}, 2023.

\bibitem{momentrusting}
Ali Momen, Ewart~J de~Visser, Kyle Wolsten, Katrina Cooley, James Walliser, and
  Chad~C Tossell.
\newblock Trusting the moral judgments of a robot: Perceived moral competence
  and humanlikeness of a gpt-3 enabled ai.

\bibitem{krugel2023chatgpt}
Sebastian Kr{\"u}gel, Andreas Ostermaier, and Matthias Uhl.
\newblock Chatgpt’s inconsistent moral advice influences users’ judgment.
\newblock {\em Scientific Reports}, 13(1):4569, 2023.

\bibitem{spitale2023ai}
Giovanni Spitale, Nikola Biller-Andorno, and Federico Germani.
\newblock Ai model gpt-3 (dis) informs us better than humans.
\newblock {\em arXiv preprint arXiv:2301.11924}, 2023.

\bibitem{ye2023improved}
Yang Ye, Hengxu You, and Jing Du.
\newblock Improved trust in human-robot collaboration with chatgpt.
\newblock {\em arXiv preprint arXiv:2304.12529}, 2023.

\bibitem{bohm2023content}
Robert B{\"o}hm, Moritz J{\"o}rling, Leonhard Reiter, and Christoph Fuchs.
\newblock Content beats competence: People devalue chatgpt’s perceived
  competence but not its recommendations.
\newblock 2023.

\bibitem{moore2008trouble}
Don~A Moore and Paul~J Healy.
\newblock The trouble with overconfidence.
\newblock {\em Psychological review}, 115(2):502, 2008.

\bibitem{moore2007overconfidence}
Don~A Moore and Daylian~M Cain.
\newblock Overconfidence and underconfidence: When and why people underestimate
  (and overestimate) the competition.
\newblock {\em Organizational Behavior and Human Decision Processes},
  103(2):197--213, 2007.

\bibitem{chen2022close}
Yangyi Chen, Lifan Yuan, Ganqu Cui, Zhiyuan Liu, and Heng Ji.
\newblock A close look into the calibration of pre-trained language models.
\newblock {\em arXiv preprint arXiv:2211.00151}, 2022.

\bibitem{kadavath2022language}
Saurav Kadavath, Tom Conerly, Amanda Askell, Tom Henighan, Dawn Drain, Ethan
  Perez, Nicholas Schiefer, Zac~Hatfield Dodds, Nova DasSarma, Eli
  Tran-Johnson, et~al.
\newblock Language models (mostly) know what they know.
\newblock {\em arXiv preprint arXiv:2207.05221}, 2022.

\bibitem{jiang2021can}
Zhengbao Jiang, Jun Araki, Haibo Ding, and Graham Neubig.
\newblock How can we know when language models know? on the calibration of
  language models for question answering.
\newblock {\em Transactions of the Association for Computational Linguistics},
  9:962--977, 2021.

\bibitem{bai2022training}
Yuntao Bai, Andy Jones, Kamal Ndousse, Amanda Askell, Anna Chen, Nova DasSarma,
  Dawn Drain, Stanislav Fort, Deep Ganguli, Tom Henighan, et~al.
\newblock Training a helpful and harmless assistant with reinforcement learning
  from human feedback.
\newblock {\em arXiv preprint arXiv:2204.05862}, 2022.

\bibitem{chong2022human}
Leah Chong, Guanglu Zhang, Kosa Goucher-Lambert, Kenneth Kotovsky, and Jonathan
  Cagan.
\newblock Human confidence in artificial intelligence and in themselves: The
  evolution and impact of confidence on adoption of ai advice.
\newblock {\em Computers in Human Behavior}, 127:107018, 2022.

\bibitem{hendrycks2020measuring}
Dan Hendrycks, Collin Burns, Steven Basart, Andy Zou, Mantas Mazeika, Dawn
  Song, and Jacob Steinhardt.
\newblock Measuring massive multitask language understanding.
\newblock {\em arXiv preprint arXiv:2009.03300}, 2020.

\bibitem{ouyang2022training}
Long Ouyang, Jeffrey Wu, Xu~Jiang, Diogo Almeida, Carroll Wainwright, Pamela
  Mishkin, Chong Zhang, Sandhini Agarwal, Katarina Slama, Alex Ray, et~al.
\newblock Training language models to follow instructions with human feedback.
\newblock {\em Advances in Neural Information Processing Systems},
  35:27730--27744, 2022.

\bibitem{wei2022chain}
Jason Wei, Xuezhi Wang, Dale Schuurmans, Maarten Bosma, Ed~Chi, Quoc Le, and
  Denny Zhou.
\newblock Chain of thought prompting elicits reasoning in large language
  models.
\newblock {\em arXiv preprint arXiv:2201.11903}, 2022.

\bibitem{sniezek2001trust}
Janet~A Sniezek and Lyn~M Van~Swol.
\newblock Trust, confidence, and expertise in a judge-advisor system.
\newblock {\em Organizational behavior and human decision processes},
  84(2):288--307, 2001.

\bibitem{rufibach2010use}
Kaspar Rufibach.
\newblock Use of brier score to assess binary predictions.
\newblock {\em Journal of clinical epidemiology}, 63(8):938--939, 2010.

\bibitem{roulston2007performance}
Mark~S Roulston.
\newblock Performance targets and the brier score.
\newblock {\em Meteorological Applications: A journal of forecasting, practical
  applications, training techniques and modelling}, 14(2):185--194, 2007.

\bibitem{gunes2021strategic}
Taha Gunes et~al.
\newblock {\em Strategic and Adaptive Behaviours in Trust Systems}.
\newblock PhD thesis, University of Southampton, 2021.

\bibitem{brunner2000nonparametric}
Edgar Brunner and Ullrich Munzel.
\newblock The nonparametric behrens-fisher problem: asymptotic theory and a
  small-sample approximation.
\newblock {\em Biometrical Journal: Journal of Mathematical Methods in
  Biosciences}, 42(1):17--25, 2000.

\bibitem{benjamini1995controlling}
Yoav Benjamini and Yosef Hochberg.
\newblock Controlling the false discovery rate: a practical and powerful
  approach to multiple testing.
\newblock {\em Journal of the Royal statistical society: series B
  (Methodological)}, 57(1):289--300, 1995.

\bibitem{choudhury2023investigating}
Avishek Choudhury and Hamid Shamszare.
\newblock Investigating the impact of user trust on adoption and use of
  chatgpt: A survey analysis.
\newblock {\em Tellus}, 2023.

\bibitem{bowman2022measuring}
Samuel~R Bowman, Jeeyoon Hyun, Ethan Perez, Edwin Chen, Craig Pettit, Scott
  Heiner, Kamile Lukosuite, Amanda Askell, Andy Jones, Anna Chen, et~al.
\newblock Measuring progress on scalable oversight for large language models.
\newblock {\em arXiv preprint arXiv:2211.03540}, 2022.

\bibitem{sempere2021alignment}
Nu{\~n}o Sempere and Alex Lawsen.
\newblock Alignment problems with current forecasting platforms.
\newblock {\em arXiv preprint arXiv:2106.11248}, 2021.

\bibitem{sun2023enhancing}
Jiashuo Sun, Yi~Luo, Yeyun Gong, Chen Lin, Yelong Shen, Jian Guo, and Nan Duan.
\newblock Enhancing chain-of-thoughts prompting with iterative bootstrapping in
  large language models.
\newblock {\em arXiv preprint arXiv:2304.11657}, 2023.

\bibitem{wang2022self}
Xuezhi Wang, Jason Wei, Dale Schuurmans, Quoc Le, Ed~Chi, and Denny Zhou.
\newblock Self-consistency improves chain of thought reasoning in language
  models.
\newblock {\em arXiv preprint arXiv:2203.11171}, 2022.

\bibitem{vodrahalli2022uncalibrated}
Kailas Vodrahalli, Tobias Gerstenberg, and James~Y Zou.
\newblock Uncalibrated models can improve human-ai collaboration.
\newblock {\em Advances in Neural Information Processing Systems},
  35:4004--4016, 2022.

\bibitem{srivastava2022beyond}
Aarohi Srivastava, Abhinav Rastogi, Abhishek Rao, Abu Awal~Md Shoeb, Abubakar
  Abid, Adam Fisch, Adam~R Brown, Adam Santoro, Aditya Gupta, Adri{\`a}
  Garriga-Alonso, et~al.
\newblock Beyond the imitation game: Quantifying and extrapolating the
  capabilities of language models.
\newblock {\em arXiv preprint arXiv:2206.04615}, 2022.

\bibitem{papineni2002bleu}
Kishore Papineni, Salim Roukos, Todd Ward, and Wei-Jing Zhu.
\newblock Bleu: a method for automatic evaluation of machine translation.
\newblock In {\em Proceedings of the 40th annual meeting of the Association for
  Computational Linguistics}, pages 311--318, 2002.

\bibitem{ghosh2019generating}
Shalini Ghosh, Giedrius Burachas, Arijit Ray, and Avi Ziskind.
\newblock Generating natural language explanations for visual question
  answering using scene graphs and visual attention.
\newblock {\em arXiv preprint arXiv:1902.05715}, 2019.

\bibitem{salewski2022clevr}
Leonard Salewski, A~Sophia Koepke, Hendrik~PA Lensch, and Zeynep Akata.
\newblock Clevr-x: A visual reasoning dataset for natural language
  explanations.
\newblock In {\em xxAI-Beyond Explainable AI: International Workshop, Held in
  Conjunction with ICML 2020, July 18, 2020, Vienna, Austria, Revised and
  Extended Papers}, pages 69--88. Springer, 2022.

\bibitem{chadha2021ireason}
Aman Chadha and Vinija Jain.
\newblock ireason: Multimodal commonsense reasoning using videos and natural
  language with interpretability.
\newblock {\em arXiv preprint arXiv:2107.10300}, 2021.

\bibitem{yu2020long}
Ting Yu, Jun Yu, Zhou Yu, Qingming Huang, and Qi~Tian.
\newblock Long-term video question answering via multimodal hierarchical memory
  attentive networks.
\newblock {\em IEEE Transactions on Circuits and Systems for Video Technology},
  31(3):931--944, 2020.

\bibitem{zarifhonarvar2023economics}
Ali Zarifhonarvar.
\newblock Economics of chatgpt: A labor market view on the occupational impact
  of artificial intelligence.
\newblock {\em Available at SSRN 4350925}, 2023.

\bibitem{eloundou2023gpts}
Tyna Eloundou, Sam Manning, Pamela Mishkin, and Daniel Rock.
\newblock Gpts are gpts: An early look at the labor market impact potential of
  large language models.
\newblock {\em arXiv preprint arXiv:2303.10130}, 2023.

\bibitem{bailey2022meta}
Phoebe~E Bailey, Tarren Leon, Natalie~C Ebner, Ahmed~A Moustafa, and Gabrielle
  Weidemann.
\newblock A meta-analysis of the weight of advice in decision-making.
\newblock {\em Current Psychology}, pages 1--26, 2022.

\bibitem{galesic2009effects}
Mirta Galesic and Michael Bosnjak.
\newblock Effects of questionnaire length on participation and indicators of
  response quality in a web survey.
\newblock {\em Public opinion quarterly}, 73(2):349--360, 2009.

\end{thebibliography}

%%%%%%%%%%%%%%%%%%%%%%%
% Appendix
%%%%%%%%%%%%%%%%%%%%%%%
\newpage
\appendix
\section{Procedure Appendix}
\label{sec:appendix_procedure}

%%%%%%%%%%%%%%%%%%%
% DATASET
%%%%%%%%%%%%%%%%%%%

\subsection{Dataset}
\label{sec:appendix_dataset}

\paragraph{Question selection}
To select topics, the authors attempted questions within each topic and used intuition to downsize the benchmark to 25 topics. The \verb|dev| and \verb|train| splits are used to sample questions. All 5 questions in each \verb|dev| set and up to 30 questions from the \verb|train| set are included. A smaller set of questions is preferred in order to better capture question-specific random effects. The much larger \verb|test| set is reserved for model evaluation and follow-up studies. For model evaluation, 50 questions are sampled uniformly from all splits for the topics.

\begin{table}[H]
    \centering
    \caption{Question counts and descriptions \cite{hendrycks2020measuring} by topic.}
    \begin{tabular}{rcl}
    \toprule
    \textbf{Topic}                      & \textbf{Questions} &     \textbf{Description}                                               \\
    \midrule
    Clinical Knowledge                  & 34                 & Spot diagnosis, joints, abdominal examination, ...                 \\
    Conceptual Physics                  & 31                 & Electromagnetism, thermodynamics, special relativity, ...          \\
    Elementary Mathematics              & 35                 & Word problems, multiplication, remainders, rounding, ...           \\
    Formal Logic                        & 19                 & Propositions, predicate logic, first-order logic, ...              \\
    Global Facts                        & 15                 & Extreme poverty, literacy rates, life expectancy, ...              \\
    High School Biology                 & 35                 & Cellular structure, molecular biology, ecology, ...                \\
    High School Chemistry               & 27                 & Analytical, organic, inorganic, physical, ...                      \\
    High School Computer Science        & 14                 & Algorithms, systems, graphs, recursion, ...                        \\
    High School European History        & 23                 & Renaissance, reformation, industrialization, ...                   \\
    High School Geography               & 27                 & Population migration, rural land-use, urban processes, ...         \\
    High School Government and Politics & 26                 & Branches of government, civil liberties, political ideologies, ... \\
    High School Macroeconomics          & 35                 & Economic indicators, national income, international trade, ...     \\
    High School Microeconomics          & 31                 & Supply and demand, imperfect competition, market failure, ...      \\
    High School Physics                 & 22                 & Kinematics, energy, torque, fluid pressure, ...                    \\
    High School Psychology              & 35                 & Behavior, personality, emotions, learning, ...                     \\
    High School Statistics              & 28                 & Random variables, sampling distributions, chi-square tests, ...    \\
    High School U.S. History            & 27                 & Civil War, the Great Depression, The Great Society, ...            \\
    High School World History           & 31                 & Ottoman empire, economic imperialism, World War I, ...             \\
    Human Aging                         & 28                 & Senescence, dementia, longevity, personality changes, ...          \\
    Human Sexuality                     & 17                 & Pregnancy, sexual differentiation, sexual orientation, ...         \\
    Miscellaneous Topics                & 35                 & Agriculture, Fermi estimation, pop culture, ...                    \\
    Nutrition                           & 35                 & Metabolism, water-soluble vitamins, diabetes, ...                  \\
    Philosophy                          & 35                 & Skepticism, phronesis, skepticism, Singer’s Drowning Child, ..     \\
    Sociology                           & 27                 & Socialization, cities and community, inequality and wealth, ...    \\
    U.S. Foreign Policy                 & 16                 & Soft power, Cold War foreign policy, isolationism, ...         \\
    \bottomrule
\end{tabular}
    \label{tab:topic_descriptions}
\end{table}

\paragraph{Question order}

\begin{figure}[H]
  \centering
  \includegraphics[width=.5\textwidth]{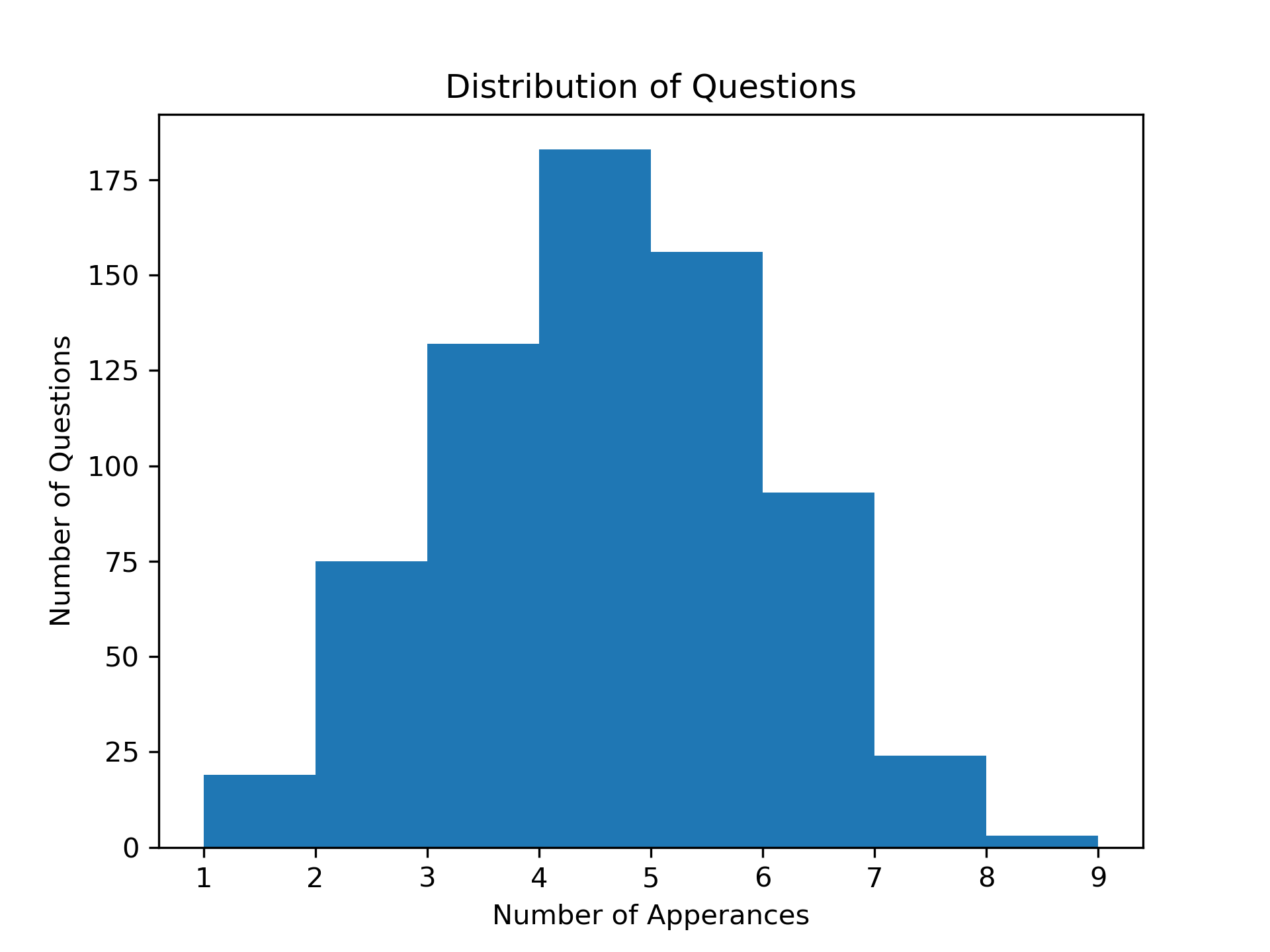}
  \caption{\textbf{Question distribution.} Every question is presented at least once. The modal number of appearances is 4 with a maximum of 8. The distribution is consistent with the result of a uniform selection process.}
  \label{fig:question_distribution}
\end{figure}

The dataset is reordered twice during survey administration to overcome a limitation in Qualtrics that presents questions in increasing order of index. In the first reordering, the first and second halves of the questions are switched. The second reordering, the questions are shuffled entirely. As a result, the distribution of question appearances is roughly normal with no apparent bias (Figure \ref{fig:question_distribution}). The procedure for shuffling questions is documented in the accompanying \href{https://github.com/petezh/ChatGPT-Advice/}{repository}.

%%%%%%%%%%%%%%%%%%%
% MODEL EVALUATION
%%%%%%%%%%%%%%%%%%%

\subsection{Model Evaluation}
\label{sec:appendix_model_evaluation}

\begin{figure}[H]
  \centering
  \includegraphics[width=.8\textwidth]{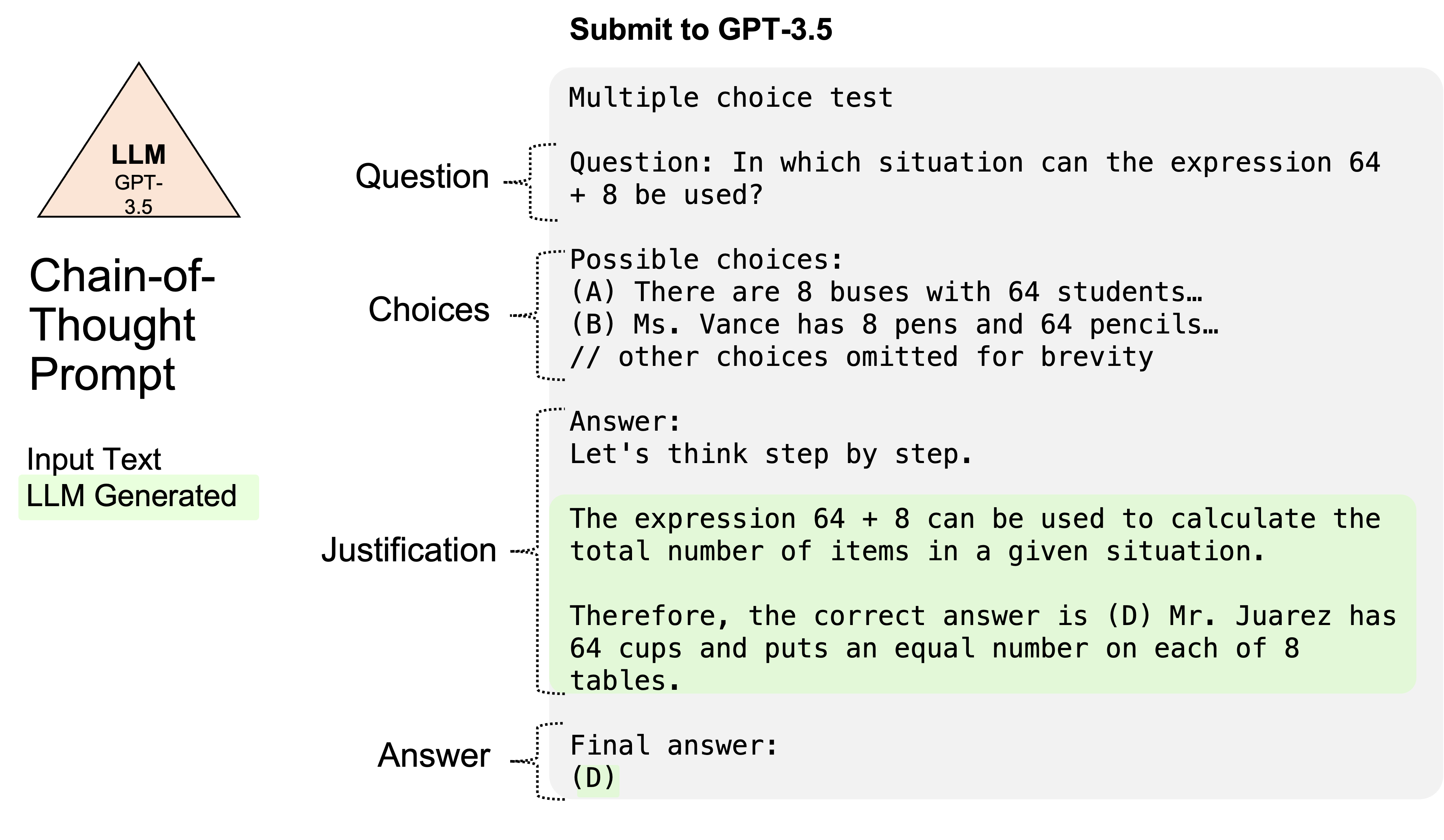}
  \caption{\textbf{Chain-of-thought prompt.} The model is prompted to think "step-by-step" and provide a reasoning. The final answer is treated as the model's answer.}
  \label{fig:prompt_cot}
\end{figure}

\begin{figure}[H]
  \centering
  \includegraphics[width=.8\textwidth]{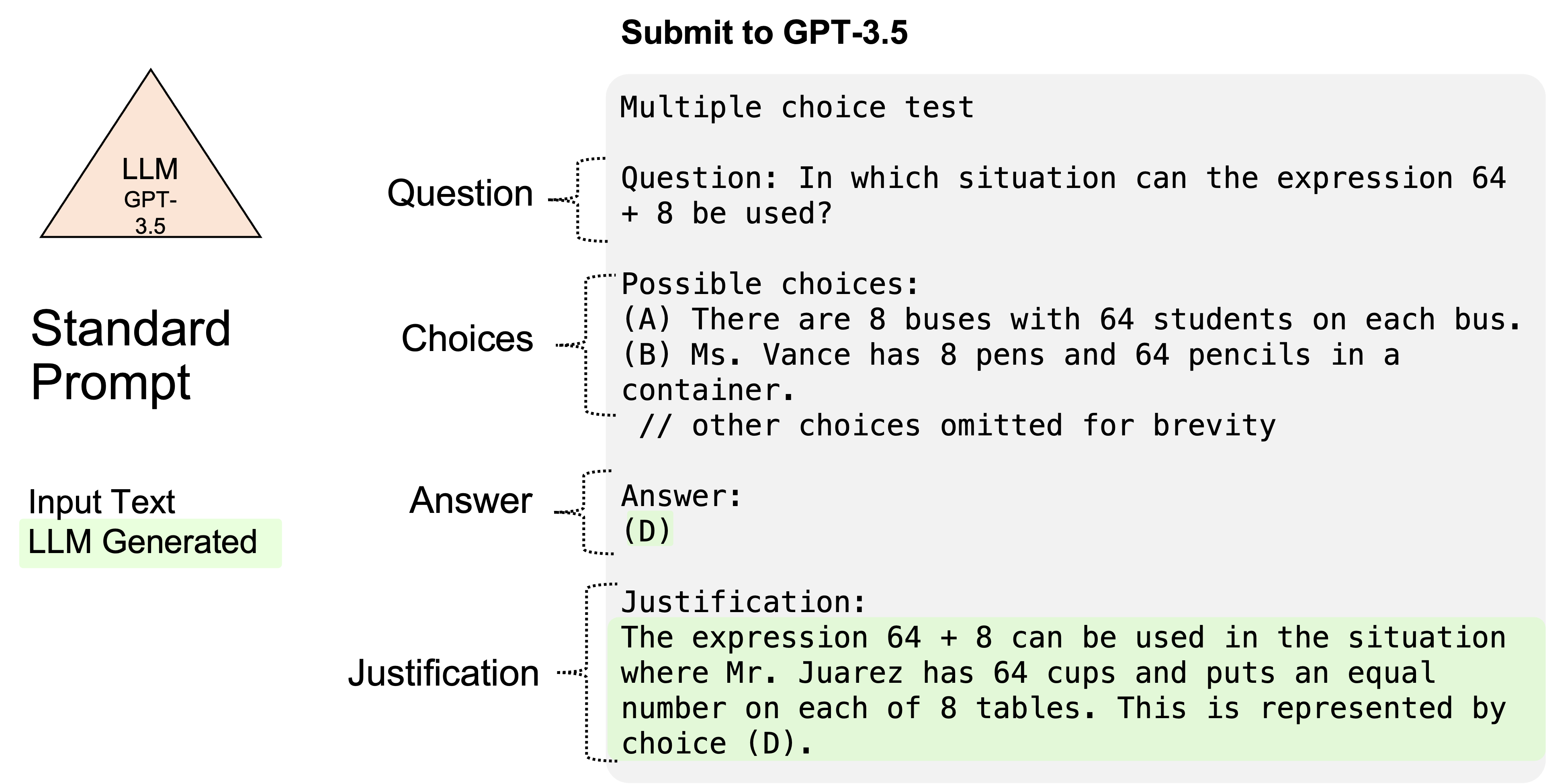}
  \caption{\textbf{Standard prompt.} The model is first prompted to immediately answer the question. Then it is prompted to provide a justification.}
  \label{fig:prompt_standard}
\end{figure}

All model evaluations use the Completions API with \verb|engine=”text-davinci-003”|, \verb|temperature=0|, and control the completion length. The CoT prompt (Figure \ref{fig:prompt_cot}) is compared against the standard prompt (Figure \ref{fig:prompt_standard}). In a reversal of the CoT approach, the model is first forced to output a response and then provide a justification.

\begin{figure}[H]
  \centering
\begin{subfigure}{.65\textwidth}
  \includegraphics[width=\textwidth]{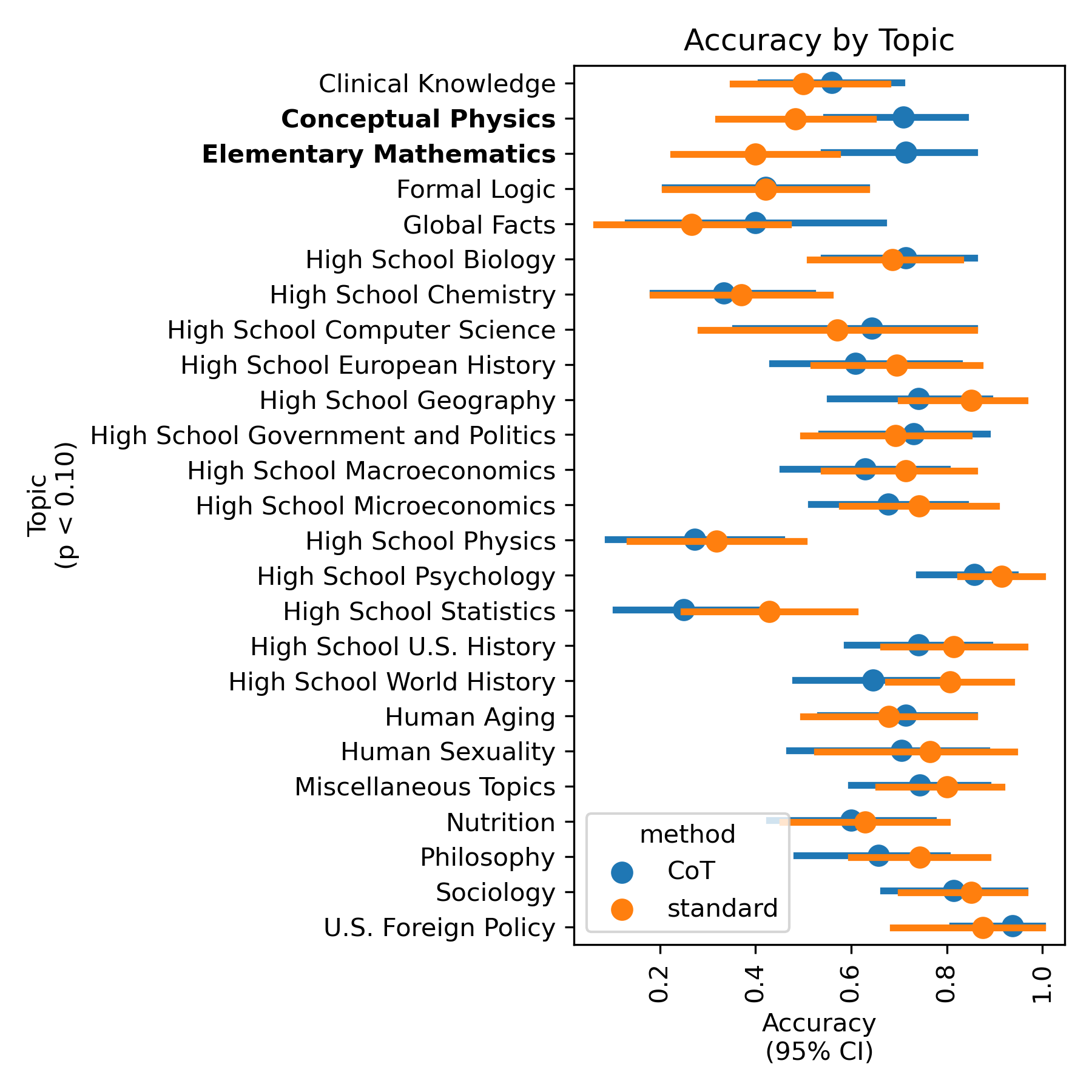}
  \caption{Accuracy by topic.}
  \label{fig:model_accuracy_by_topic}
\end{subfigure}%
\begin{subfigure}{.35\textwidth}
  \includegraphics[width=\textwidth]{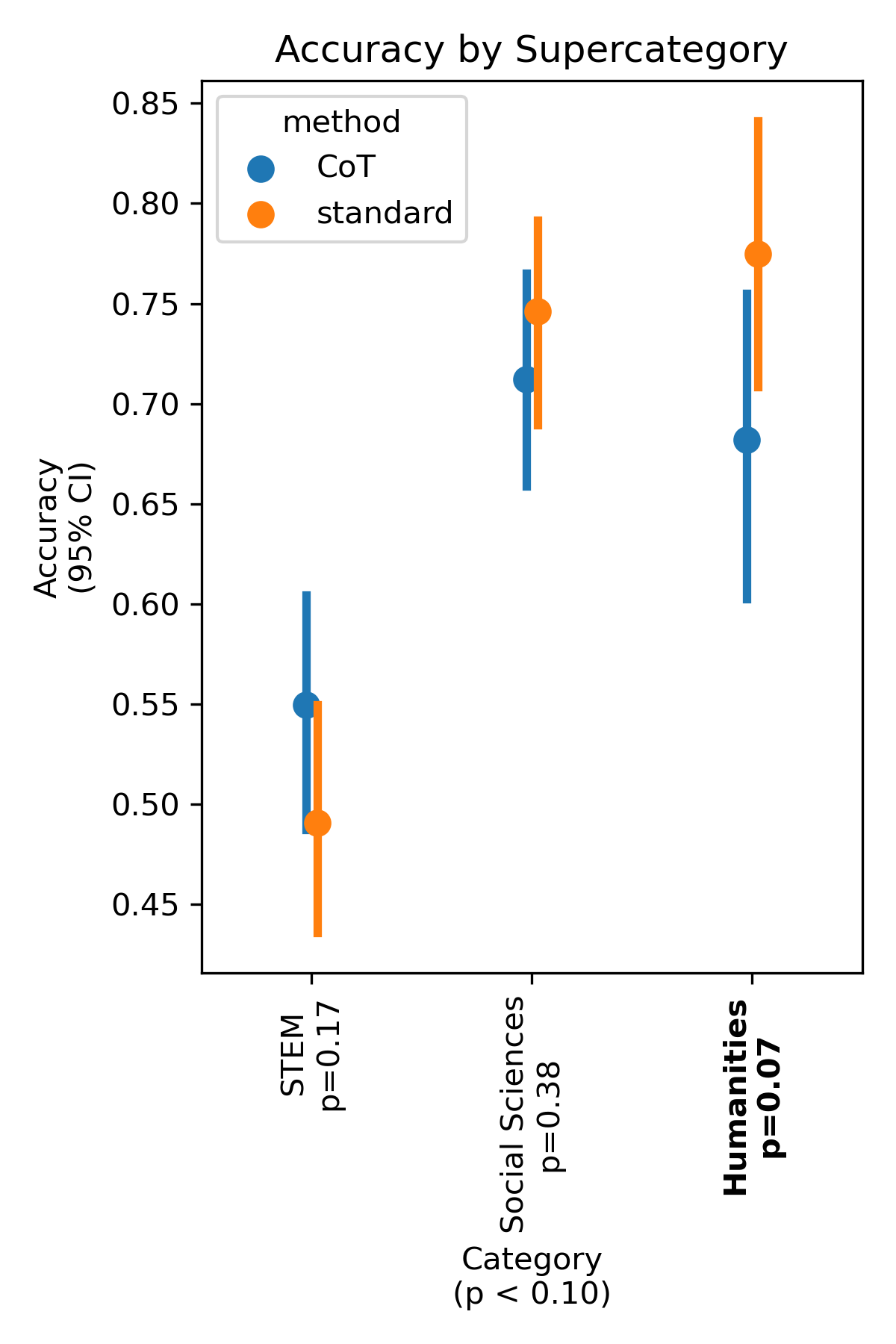}
  \caption{Accuracy by category.}
  \label{fig:model_accuracy_by_category}
\end{subfigure}
  \caption{\textbf{Accuracy by topic and category.} Error bars are 95\% confidence intervals. Topics and categories where prompts diverged significantly (at a 90\% level) are \textbf{bolded}.}
\end{figure}

Both prompts achieved an accuracy of $62.9\% \pm 3.4\%$ but their responses are only moderately correlated (Spearman’s $r$=0.573, $p$<1e-68). Prompt accuracy differs significantly across topics (Figure \ref{fig:model_accuracy_by_topic}) and categories (Figure \ref{fig:model_accuracy_by_category}). Generally, CoT achieves improvements on procedural topics (Conceptual Physics, Elementary Mathematics) at the expense of performance losses on fact-based humanities topics (history topics, Philosophy).

The CoT prompt is preferred due to the quality of justifications. Whereas the standard prompt often leads models to adopt awkward justifications, CoT generates a coherent stream of reasoning that resolves to an answer. One limitation of this choice is the possibility of invalid outputs. For five questions (< 1\% of the dataset), the model justification rejects all of the answer choices and answers \verb|"None"| or \verb|"E"|. In the evaluation, these answers are as treated as incorrect with a weight on advice of 0. These "errors" are preserved in order to approximate a real-world analog in which an AI chatbot might similarly reject all options.

%%%%%%%%%%%%%%%%%%%
% PROCESSING
%%%%%%%%%%%%%%%%%%%

\subsection{Data Processing}
\label{sec:appendix_processing}

\paragraph{Weight on advice}
Most studies \cite{bailey2022meta} compute WoA as the the difference between the initial and adjusted judgment divided by the difference between the initial judgment and the advice. The advice is assumed to correspond to 100\% belief in the advised answer. In the context of probabilities, this approach might conflict with intuitions about scale in the context of probability sums. If a participant is initially 99\% confident in the advisor’s answer, then an adjustment to 100\% is not a large change. Yet whereas the absolute change is small ($WoA=0.01$), proportionate change is the largest possible value ($WoA =1$). This property can lead to noisy results for WoA when many participants opt to “round” high confidence levels (80-99\%) up to full confidence (100\%).

\paragraph{Topic familiarity}
Participants are asked “How comfortable do you feel with the following topic areas?” and provided the options "\textcolor{Blue}{Uncomfortable}", "Neutral", and "\textcolor{Maroon}{Comfortable}.” These are mapped to \textcolor{Blue}{-1}, 0, and \textcolor{Maroon}{1} respectively in summary statistics and use indicators in the regression, as preregistered.  to familiarity areas using the mapping in Table \ref{tab:topic_to_familiarity}. These areas are created to roughly capture the skills and knowledge required in each topic area as an alternative to tedious topic-by-topic ratings of comfort level.

\begin{table}[H]
    \centering
    \caption{Linking topics to familiarity areas.}
    \begin{tabular}{lcr}
\toprule
                         Topic/Task &  Questions &         Familiarity Area \\
\midrule
                 Clinical Knowledge &        34 & Biological Sciences \\
                 Conceptual Physics &        31 &             Physics \\
             Elementary Mathematics &        35 &         Mathematics \\
                       Formal Logic &        19 &         Mathematics \\
                       Global Facts &        15 &              Trivia \\
                High School Biology &        35 & Biological Sciences \\
              High School Chemistry &        27 & Biological Sciences \\
       High School Computer Science &        14 &    Computer Science \\
       High School European History &        23 &             History \\
              High School Geography &        27 &              Trivia \\
High School Government and Politics &        26 &           Economics \\
         High School Macroeconomics &        35 &           Economics \\
         High School Microeconomics &        31 &           Economics \\
                High School Physics &        22 &             Physics \\
             High School Psychology &        35 & Biological Sciences \\
             High School Statistics &        28 &         Mathematics \\
           High School U.S. History &        27 &             History \\
          High School World History &        31 &             History \\
                        Human Aging &        28 & Biological Sciences \\
                    Human Sexuality &        17 & Biological Sciences \\
               Miscellaneous Topics &        35 &              Trivia \\
                          Nutrition &        35 & Biological Sciences \\
                         Philosophy &        35 &          Literature \\
                          Sociology &        27 &          Literature \\
                U.S. Foreign Policy &        16 &             History \\
\bottomrule
\end{tabular}

    \label{tab:topic_to_familiarity}
\end{table}

\paragraph{Past usage}
Participants answer four progressive questions about usage of AI chatbots. Each question is only displayed if they respond "yes" to the previous:
\begin{enumerate}
    \item Have you heard of AI chatbots before?
    \item Have you used AI chatbots before?
    \item Have you used AI chatbots in a classroom setting before? (homework, quiz, studying, etc.)
    \item Have you used AI chatbots to answer multiple choice questions before?
\end{enumerate}

The number of questions to which they respond "yes", which may vary from 0 to 4, is summed and used to measure past usage.

\paragraph{Experience}
Participant beliefs about the accuracy of advice are modeled with a Beta-Bernoulli. A fairly weak prior of $\textrm{Beta}(\alpha=0.5,\beta=0.5)$ is chosen. Upon receiving feedback on a question, it is assumed that participants update to $\textrm{Beta}(\alpha+1,\beta)$ if the advice is correct and $\textrm{Beta}(\alpha,\beta+1)$ if the advice is wrong. This is a fairly strong assumption since it assumes that participants can accurately track and evenly weight past performance.

\newpage
\section{Analysis Appendix}
\label{sec:appendix_analysis}

%%%%%%%%%%%%%%%%%%%
% SUMMARY STATS
%%%%%%%%%%%%%%%%%%%

\subsection{Descriptive Statistics}
\label{sec:appendix_descriptive_stats}

\paragraph{Summary statistics}
Summary statistics are displayed in Table \ref{tab:summary_stats}.

\begin{table}[H]
    \centering
    \caption{Summary statistics for key variables.}
    \begin{tabular}{lrrrrrrr}
\toprule
{} &  weight\_on\_advice &  init\_advice\_confidence &  advice\_confidence &  advice\_is\_correct \\
\midrule
count &          2828.000 &                2828.000 &           2828.000 &           2828.000 \\
mean  &             0.337 &                   0.359 &              0.587 &              0.639 \\
std   &             0.395 &                   0.309 &              0.357 &              0.480 \\
min   &             0.000 &                   0.000 &              0.000 &              0.000 \\
25\%   &             0.000 &                   0.156 &              0.278 &              0.000 \\
50\%   &             0.149 &                   0.250 &              0.571 &              1.000 \\
75\%   &             0.672 &                   0.500 &              1.000 &              1.000 \\
max   &             1.000 &                   1.000 &              1.000 &              1.000 \\
\bottomrule\\
\toprule
{} &  net\_familiarity &  uncomfortable &   neutral &  comfortable \\
\midrule
count &         2828.000 &       2828.000 &  2828.000 &     2828.000 \\
mean  &            0.056 &          0.269 &     0.405 &        0.326 \\
std   &            0.770 &          0.444 &     0.491 &        0.469 \\
min   &           -1.000 &          0.000 &     0.000 &        0.000 \\
25\%   &           -1.000 &          0.000 &     0.000 &        0.000 \\
50\%   &            0.000 &          0.000 &     0.000 &        0.000 \\
75\%   &            1.000 &          1.000 &     1.000 &        1.000 \\
max   &            1.000 &          1.000 &     1.000 &        1.000 \\
\bottomrule\\
\toprule
{} &  usage\_level &  heard\_of &      used &  used\_in\_class &  answered\_mc \\
\midrule
count &     2828.000 &  2828.000 &  2828.000 &       2828.000 &     2828.000 \\
mean  &        2.678 &     0.986 &     0.832 &          0.595 &        0.264 \\
std   &        1.068 &     0.120 &     0.374 &          0.491 &        0.441 \\
min   &        0.000 &     0.000 &     0.000 &          0.000 &        0.000 \\
25\%   &        2.000 &     1.000 &     1.000 &          0.000 &        0.000 \\
50\%   &        3.000 &     1.000 &     1.000 &          1.000 &        0.000 \\
75\%   &        4.000 &     1.000 &     1.000 &          1.000 &        1.000 \\
max   &        4.000 &     1.000 &     1.000 &          1.000 &        1.000 \\
\bottomrule\\
\toprule
{} &  question\_num &  correct\_advice\_count &  incorrect\_advice\_count &  init\_time &  adjusted\_time \\
\midrule
count &      2828.000 &              2828.000 &                2828.000 &   2828.000 &       2828.000 \\
mean  &        13.099 &                 7.826 &                   4.273 &     28.292 &         10.728 \\
std   &         8.151 &                 5.482 &                   3.322 &     20.390 &          8.608 \\
min   &         1.000 &                 0.000 &                   0.000 &      5.458 &          5.115 \\
25\%   &         6.000 &                 3.000 &                   2.000 &     13.497 &          6.126 \\
50\%   &        12.000 &                 7.000 &                   4.000 &     22.524 &          7.536 \\
75\%   &        18.000 &                11.000 &                   6.000 &     36.145 &         11.388 \\
max   &        40.000 &                27.000 &                  18.000 &     90.141 &         90.038 \\
\bottomrule
\end{tabular}

    \label{tab:summary_stats}
\end{table}

\paragraph{Conditions}
Participants ($n=188$) were assigned randomly to a 2x2 condition (see Figure \ref{tab:conditions}). Notably, significantly more participants (66) were assigned to receive a justification than not (52). The ability to evenly assign participants was limited by the survey platform and administration method, which led to many “in-progress” surveys that were not completed.

\begin{table}[H]
 \caption{Conditions assignments.}
  \centering
  \begin{tabular}{ccc}
\toprule
 & \multicolumn{2}{c}{\textit{Justification}}                   \\
\cmidrule(r){2-3}
\textit{Advisor}     & Yes          & No   \\
\midrule
AI chatbot & 32         & 24     \\
expert     & 34         & 28     \\
\bottomrule
\end{tabular}
  \label{tab:conditions}
\end{table}

\begin{figure}[H]
  \centering
  \includegraphics[width=.5\textwidth]{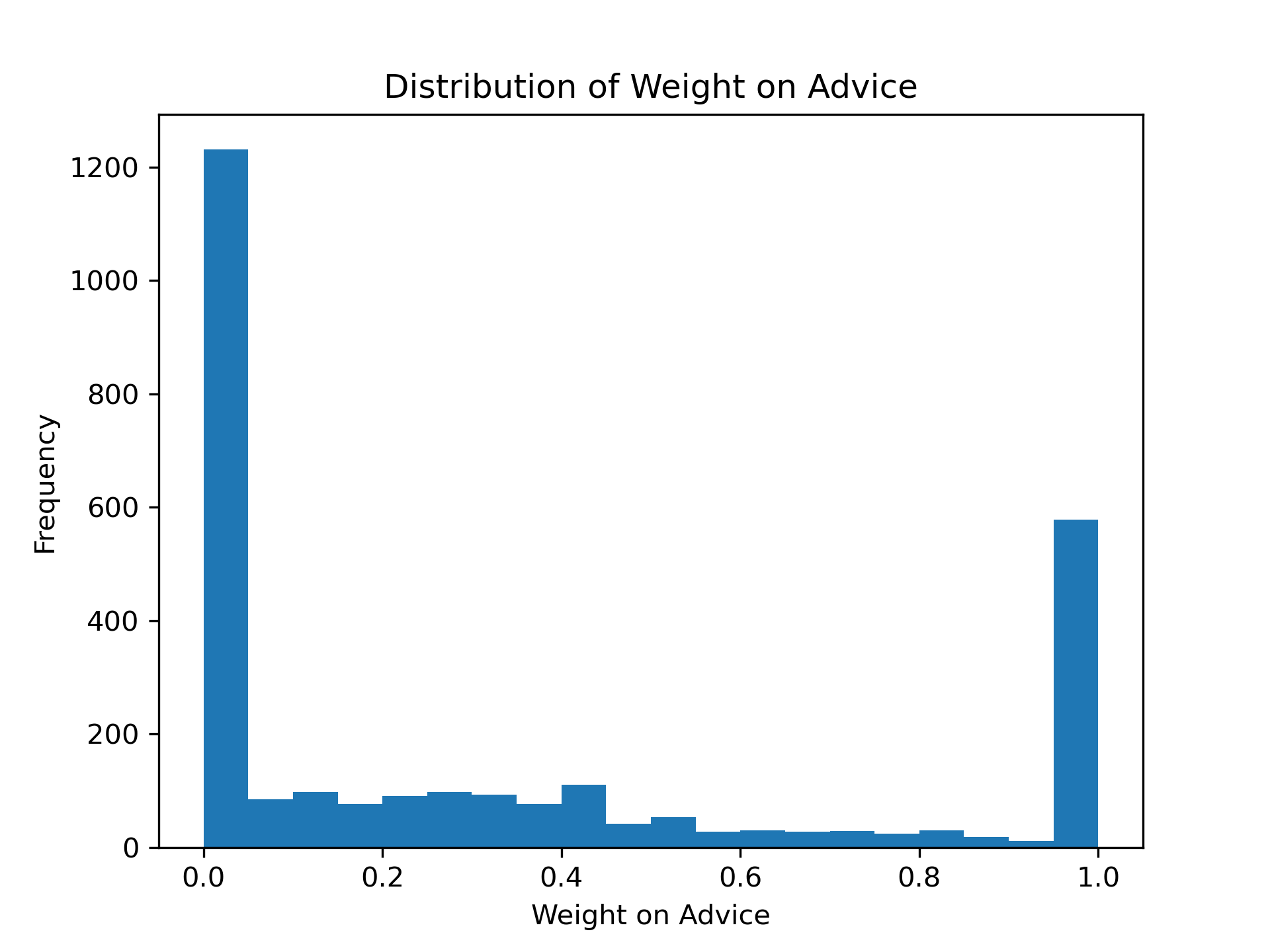}
  \caption{\textbf{Distribution of weight on advice.} The distribution is bimodal at 0 and 1. Weight on advice is depicted post-winsorization. Notice the absence of values between 80\% and 95\%, suggesting a tendency to "round up" high confidence to full confidence.}
  \label{fig:weight_on_advice_dist}
\end{figure}

\paragraph{Weight on advice}
Average $\WoA$ was 0.237 with a standard deviation of 0.300 (Figure \ref{fig:weight_on_advice_dist}). In the sample, 40.45\% of answers placed no weight on advice ($\WoA \leq 0$).

\begin{figure}[H]
  \centering
  \includegraphics[width=.6\textwidth]{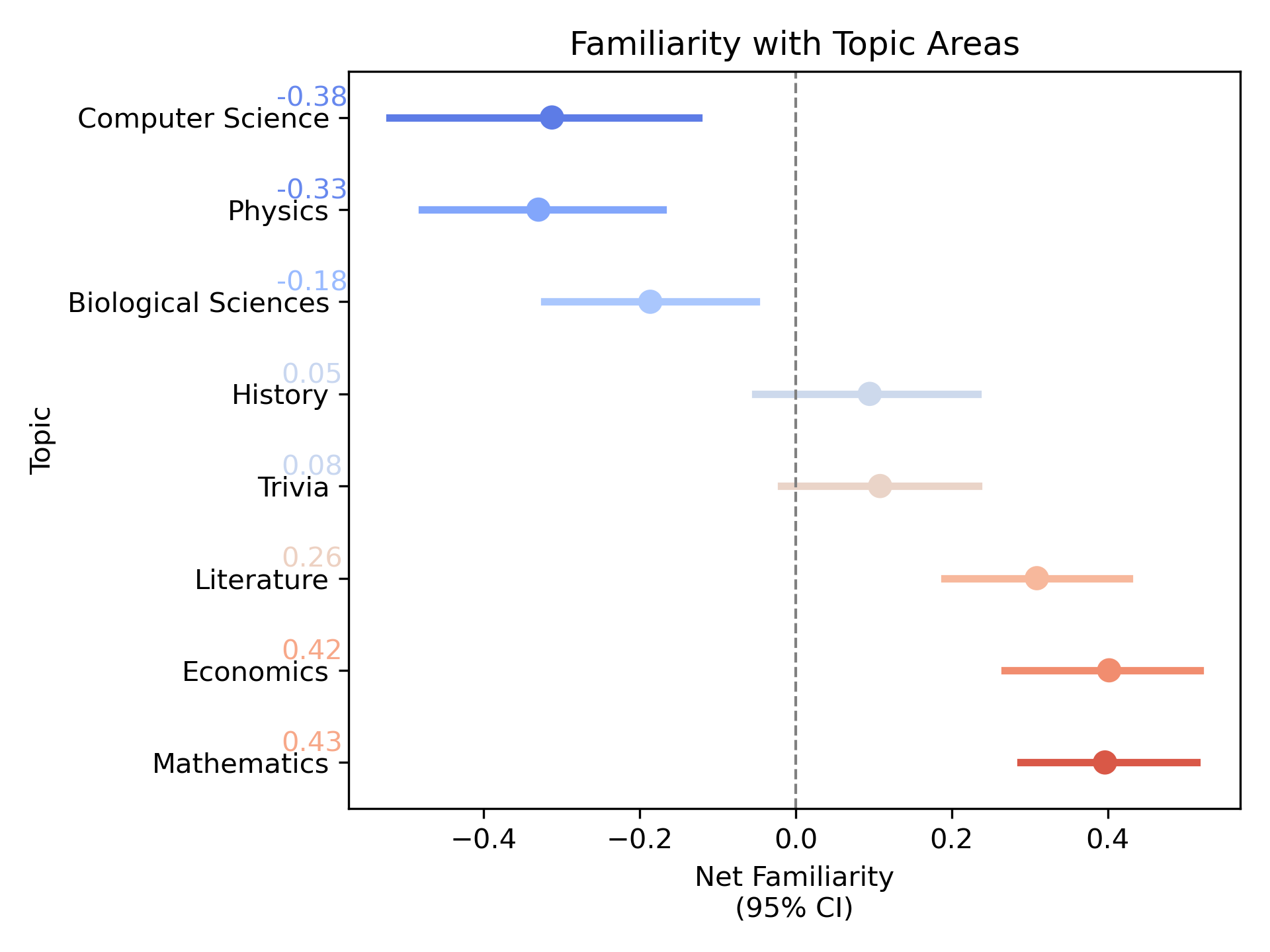}
  \caption{\textbf{Familiarity in topic areas.} Net familiarity is annotated above each topic label. Error bars are 95\% confidence intervals.}
  \label{fig:familiarity}
\end{figure}

\paragraph{Topic familiarity}
Familiarity varies significantly across topics (Figure \ref{fig:familiarity}). Participants are most familiar with economics and mathematics, which is sensible for a business degree program. Participants rated themselves least familiar with computer science, for which there is also the most variation. This seems to be due to the substantial number of participants majoring in Computer Science ($n=10$), EECS ($n=9$), or data science ($n=19$).

\begin{figure}[H]
  \centering
\begin{subfigure}{.5\textwidth}
  \includegraphics[width=\textwidth]{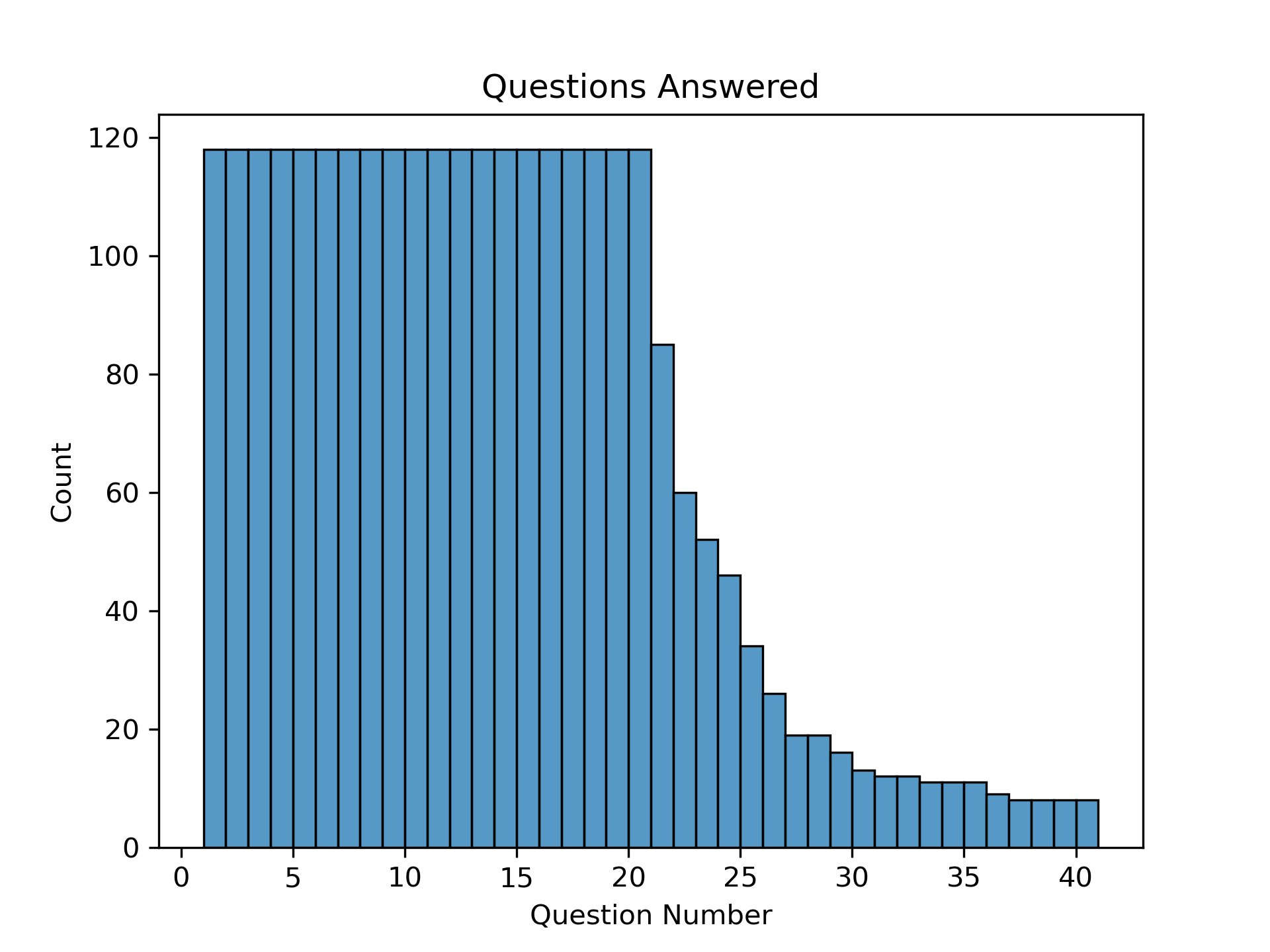}
  \caption{Number of questions answered.}
  \label{fig:questions_answered}
\end{subfigure}%
\begin{subfigure}{.5\textwidth}
  \includegraphics[width=\textwidth]{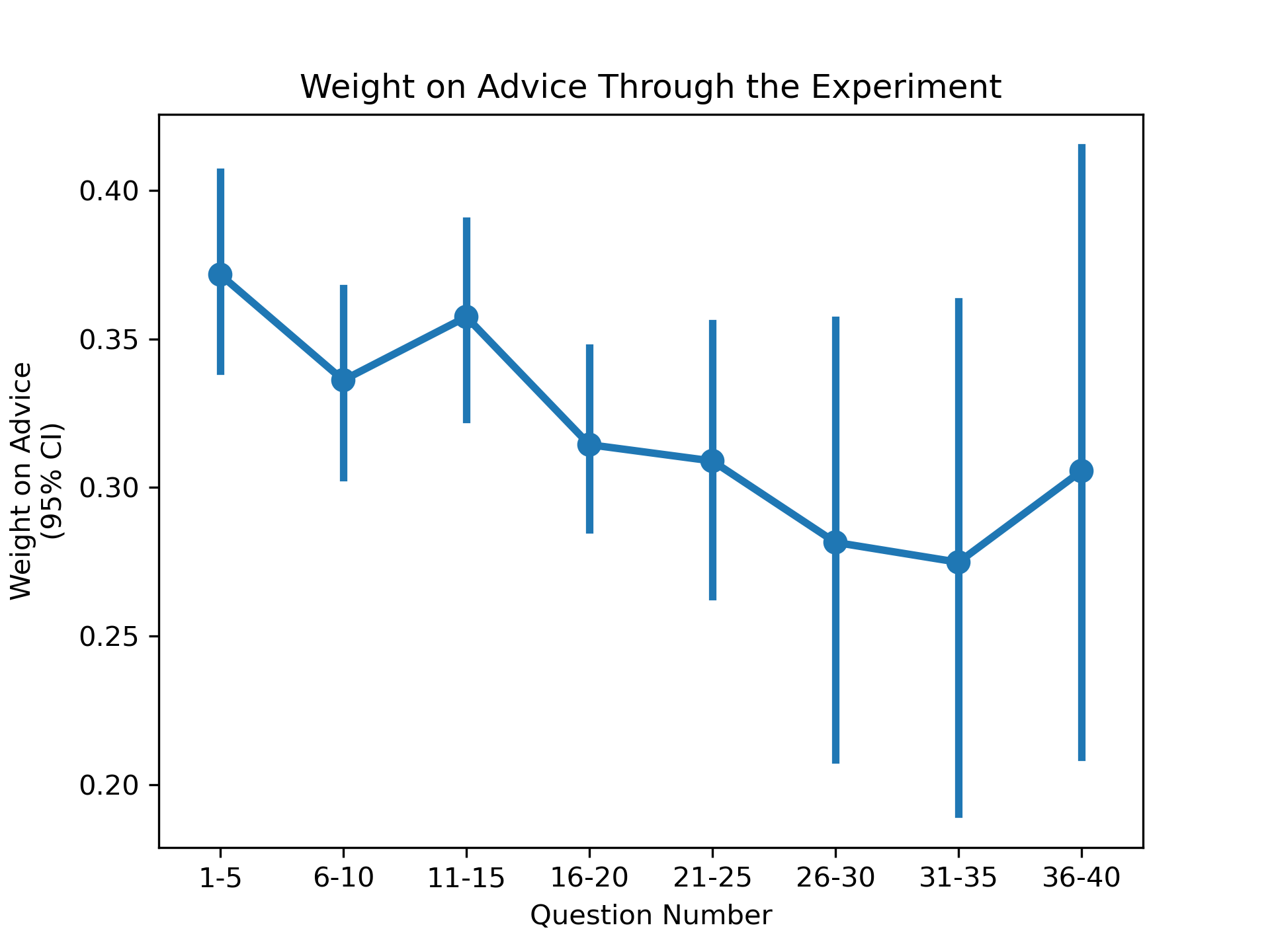}
  \caption{Weight on advice by question group.}
  \label{fig:over_questions}
\end{subfigure}
  \caption{\textbf{Questions answered.} Error bars are 95\% confidence intervals. Error bars widen past 20 questions because fewer participants voluntarily answered additional questions.}
\end{figure}

\paragraph{Questions answered.}
Most participants (92/118) answered less than 25 questions. There is a significant dropoff after around questions 20-22 once participants notice that they have the option to opt-out (Figure \ref{fig:questions_answered}.

Weight on advice appears to decline throughout the experiment (Figure \ref{fig:over_questions}). Conducting an type-2 ANOVA analysis of regressing weight on advice question groups yields a slightly significant result ($p$=0.0731). In the section \ref{sec:analysis}, that question number is a significant predictor of weight on advice. Note that because answering questions 21-40 is voluntary, the effect might be confounded by participant attributes (e.g. diligence or competitiveness).

All but one participant heard of ChatGPT and other AI chatbots. Among participants in the AI chatbot condition, participants appear to place much greater weight on advice if they have used ChatGPT, and specifically if they have used it to answer multiple choice questions. Note that Figure X is plotted without accounting for random effects so the standard errors may not be reflective.

\begin{figure}[H]
  \centering
\begin{subfigure}{.5\textwidth}
  \includegraphics[width=\textwidth]{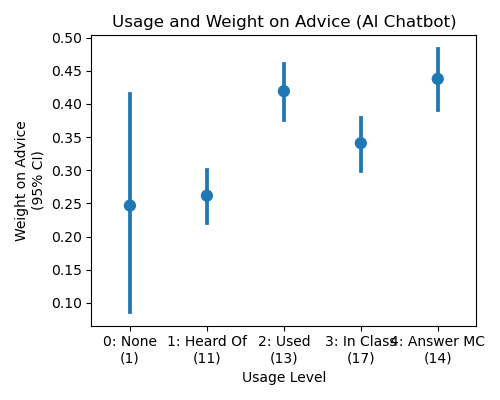}
  \caption{AI chatbot condition.}
  \label{fig:woa_by_advisor_AI}
\end{subfigure}%
\begin{subfigure}{.5\textwidth}
  \includegraphics[width=\textwidth]{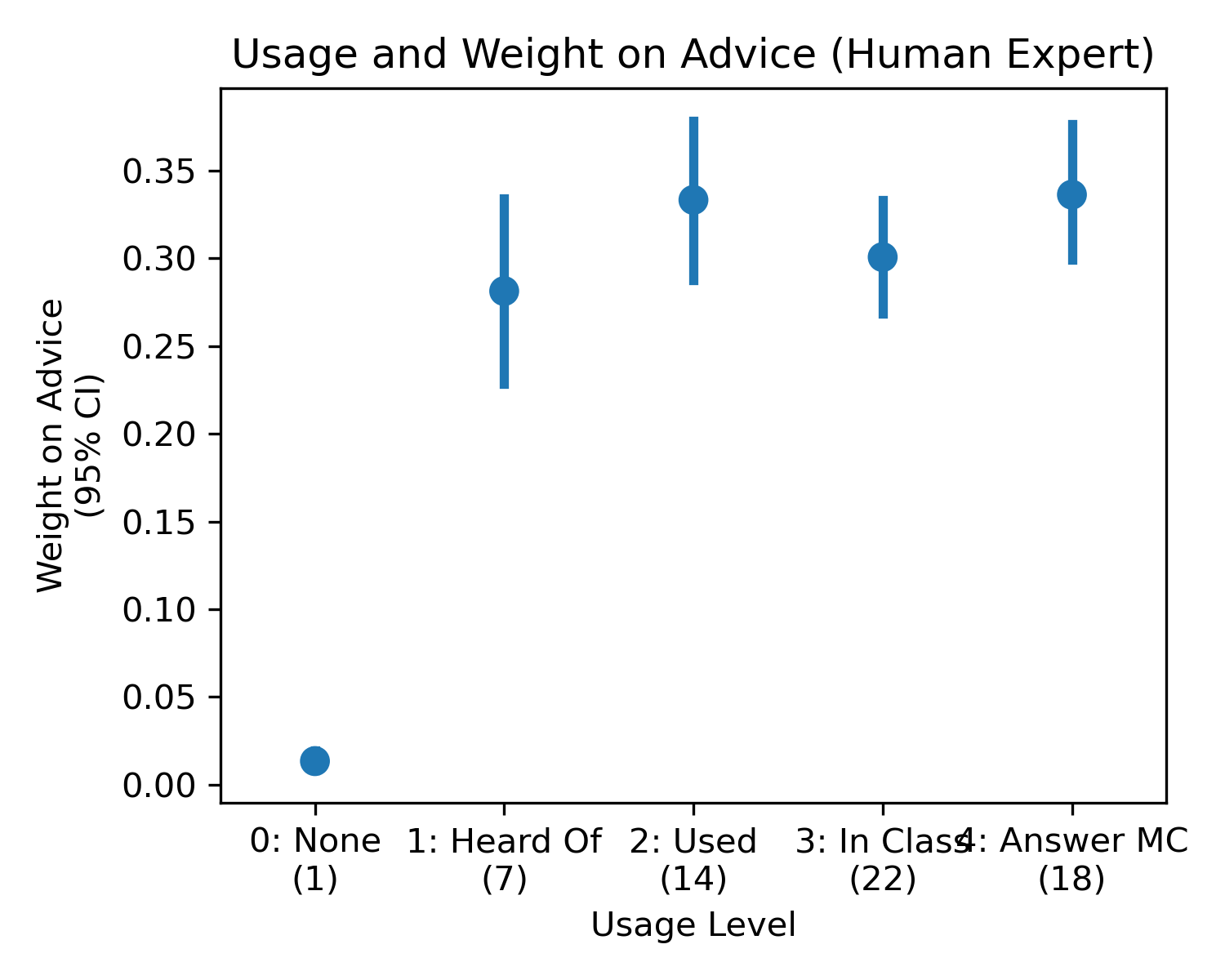}
  \caption{Expert condition.}
  \label{fig:woa_by_advisor_expert}
\end{subfigure}
  \caption{\textbf{Weight on advice by usage level.} Error bars are 95\% confidence intervals.}
\end{figure}

\paragraph{Past usage}
 All but one participant heard of ChatGPT and other AI chatbots. Among participants in the AI chatbot condition, participants appear to place much greater weight on advice if they have used ChatGPT, and specifically if they have used it to answer multiple choice questions. Note that Figure \ref{fig:woa_by_advisor_AI} is plotted without accounting for random effects so the standard errors may not be reflective.

These effects are much smaller but directionally similar in the expert condition (Figure \ref{fig:woa_by_advisor_expert}).

\paragraph{Advice correctness}

\begin{figure}[H]
  \centering
  \includegraphics[width=.6\textwidth]{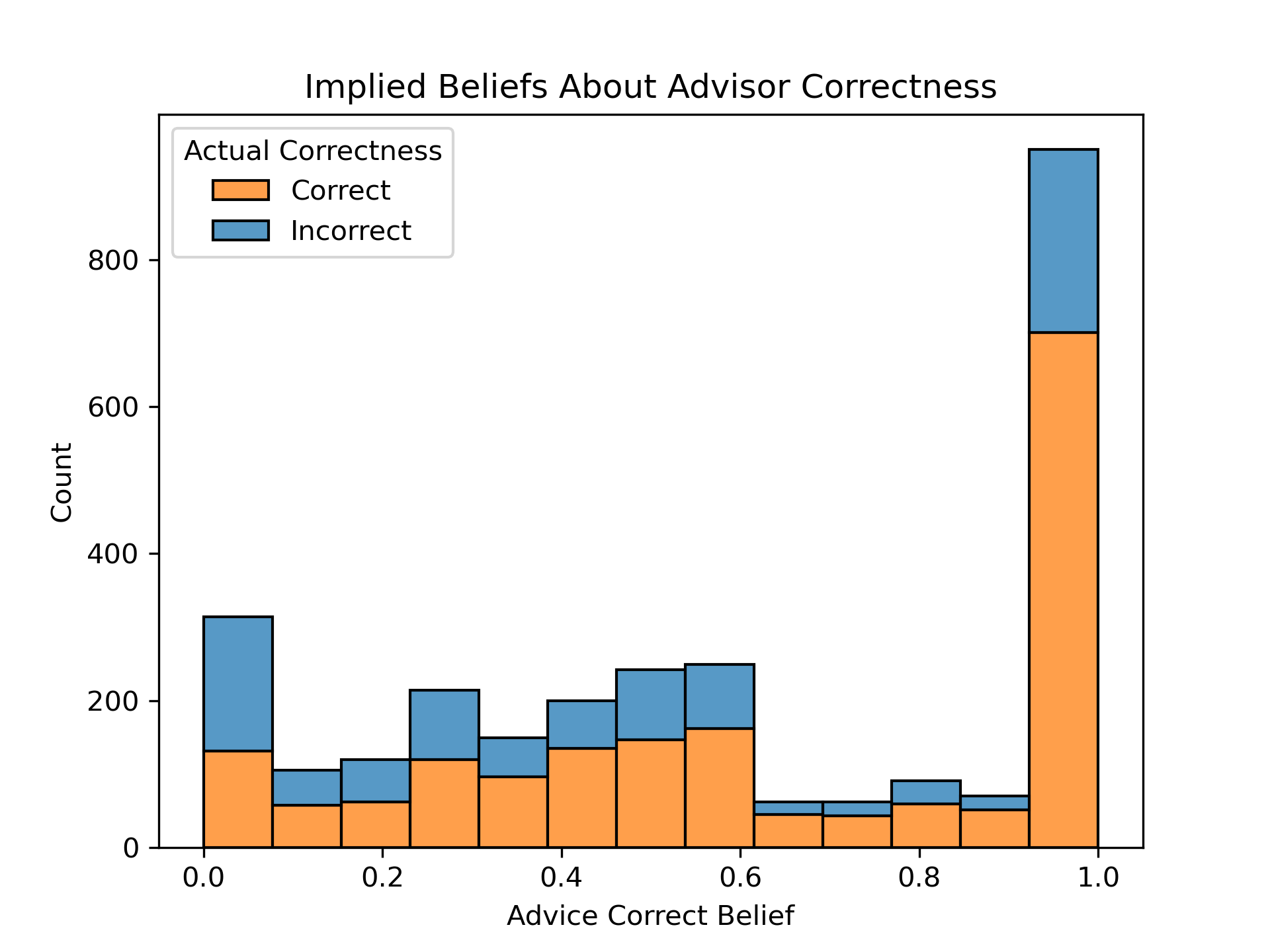}
  \caption{\textbf{Advice correctness beliefs.} Participants are apparently capable of discerning correct and incorrect advice. When advice is correct, a greater proportion of participants judge the advised answer as very likely to be correct, and vice versa.
}
  \label{fig:advice_correctness_belief}
\end{figure}

The advice is correct for 63.9\% of samples. Participants are somewhat able to discern the correctness of advice (Figure \ref{fig:advice_correctness_belief}). In general, judgements about advice correctness are bimodal near 0 and 1. There is a noticeable absence of beliefs in the 60\% to 90\% confidence range, suggesting that when participants are “pretty sure” that the advice is correct, they round up to 100\%.

\begin{figure}[H]
  \centering
  \includegraphics[width=.6\textwidth]{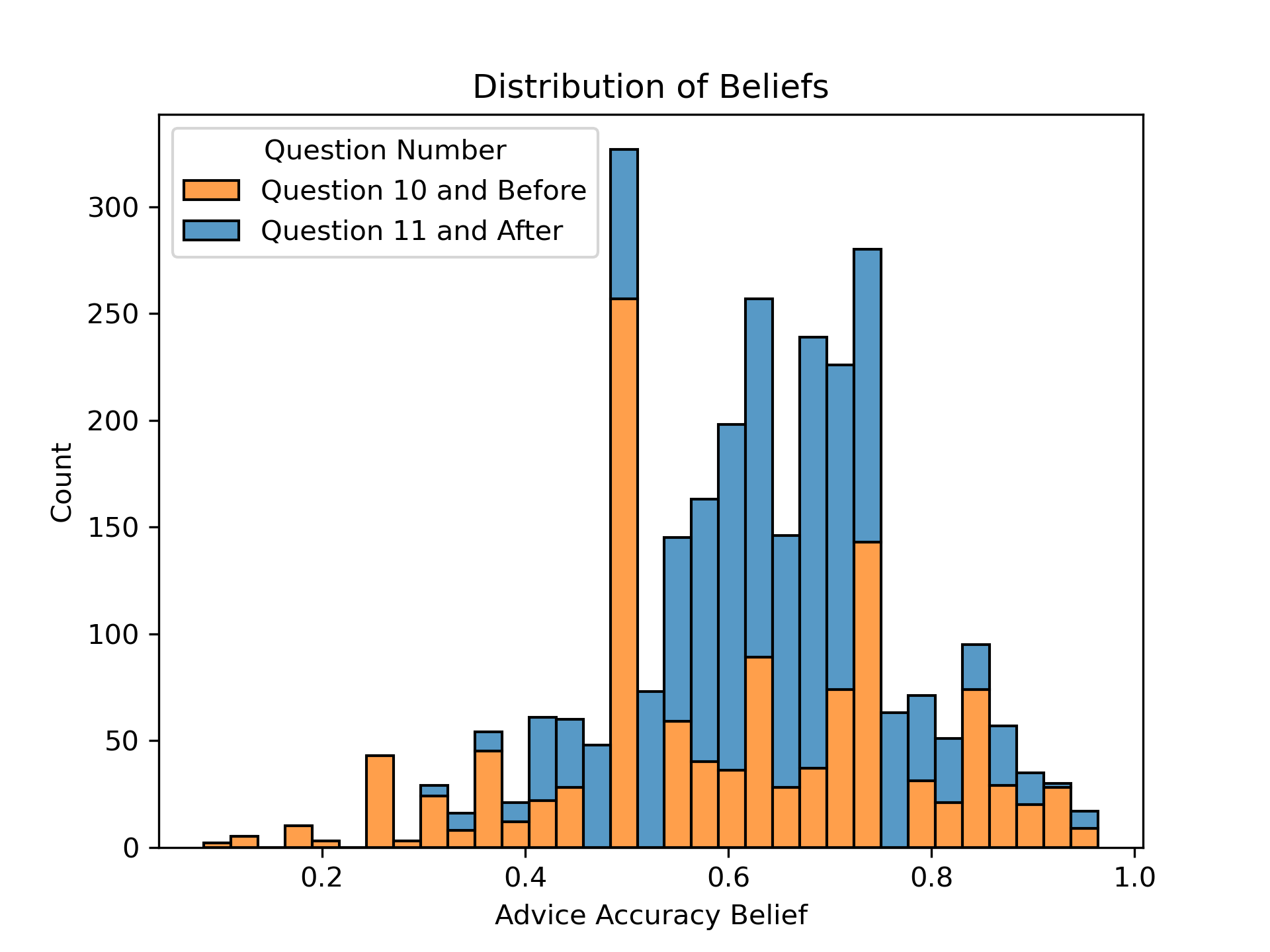}
  \caption{\textbf{Model advice accuracy beliefs} Beliefs have much greater spread in the first 10 questions as expected due to the greater probability of receiving mostly correct or incorrect answers. After, beliefs tend to be more centered around the true accuracy of 64.9\%.}
  \label{fig:advice_accuracy_belief}
\end{figure}

\paragraph{Experiences}
Beliefs about advice accuracy are roughly normally distributed with a mode at uninformative prior, 0.5 (Figure \ref{fig:advice_accuracy_belief}. The mean belief in accuracy is 60.5\% with a standard deviation of 13.1\%.

%%%%%%%%%%%%%%%%%%%
% OPTIONALITY
%%%%%%%%%%%%%%%%%%%
\newcommand{\Ai}{\textbf{\textcolor{BrickRed}{A2}}}
\newcommand{\Aii}{\textbf{\textcolor{Red}{A3}}}

\subsection{Optionality}
\label{sec:appendix_optionality}

The survey is fairly long, taking most particpants at least 20 minutes. Longer questionnaires are known to suffer from reduced response quality, particularly for later questions \cite{galesic2009effects}. This may, for example, explain and confound the significant decline in weight on advice over time we find in Table \ref{tab:main_regression_results}. But one unique feature of our survey is the opportunity to complete \textit{optional} questions for additional points. If a participant voluntarily answers additional questions, it might suggest greater engagement with the survey and improved response quality.

The regressions in Table \ref{tab:optionality_regression_results} consider whether the optionality of a question interacts with the interventions. In regression \Ai, we regress on both the question number and on \verb|is\_optional|, whether the question number is greater than 20. Interestingly, the interaction terms recover the hypothesized effects. For optional questions (compared to non-optional questions):
\begin{itemize}
    \item Advice with no justification is given 14.0\% 95 CI[1.0\%, 26.9\%] greater weight if it comes from an advisor.
    \item When justifications are provided, advice from the AI advisor is given an insignificant 2.1\% greater weight on advice.
\end{itemize}

These effects largely persist after adding all of the other controls in \Aii. Taken together, this exploratory analysis suggests that engagement mediates the effects of interest.

\begin{table}[H]
    \centering
    \caption{Results of additional regressions on question optionality.}
    \begin{center}
\begin{tabular}{llll}
\toprule
                                                                  & \A        & \Ai       & \Aii         \\
\midrule
Intercept                                                         & 0.329*** & 0.323*** & -0.174*    \\
                                                                  & (0.044)  & (0.044)  & (0.096)    \\
advice\_accuracy\_belief                                          &          &          & 0.619***   \\
                                                                  &          &          & (0.089)    \\
usage\_level                                                      &          &          & 0.046*     \\
                                                                  &          &          & (0.025)    \\
topic\_familiarity[T.Uncomfortable]                               &          &          & 0.065***   \\
                                                                  &          &          & (0.020)    \\
topic\_familiarity[T.Neutral]                                     &          &          & 0.027      \\
                                                                  &          &          & (0.017)    \\
question\_num                                                     &          &          & -0.005***  \\
                                                                  &          &          & (0.001)    \\
question\_id Var                                                  & 0.044**  & 0.046**  & 0.048***   \\
                                                                  & (0.018)  & (0.018)  & (0.018)    \\
participant\_id Var                                               & 0.364*** & 0.367*** & 0.340***   \\
                                                                  & (0.056)  & (0.056)  & (0.053)    \\
is\_optional[T.True]                                              &          & 0.080    & 0.166***   \\
                                                                  &          & (0.056)  & (0.057)    \\
give\_justification[T.yes]:is\_optional[T.True]                   &          & -0.147** & -0.134**   \\
                                                                  &          & (0.067)  & (0.066)    \\
give\_justification[T.yes]                                        & 0.057    & 0.072    & 0.084      \\
                                                                  & (0.058)  & (0.058)  & (0.056)    \\
advisor[T.expert]:is\_optional[T.True]                            &          & -0.140** & -0.140**   \\
                                                                  &          & (0.066)  & (0.065)    \\
advisor[T.expert]:give\_justification[T.yes]:is\_optional[T.True] &          & 0.161*   & 0.115      \\
                                                                  &          & (0.083)  & (0.082)    \\
advisor[T.expert]:give\_justification[T.yes]                      & -0.027   & -0.046   & -0.029     \\
                                                                  & (0.079)  & (0.080)  & (0.077)    \\
advisor[T.expert]                                                 & -0.027   & -0.011   & 0.080      \\
                                                                  & (0.059)  & (0.060)  & (0.138)    \\
advice\_is\_correct                                               &          &          & 0.030**    \\
                                                                  &          &          & (0.015)    \\
advice\_accuracy\_belief:advisor[T.expert]                        &          &          & -0.101     \\
                                                                  &          &          & (0.120)    \\
usage\_level:advisor[T.expert]                                    &          &          & -0.020     \\
                                                                  &          &          & (0.035)    \\
\bottomrule
\end{tabular}
\end{center}
    \label{tab:optionality_regression_results}
\end{table}

%%%%%%%%%%%%%%%%%%%
% TOPIC FAMILIARITY
%%%%%%%%%%%%%%%%%%%
\newcommand{\Bi}{\textbf{\textcolor{Periwinkle}{B2}}}
\newcommand{\Bii}{\textbf{\textcolor{MidnightBlue}{B3}}}

\subsection{Topic Familiarity}
\label{sec:appendix_topic_familiarity}

\begin{figure}[H]
  \centering
  \includegraphics[width=.6\textwidth]{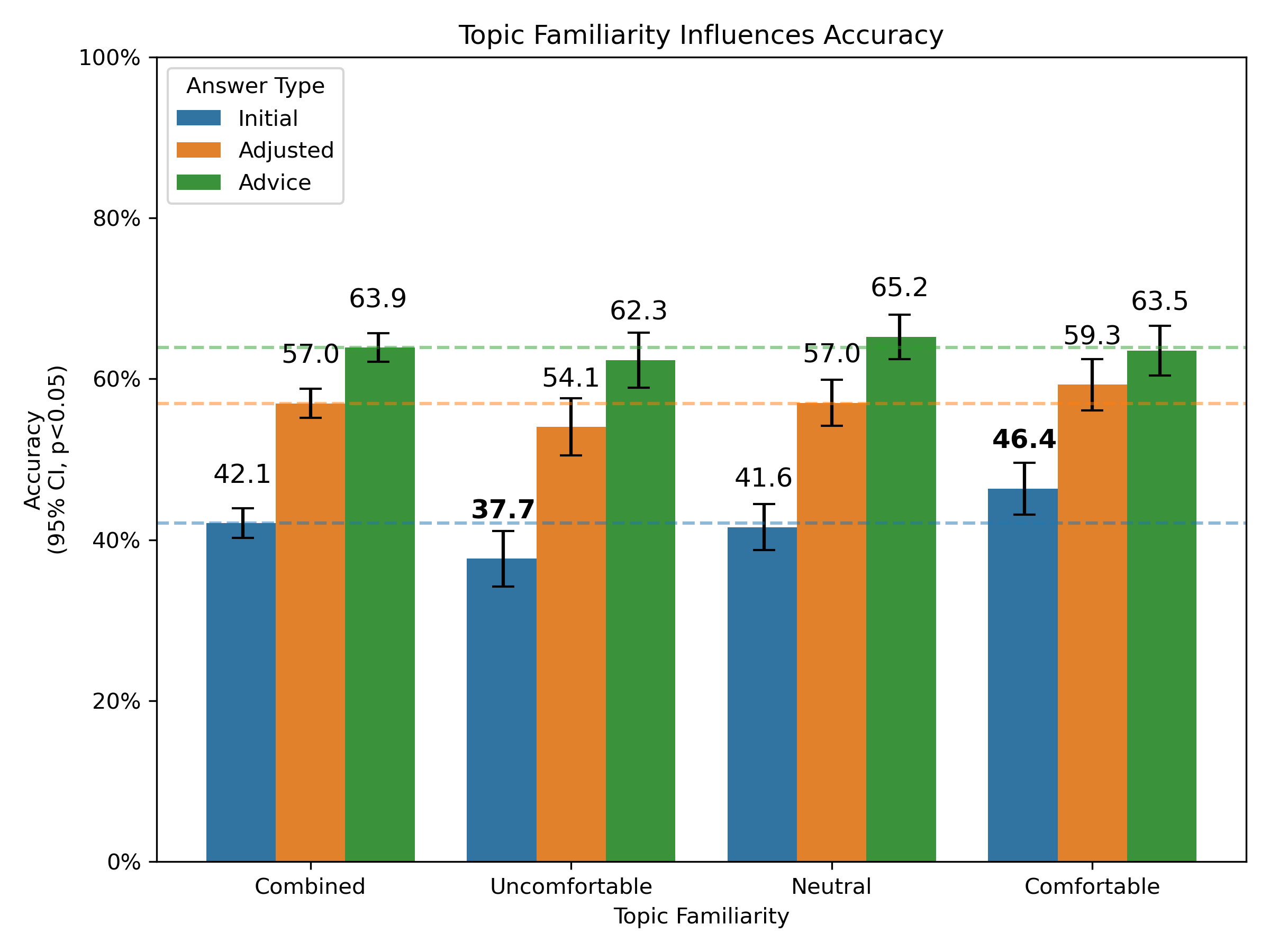}
  \caption{\textbf{Effect of topic familiarity on accuracy.} Dotted line corresponds to combined accuracy. \textbf{Bolded} values are significantly different at a 95\% significance level than the combined accuracy in a $t$-test.}
  \label{fig:accuracy_by_answer_type_and_comfort_level}
\end{figure}

Topic familiarity tangibly affects participant performance. Participants are assumed to pick their highest confidence answer and randomly pick when there is a tie. Across all questions, participants achieve an accuracy of 42.1\% before advice that improves to 57\% after receiving advice. Both significantly underperform GPT’s 63.9\% accuracy. When divided by topic familiarity, participant initial accuracy on uncomfortable topics (37.7\%) and comfortable topics (46.4\%) is significantly worse and better than the combined baseline, respectively (Figure \ref{fig:accuracy_by_answer_type_and_comfort_level}). These results suggest that participant’s subjective judgements of their familiarity are predictive.

Two additional regressions explore the effect of topic familiarity (Table \ref{tab:familiarity_regression_results}).  specification \Bi tests whether the effect of topic familiarity is captured by differences in weight on advice for each topic.The effect is attenuated but continues to be significant. Specification \Bii further examines whether being uncomfortable with a topic causes greater weight on AI chatbot advice. Topic familiarity drives greater weight in advice more in the AI chatbot condition than in the human expert condition, but not significantly so.

\begin{table}[H]
    \centering
    \caption{Results of additional regressions on topic familiarity.}
    \begin{tabular}{llll}
\toprule
                                                      & \B        & \Bi       & \Bii        \\
\midrule
Intercept                                             & 0.302*** & 0.311*** & 0.299***  \\
                                                      & (0.045)  & (0.052)  & (0.048)   \\
advisor[T.expert]                                     & -0.028   & -0.028   & -0.012    \\
                                                      & (0.059)  & (0.066)  & (0.064)   \\
advisor[T.expert]:give\_justification[T.yes]          & -0.021   & -0.024   & -0.025    \\
                                                      & (0.079)  & (0.088)  & (0.080)   \\
give\_justification[T.yes]                            & 0.056    & 0.060    & 0.058     \\
                                                      & (0.058)  & (0.064)  & (0.058)   \\
participant\_id Var                                   & 0.362*** & 0.479*** & 0.367***  \\
                                                      & (0.055)  & (0.091)  & (0.056)   \\
question\_id Var                                      & 0.040**  & 0.055*** & 0.026     \\
                                                      & (0.018)  & (0.021)  & (0.017)   \\
topic Var                                             &          & 0.051    & 0.014**   \\
                                                      &          & (nan)    & (0.007)   \\
topic\_familiarity[T.Neutral]                         & 0.026    & 0.020    & 0.028     \\
                                                      & (0.018)  & (0.018)  & (0.026)   \\
topic\_familiarity[T.Neutral]:advisor[T.expert]       &          &          & -0.010    \\
                                                      &          &          & (0.035)   \\
topic\_familiarity[T.Uncomfortable]                   & 0.061*** & 0.041*   & 0.069**   \\
                                                      & (0.020)  & (0.021)  & (0.029)   \\
topic\_familiarity[T.Uncomfortable]:advisor[T.expert] &          &          & -0.040    \\
                                                      &          &          & (0.040)   \\
\bottomrule
\end{tabular}
    \label{tab:familiarity_regression_results}
\end{table}

%%%%%%%%%%%%%%%%%%%
% PAST USAGE}
%%%%%%%%%%%%%%%%%%%
\newcommand{\Ci}{\textbf{\textcolor{LimeGreen}{C2}}}
\newcommand{\Cii}{\textbf{\textcolor{PineGreen}{C3}}}
\newcommand{\Ciii}{\textbf{\textcolor{ForestGreen}{C4}}}

\subsection{Past Usage}
\label{sec:appendix_past_usage}

Several additional regressions test the strength of the effect of past usage (Table \ref{tab:past_usage_regression_results}). Specification \Ci removes the interaction term with the advisor. The effect is still significant but is diminished in size. In specification \Cii, indicators are added for each usage level. While the coefficients are directionally sensible, none are significant. Specification \Ciii tests for the effect of having used AI chatbots before (\verb|used|). There is a large and significant effect on weight on advice, suggesting that using chatbots before leads participants in the AI chatbot condition to place 0.149 greater weight on advice.

\begin{table}[H]
    \centering
    \caption{Results of additional regressions on past usage.}
    \begin{tabular}{lllll}
\toprule
                                             & \C        & \Ci       & \Cii       & \Ciii        \\
\midrule
C(usage\_level)[T.1]                         &          &          & 0.117    &           \\
                                             &          &          & (0.160)  &           \\
C(usage\_level)[T.2]                         &          &          & 0.230    &           \\
                                             &          &          & (0.157)  &           \\
C(usage\_level)[T.3]                         &          &          & 0.189    &           \\
                                             &          &          & (0.156)  &           \\
C(usage\_level)[T.4]                         &          &          & 0.241    &           \\
                                             &          &          & (0.156)  &           \\
Intercept                                    & 0.169**  & 0.211*** & 0.104    & 0.183***  \\
                                             & (0.081)  & (0.066)  & (0.159)  & (0.071)   \\
advisor[T.expert]                            & 0.056    & -0.036   & -0.036   & 0.036     \\
                                             & (0.118)  & (0.059)  & (0.060)  & (0.107)   \\
advisor[T.expert]:give\_justification[T.yes] & -0.025   & -0.019   & -0.020   & -0.020    \\
                                             & (0.079)  & (0.078)  & (0.079)  & (0.078)   \\
give\_justification[T.yes]                   & 0.064    & 0.061    & 0.062    & 0.057     \\
                                             & (0.057)  & (0.057)  & (0.057)  & (0.057)   \\
participant\_id Var                          & 0.354*** & 0.353*** & 0.354*** & 0.350***  \\
                                             & (0.055)  & (0.054)  & (0.055)  & (0.054)   \\
question\_id Var                             & 0.040**  & 0.040**  & 0.040**  & 0.040**   \\
                                             & (0.018)  & (0.018)  & (0.018)  & (0.018)   \\
topic\_familiarity[T.Neutral]                & 0.027    & 0.027    & 0.026    & 0.026     \\
                                             & (0.018)  & (0.018)  & (0.018)  & (0.018)   \\
topic\_familiarity[T.Uncomfortable]          & 0.062*** & 0.062*** & 0.061*** & 0.063***  \\
                                             & (0.020)  & (0.020)  & (0.020)  & (0.020)   \\
usage\_level                                 & 0.050*   & 0.034*   &          &           \\
                                             & (0.025)  & (0.018)  &          &           \\
usage\_level:advisor[T.expert]               & -0.033   &          &          &           \\
                                             & (0.036)  &          &          &           \\
used[T.True]                                 &          &          &          & 0.149**   \\
                                             &          &          &          & (0.069)   \\
used[T.True]:advisor[T.expert]               &          &          &          & -0.088    \\
                                             &          &          &          & (0.105)   \\
\bottomrule
\end{tabular}
    \label{tab:past_usage_regression_results}
\end{table}

%%%%%%%%%%%%%%%%%%%
% ADVICE QUALITY
%%%%%%%%%%%%%%%%%%%

\newcommand{\Di}{\textbf{\textcolor{Dandelion}{D2}}}
\newcommand{\Dii}{\textbf{\textcolor{Tan}{D3}}}
\newcommand{\Diii}{\textbf{\textcolor{RawSienna}{D4}}}
\newcommand{\Div}{\textbf{\textcolor{Mahogany}{D5}}}

\subsection{Advice Quality}
\label{sec:appendix_advice_quality}

The effect of advice quality/correctness may masked by misplaced trust in the advice answer. Suppose participants are modest and tend to put between 80\% confidence in the advisor’s answer regardless of their initial answer. Further suppose that participants are capable and tend to pick the same answer as the advisor. When the advice is correct, they are likely to place greater weight on the advice answer initially and under-adjust proportionately. When the advice is incorrect, they are likely to move more in the direction of the advice answer. Perversely, this would suggest that advice correctness is negatively correlated with weight on advice.

This is controlled for by directly regressing on initial confidence in the advice answer (\verb|init_advice_confidence|) in a series of regressions (Table \ref{tab:correctness_regression_results}). Compared to the original specification \D, specification \Di\ adds a term that controls for initial advice confidence. Question random effects are no longer significant after controlling for initial confidence, suggesting that initial confidence indeed controls for question difficulty. Next, adding a quadratic term is justified, as the coefficients in \Dii\ remain significant. An interaction term with giving justifications in \Diii suggests that the effect of advice correctness depends on giving justifications.

These results suggest a natural explanation in which correct advice is more convincing because the justifications are more coherent. If true, one would expect people to spend more time reading when there is a justification. Indeed, participants that receive justifications spend 2.5 95\% CI[1.24\%, 3.76\%] more seconds on adjusting answers, `adjusted\_time`. Specification \Div regresses on `adjusted\_time` with an interaction term  with `advice\_is\_correct`. Consistent with the hypothesis, the interaction of advice being correct and time spent incorporating advice explains most of the effect. Every additional 10 seconds spent incorporating advice is associated with a 3.95\% 95\% CI[1.16\%, 6.74\%] increase in weight on advice.

\begin{table}[H]
    \centering
    \caption{Results of additional regressions on advice correctness.}
    \begin{tabular}{llllll}
\toprule
                                                       & \D         & \Di        & \Dii        & \Diii      & \Div   \\
\midrule
I(init\_advice\_confidence ** 2)               &          &           & -1.260*** & -1.261*** & -1.250***  \\
                                               &          &           & (0.071)   & (0.071)   & (0.071)    \\
Intercept                                      & 0.161**  & 0.234***  & 0.098     & 0.118     & 0.134      \\
                                               & (0.082)  & (0.084)   & (0.094)   & (0.092)   & (0.092)    \\
adjusted\_time                                 &          &           &           &           & -0.002     \\
                                               &          &           &           &           & (0.001)    \\
advice\_is\_correct                            & 0.013    & 0.053***  & 0.047***  & 0.018     & -0.020     \\
                                               & (0.015)  & (0.014)   & (0.013)   & (0.019)   & (0.024)    \\
advice\_is\_correct:adjusted\_time             &          &           &           &           & 0.004***   \\
                                               &          &           &           &           & (0.001)    \\
advice\_is\_correct:give\_justification[T.yes] &          &           &           & 0.051**   & 0.041      \\
                                               &          &           &           & (0.025)   & (0.025)    \\
advisor[T.expert]                              & 0.054    & 0.051     & 0.044     & 0.044     & 0.043      \\
                                               & (0.118)  & (0.120)   & (0.136)   & (0.131)   & (0.131)    \\
advisor[T.expert]:give\_justification[T.yes]   & -0.025   & -0.031    & -0.058    & -0.059    & -0.058     \\
                                               & (0.079)  & (0.080)   & (0.091)   & (0.088)   & (0.088)    \\
give\_justification[T.yes]                     & 0.064    & 0.065     & 0.071     & 0.039     & 0.043      \\
                                               & (0.057)  & (0.058)   & (0.066)   & (0.066)   & (0.066)    \\
init\_advice\_confidence                       &          & -0.274*** & 0.984***  & 0.984***  & 0.981***   \\
                                               &          & (0.022)   & (0.074)   & (0.074)   & (0.074)    \\
participant\_id Var                            & 0.353*** & 0.389***  & 0.577***  & 0.528***  & 0.531***   \\
                                               & (0.055)  & (0.060)   & (0.093)   & (0.080)   & (0.080)    \\
question\_id Var                               & 0.040**  & 0.027     & 0.021     & 0.005     & 0.007      \\
                                               & (0.018)  & (0.017)   & (0.018)   & (0.016)   & (0.016)    \\
topic\_familiarity[T.Neutral]                  & 0.027    & 0.025     & 0.012     & 0.011     & 0.010      \\
                                               & (0.018)  & (0.017)   & (0.016)   & (0.016)   & (0.016)    \\
topic\_familiarity[T.Uncomfortable]            & 0.062*** & 0.050**   & 0.019     & 0.019     & 0.019      \\
                                               & (0.020)  & (0.020)   & (0.019)   & (0.019)   & (0.019)    \\
usage\_level                                   & 0.049*   & 0.051*    & 0.045     & 0.045     & 0.045      \\
                                               & (0.025)  & (0.026)   & (0.029)   & (0.028)   & (0.028)    \\
usage\_level:advisor[T.expert]                 & -0.033   & -0.031    & -0.025    & -0.025    & -0.025     \\
                                               & (0.036)  & (0.037)   & (0.042)   & (0.041)   & (0.041)    \\
\bottomrule
\end{tabular}
    \label{tab:correctness_regression_results}
\end{table}

%%%%%%%%%%%%%%%%%%%
% EXPERIENCE
%%%%%%%%%%%%%%%%%%%

\newcommand{\Ei}{\textbf{\textcolor{Lavender}{E2}}}
\newcommand{\Eii}{\textbf{\textcolor{Orchid}{E3}}}
\newcommand{\Eiii}{\textbf{\textcolor{RedViolet}{E4}}}
\newcommand{\Eiv}{\textbf{\textcolor{Thistle}{E5}}}
\newcommand{\Ev}{\textbf{\textcolor{RubineRed}{E6}}}
\newcommand{\Evi}{\textbf{\textcolor{WildStrawberry}{E7}}}
\newcommand{\Evii}{\textbf{\textcolor{Violet}{E8}}}

\subsection{Experience}
\label{sec:appendix_experience}

Several additional regressions are conducted to assess how sensitive the finding is to the model of participant beliefs (Table \ref{tab:experience_specifications}). The original specification \E\ uses a uninformative prior $\textrm{Beta}(\alpha=0.5,\beta=0.5)$ to model experiences. Specification \Ei tests a term that indicates whether the last piece of advice given is correct (\verb|last_advice_is_correct|). The  advice accuracy belief term is reintroduced in \Eii. This specification tests whether recency bias leads participants to overweight recent advice. Specifications \Eiii-\Ev vary the strength of the uninformative prior. Specifications \Evi\ and \Evii modify the prior's estimate to correspond to accuracy beliefs of 25\% and 75\%, respectively.

\begin{table}[H]
    \centering
    \caption{Specifications for additional regressions on participant beliefs.}
    \begin{tabular}{ccccccccc}
\toprule
 & \multicolumn{8}{c}{\textit{Specification}}                   \\
\cmidrule(r){2-9}
           & \E          & \Ei          & \Eii         & \Eiii       & \Eiv       & \Ev      & \Evi & \Evii\\
\midrule
$\alpha$ prior & 0.5   & \NA   & 0.5  &  0.05 & 0.2   & 1  & 0.25   & 0.75    \\
$\beta$ prior& 0.5   & \NA   & 0.5  &  0.05 & 0.2   & 1  & 0.75   & 0.25 \\
Last question? & \xmark & \cmark  & \cmark  & \xmark & \xmark & \xmark & \xmark  & \xmark \\
Beliefs? & \cmark & \xmark  & \cmark  & \cmark & \cmark & \cmark & \cmark  & \cmark \\
\bottomrule
\end{tabular}
    \label{tab:experience_specifications}
\end{table}

The results of the regressions are displayed in Table \ref{tab:experience_regression_results}.If the last advice given was correct, participants place 8.7\% 95 CI[4.8\%, 12.4\%] greater weight on advice in the AI advisor condition, but only 4.2\% 95 CI[0.6\%, 7.8\%] in the human expert advisor condition. These results are subsumed after reintroducing the beliefs term. The results are robust to last advice correctness, prior strength, and prior estimate. Regardless of the prior, the coefficient on beliefs is lower for the human expert condition, corroborating the previous finding.

\newpage
\begin{landscape}
\begin{table}[H]
    \centering
    \caption{Specifications for additional regressions on participant beliefs.}
    \begin{tabular}{lllllllll}
\toprule
                                                    & \E        & \Ei       & \Eii       & \Eiii       & \Eiv       & \Ev       & \Evi       & \Evii \\
\midrule
Intercept                                           & -0.159*   & 0.108    & -0.174*  & -0.097   & -0.139   & -0.252** & -0.050   & -0.247**  \\
                                                    & (0.096)   & (0.082)  & (0.097)  & (0.089)  & (0.092)  & (0.101)  & (0.090)  & (0.098)   \\
advice\_accuracy\_belief                            & 0.602***  &          & 0.502*** & 0.405*** & 0.475*** & 0.674*** & 0.361*** & 0.616***  \\
                                                    & (0.088)   &          & (0.099)  & (0.067)  & (0.075)  & (0.104)  & (0.073)  & (0.089)   \\
advice\_accuracy\_belief:advisor[T.expert]          & -0.097    &          & -0.043   & -0.047   & -0.077   & -0.201   & -0.146   & -0.033    \\
                                                    & (0.120)   &          & (0.134)  & (0.091)  & (0.102)  & (0.141)  & (0.099)  & (0.121)   \\
advice\_is\_correct[T.True]                         & 0.029**   & 0.015    & 0.029**  & 0.028*   & 0.029**  & 0.030**  & 0.023    & 0.032**   \\
                                                    & (0.014)   & (0.014)  & (0.015)  & (0.015)  & (0.015)  & (0.015)  & (0.015)  & (0.014)   \\
advisor[T.expert]                                   & 0.067     & 0.079    & 0.063    & 0.041    & 0.058    & 0.136    & 0.116    & 0.022     \\
                                                    & (0.137)   & (0.118)  & (0.139)  & (0.129)  & (0.132)  & (0.144)  & (0.129)  & (0.141)   \\
advisor[T.expert]:give\_justification[T.yes]        & -0.023    & -0.026   & -0.019   & -0.017   & -0.017   & -0.019   & -0.022   & -0.016    \\
                                                    & (0.076)   & (0.078)  & (0.076)  & (0.076)  & (0.076)  & (0.076)  & (0.076)  & (0.076)   \\
give\_justification[T.yes]                          & 0.074     & 0.067    & 0.070    & 0.067    & 0.068    & 0.070    & 0.067    & 0.070     \\
                                                    & (0.055)   & (0.057)  & (0.055)  & (0.055)  & (0.055)  & (0.055)  & (0.055)  & (0.055)   \\
last\_advice\_is\_correct[T.True]                   &           & 0.084*** & 0.035    &          &          &          &          &           \\
                                                    &           & (0.020)  & (0.022)  &          &          &          &          &           \\
last\_advice\_is\_correct[T.True]:advisor[T.expert] &           & -0.046*  & -0.042   &          &          &          &          &           \\
                                                    &           & (0.027)  & (0.030)  &          &          &          &          &           \\
participant\_id Var                                 & 0.337***  & 0.348*** & 0.333*** & 0.334*** & 0.333*** & 0.331*** & 0.332*** & 0.339***  \\
                                                    & (0.052)   & (0.054)  & (0.052)  & (0.052)  & (0.052)  & (0.052)  & (0.052)  & (0.053)   \\
question\_id Var                                    & 0.045**   & 0.042**  & 0.043**  & 0.044**  & 0.043**  & 0.042**  & 0.042**  & 0.043**   \\
                                                    & (0.018)   & (0.018)  & (0.018)  & (0.018)  & (0.018)  & (0.018)  & (0.018)  & (0.018)   \\
question\_num                                       & -0.004*** &          &          &          &          &          &          &           \\
                                                    & (0.001)   &          &          &          &          &          &          &           \\
topic\_familiarity[T.Neutral]                       & 0.026     & 0.029*   & 0.029    & 0.028    & 0.028    & 0.028    & 0.029*   & 0.025     \\
                                                    & (0.017)   & (0.018)  & (0.017)  & (0.017)  & (0.017)  & (0.017)  & (0.018)  & (0.017)   \\
topic\_familiarity[T.Uncomfortable]                 & 0.063***  & 0.062*** & 0.063*** & 0.064*** & 0.063*** & 0.064*** & 0.063*** & 0.063***  \\
                                                    & (0.020)   & (0.020)  & (0.020)  & (0.020)  & (0.020)  & (0.020)  & (0.020)  & (0.020)   \\
usage\_level                                        & 0.045*    & 0.049*   & 0.047*   & 0.047*   & 0.047*   & 0.046*   & 0.048*   & 0.046*    \\
                                                    & (0.024)   & (0.025)  & (0.024)  & (0.024)  & (0.024)  & (0.024)  & (0.025)  & (0.025)   \\
usage\_level:advisor[T.expert]                      & -0.019    & -0.031   & -0.022   & -0.022   & -0.022   & -0.023   & -0.027   & -0.019    \\
                                                    & (0.035)   & (0.036)  & (0.035)  & (0.035)  & (0.035)  & (0.035)  & (0.035)  & (0.035)   \\
\bottomrule
\end{tabular}
    \label{tab:experience_regression_results}
\end{table}
\end{landscape}
\newpage

%%%%%%%%%%%%%%%%%%%
% MANIPULATION EFFICACY
%%%%%%%%%%%%%%%%%%%

\subsection{Manipulation Efficacy}
\label{sec:appendix_manipulation}

\begin{figure}[H]
  \centering
  \includegraphics[width=.7\textwidth]{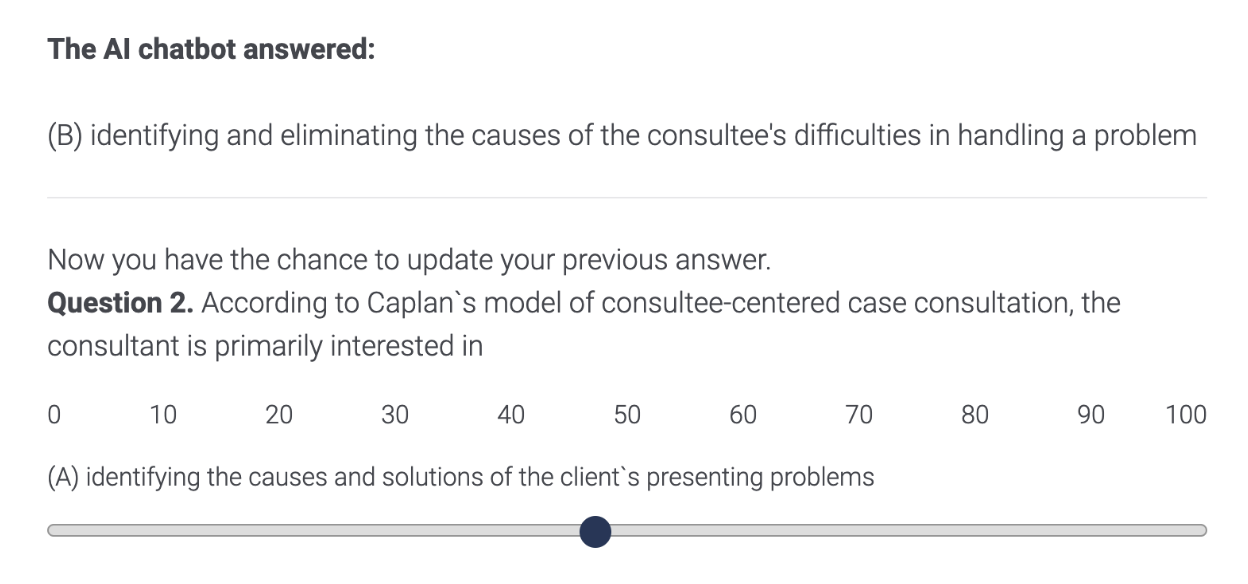}
  \caption{\textbf{Advice format.} The screenshot is taken from a Qualtrics survey and reflects what participants would've seen.}
  \label{fig:advice_format}
\end{figure}

One plausible criticism of the results is that participants may not have remembered the manipulation. The survey design attempts to mitigate this in two ways. First, participants must pass a manipulation check that confirms the identity of the advisor. Second, each feedback page explicitly repeats the advisor identity (Figure \ref{fig:advice_format}).

There is additional anecdotal evidence that the manipulation has the desired effect. One participant in the expert condition was confused about the manipulation. They emailed to say that they ``\textit{personally assumed that a different expert was asked for each question for their guess at the correct answer (for example, professors from different departments from Berkeley [...])}'', suggesting that ``\textit{the respondent will typically assume (at least that was the case for me), that the expert is human.}'' Other anecdotal comments confirm this assumption.

%%%%%%%%%%%%%%%%%%%
% MISCALIBRATION
%%%%%%%%%%%%%%%%%%%

\subsection{Persistence of Miscalibration}
\label{sec:appendix_miscalibration}

\begin{figure}[H]
  \centering
  \includegraphics[width=.6\textwidth]{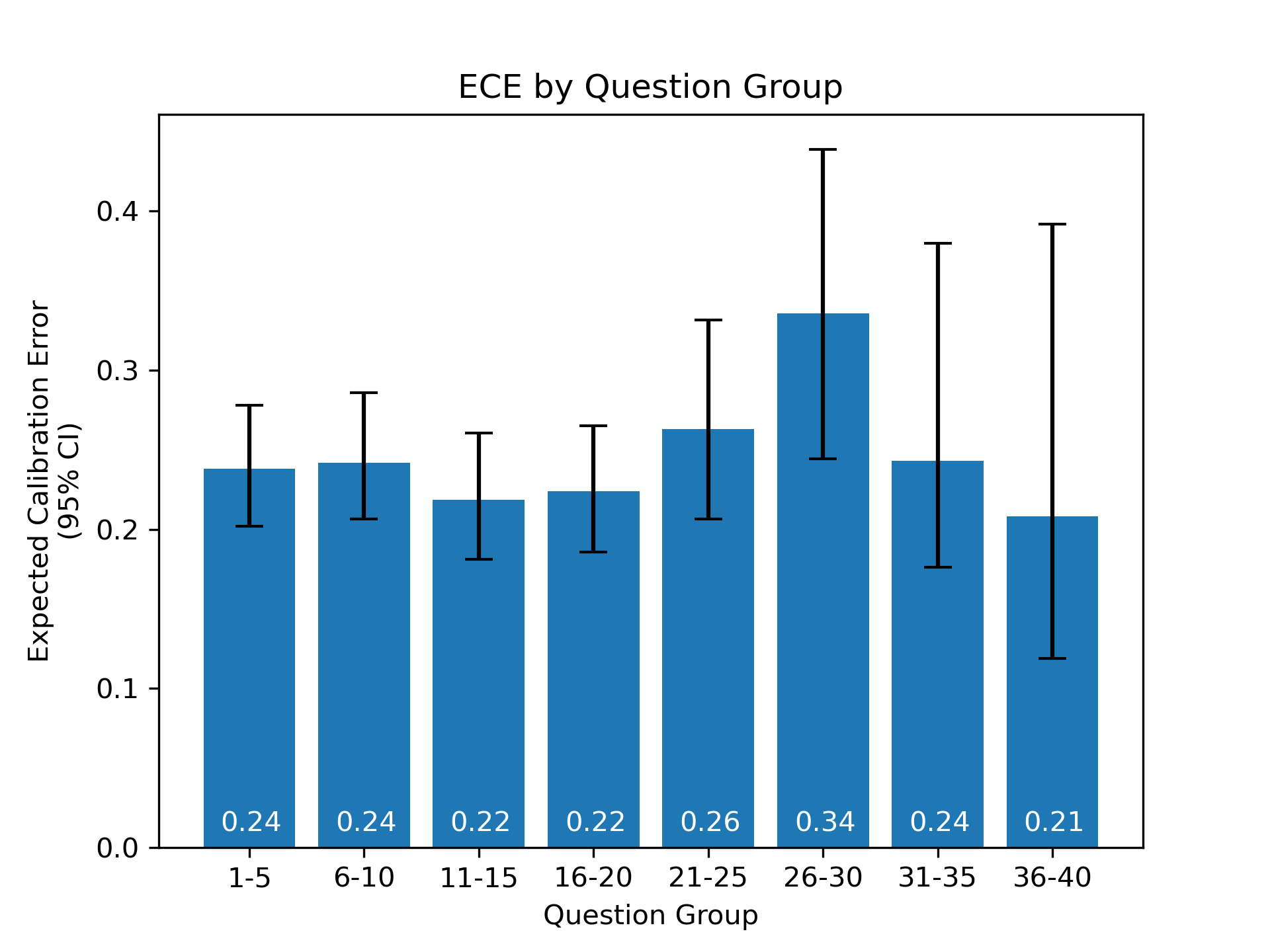}
  \caption{Expected calibration error over questions.}
  \label{fig:ece_by_question_group}
\end{figure}

The miscalibration for advised choices does not change significantly over time. Figure \ref{fig:ece_by_question_group} suggests a small decrease in $\ECE_{\mathrm{advised}}=0.201$ from questions 6-10 to questions 11-15. For the optional questions, participants are more miscalibrated, but the standard errors are much larger. None of these effects are significant at a 95\% level.

%%%%%%%%%%%%%%%%%%%
% INEFFICIENCY
%%%%%%%%%%%%%%%%%%%

\subsection{Sources of Inefficiency}
\label{sec:appendix_inefficiency}

Several habits of participants worsen their actual average score ($\overline{\BS}=0.674$) compared to a uniform baseline ($\BS \approx 0.588$) for the same $\WoA$. There are two principal mistakes.

\paragraph{Misallocation}
Misallocation is the tendency to poorly adjust other answers after receiving advice. After extending additional weight to one choice, participants often rescale the other answers in inefficient ways. For example, participants might assign zero weight to an option they initially assigned a little weight. Consider an alternative in which participants proportionately decrease their confidence in each of the other choices. Compared to the actual average Brier score $\overline{\BS}=0.674$, participants would have earned $\hat{\BS}=0.634$, a modest improvement.

\paragraph{Extremism}
Extremism is the tendency for participants put both too much and too little weight on advice. What if—holding the proportion of weight on the other adjusted answers the same—participants moved more consistently in the direction of the advice? This behavior is parameterized by a shrinking parameter $s$ which proportionately “shrinks” each question’s weight on advice closer to the sample average, $\overline{\WoA}$. Specifically, for each question i:

$$\WoA \leftarrow \WoA_i+(1-s)\overline{\WoA}$$

\begin{figure}[H]
  \centering
  \includegraphics[width=.6\textwidth]{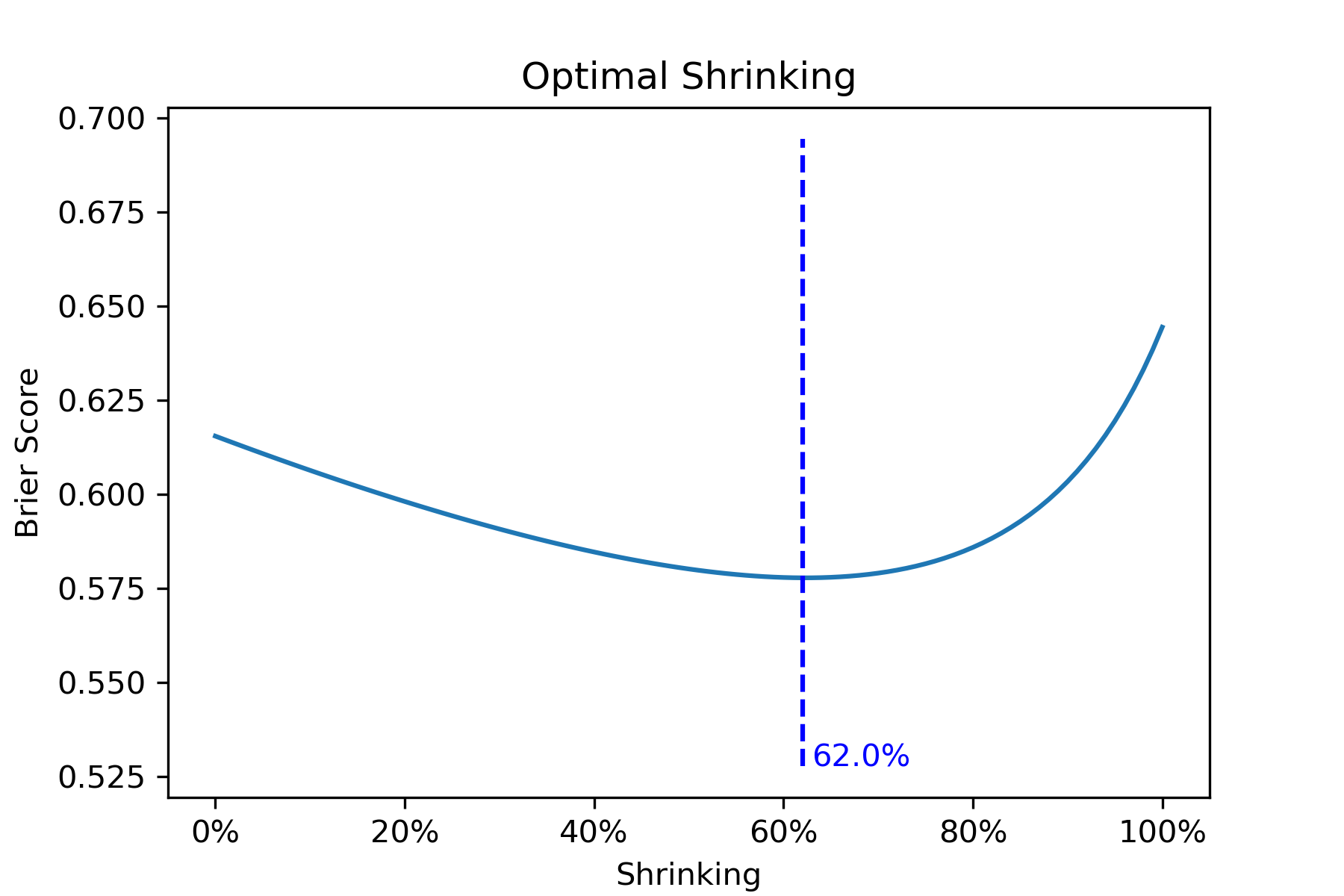}
  \caption{\textbf{Optimal shrinking.}}
  \label{fig:shrinking_brier_score}
\end{figure}

Otherwise, the relative ratio of the other adjusted choices is preserved. When originally there is full confidence in the advised answer, confidence is uniformly allocated over answers. The optimal shrinking parameter is $s^*$=0.62, achieving a score superior to uniform allocation at $\tilde{\BS}$=0.578. The relationship between scaling and Brier Score is quadratic and displayed in Figure \ref{fig:shrinking_brier_score}.

If the adjustment for misallocation is taken after depolarization, then the new optimal shrinking factor is $s^*=0$. Taken together, these results suggest that extremism drives more of the inefficiency than malapportionment.

%%%%%%%%%%%%%%%%%%%
% QUALITATIVE
%%%%%%%%%%%%%%%%%%%

\subsection{Qualitative Findings}
\label{sec:appendix_qualitative}

Participants who said they used ChatGPT in the classroom were prompted to clarify how they used the tool. Although they were assured that the information would not be identifiable, these may not be honest representations of their usage. Nonetheless, they offer a glimpse into how students use ChatGPT in practice. Among other applications, students used ChatGPT to:
\begin{itemize}
\item Explain concepts and develop understanding, for homework, projects, and studying.
\item Brainstorm, outline, and draft essays.
\item Complete coding assignments or write code snippets.
\item Complete multiple choice quizzes.
\item Summarizes lengthy content.
\item Preliminary research for essays.
\item Validate answers.
\end{itemize}

\begin{figure}[H]
  \centering
  \includegraphics[width=\textwidth]{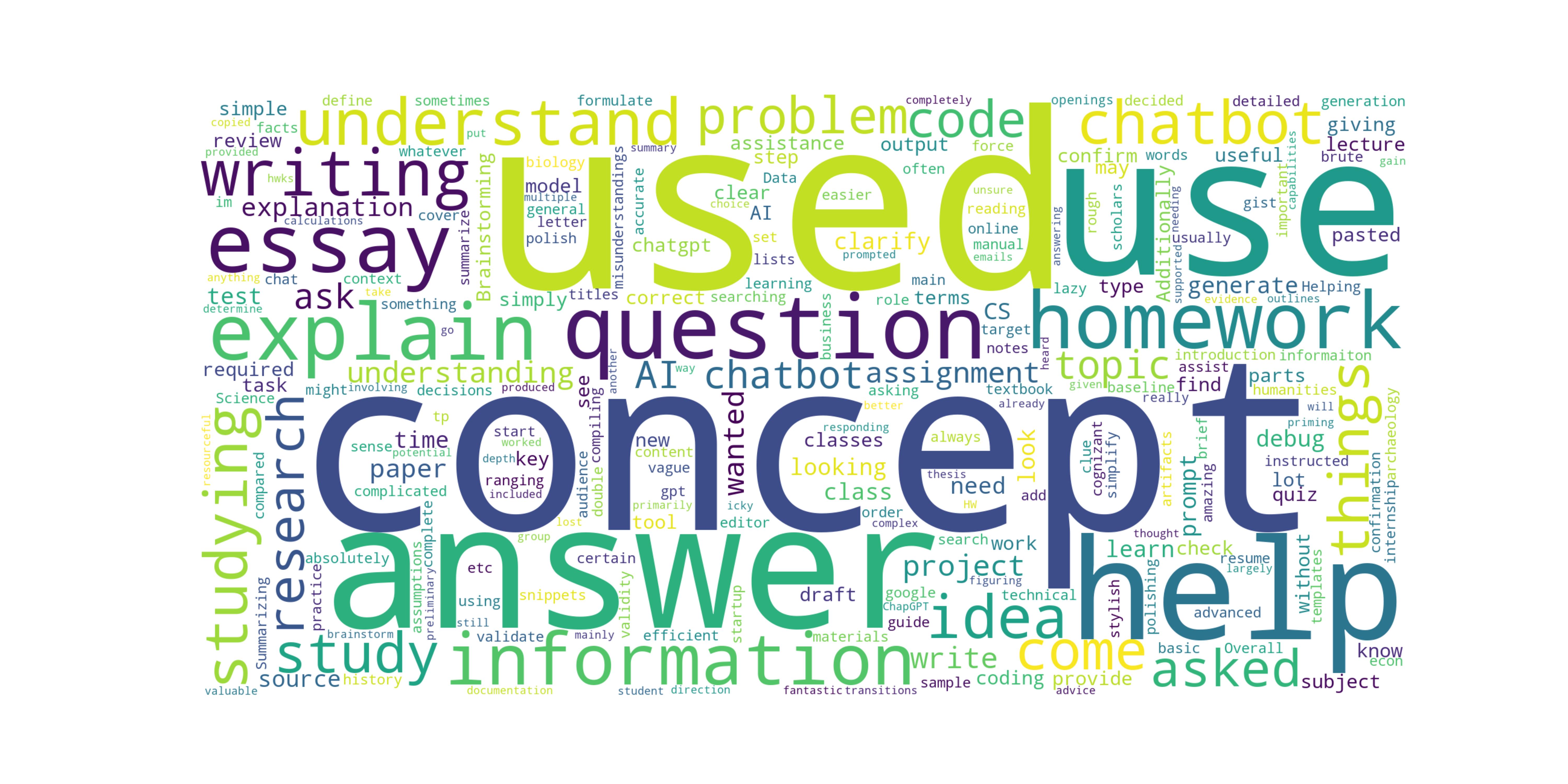}
  \caption{\textbf{Word cloud of ChatGPT usages.} The size of the phrase roughly corresponds to frequency. The most commonly mentioned use was learning "concepts."}
  \label{fig:usage_wordcloud}
\end{figure}

The range of responses suggests a broad potential for ChatGPT in the classroom. A visualization of commonly used terms is displayed in Figure \ref{fig:usage_wordcloud}.

%%%%%%%%%%%%%%%%%%%
% POWER ANALYSIS
%%%%%%%%%%%%%%%%%%%

\subsection{Power Analysis}
\label{sec:appendix_power}

A power analysis is conducted for an unpaired $t$-test on each participant’s average weight on advice across questions. Setting $\alpha$=0.05 and power $1-\beta=0.8$, the sample is able to detect an effect size of Cohen’s $d$=0.336. For the advisor condition, the effect size in the sample is $d$=0.195 requiring a sample size of $n$=412 to detect an effect. For the give justification conditions, the effect size is $d$=1.90 and it requires a sample of $n$=436 participants.

\end{document}